\documentclass[8.5pt,twoside,twocolumn,floatfix]{article}
\usepackage[super,sort&compress,comma]{natbib} 
\usepackage{mhchem}
\usepackage{times,mathptmx}
\usepackage{sectsty}
\usepackage{balance} 

\usepackage{amsmath,amssymb}
\usepackage{graphicx} 
\graphicspath{{./figs/},{./Fig/}}
\DeclareGraphicsExtensions{.pdf,.png,.jpg}
\usepackage{lastpage}
\usepackage[format=plain,justification=raggedright,singlelinecheck=false,font=small,labelfont=bf,labelsep=space]{caption} 
\usepackage{fancyhdr}
\usepackage{bm}
\newcommand{\no}[1]{}
\newcommand{\mylab}[1]{\label{#1}}
\newcommand{\diff}{{\mskip 1mu minus 1mu\mathrm{d}}}
\newcommand{\xshift}{{x_\mathrm{shift}}}
\newcommand{\Pshift}{{P_\mathrm{shift}}}
\newcommand{\xrst}{{x_\mathrm{rst}}}

\newcommand{\hpfield}{{h_\mathrm{p}}}
\newcommand{\hpioverb}{\overline {h_{\mathrm{p}_{\scriptstyle i}}}}
\newcommand{\hpitilde}{\tilde h_{\mathrm{p}_{\scriptstyle i}}}
\newcommand{\hpihat}{\hat h_{\mathrm{p}_{\scriptstyle i}}}
\begin{document}

\thispagestyle{plain}
\fancypagestyle{plain}{
\renewcommand{\headrulewidth}{1pt}}
\renewcommand{\thefootnote}{\fnsymbol{footnote}}
\renewcommand\footnoterule{\vspace*{1pt}%
\hrule width 3.4in height 0.4pt \vspace*{5pt}} 
\setcounter{secnumdepth}{5}

\makeatletter 
\def\subsubsection{\@startsection{subsubsection}{3}{10pt}{-1.25ex plus -1ex minus -.1ex}{0ex plus 0ex}{\normalsize\bf}} 
\def\paragraph{\@startsection{paragraph}{4}{10pt}{-1.25ex plus -1ex minus -.1ex}{0ex plus 0ex}{\normalsize\textit}} 
\renewcommand\@biblabel[1]{#1}            
\renewcommand\@makefntext[1]%
{\noindent\makebox[0pt][r]{\@thefnmark\,}#1}
\makeatother 
\renewcommand{\figurename}{\small{Fig.}~}
\sectionfont{\large}
\subsectionfont{\normalsize} 

\fancyfoot{}
\fancyfoot[RO]{\footnotesize{\sffamily{1--\pageref{LastPage} ~\textbar  \hspace{2pt}\thepage}}}
\fancyfoot[LE]{\footnotesize{\sffamily{\thepage~\textbar\hspace{3.45cm} 1--\pageref{LastPage}}}}
\fancyhead{}
\renewcommand{\headrulewidth}{1pt} 
\renewcommand{\footrulewidth}{1pt}
\setlength{\arrayrulewidth}{1pt}
\setlength{\columnsep}{6.5mm}
\setlength\bibsep{1pt}

\twocolumn[
  \begin{@twocolumnfalse}
\noindent\LARGE{\textbf{Modelling the formation of structured deposits at
receding contact lines of evaporating solutions and suspensions$^\dag$}}
\vspace{0.6cm}

\noindent\large{\textbf{\v{L}ubor Fra\v{s}tia,\textit{$^{a}$} Andrew J.~Archer,\textit{$^{a}$} and
Uwe Thiele $^{\ast}$\textit{$^{a}$}}}\vspace{0.5cm}

\noindent\textit{\small{\textbf{Received Xth XXXXXXXXXX 20XX, Accepted Xth XXXXXXXXX 20XX\newline
First published on the web Xth XXXXXXXXXX 200X}}}

\noindent \textbf{\small{DOI: 10.1039/b000000x}}
\vspace{0.6cm}

\noindent \normalsize{ When a film of a liquid suspension of
  nanoparticles or a polymer solution is deposited on a surface, it
  may dewet from the surface and as the solvent evaporates the solute
  particles/polymer can be deposited on the surface in regular line
  patterns. In this paper we explore a hydrodynamic model for the
  process that is based on a long-wave approximation that predicts the
  deposition of irregular and regular line patterns. This is due to a
  self-organised pinning-depinning cycle that resembles a stick-slip
  motion of the contact line. We present a detailed analysis of how
  the line pattern properties depend on quantities such as the
  evaporation rate, the solute concentration, the P\'eclet number, the
  chemical potential of the ambient vapour, the disjoining pressure,
  and the intrinsic viscosity. The results are related to several
  experiments and to depinning transitions in other soft matter
  systems.  } \vspace{0.5cm}
 \end{@twocolumnfalse}
  ]

\footnotetext{\dag~Electronic Supplementary Information (ESI) available: [details of any supplementary information available should be included here]. See DOI: 10.1039/b000000x/}


\footnotetext{\textit{$^{a}$~Department of Mathematical Sciences, Loughborough University, Loughborough, Leicestershire, LE11 3TU, UK. }}
\footnotetext{\textit{$^{\ddag}$ E-mail: u.thiele@lboro.ac.uk; web: www.uwethiele.de}}



%
%
\section{Introduction}

Many production processes employed in the chemical, pharmaceutical and other
industries, involve a wet phase where films or drops of a solution or
suspension are applied to a solid or liquid surface with the aim
of producing a homogeneous or structured layer of the solute on the
surface. Prominent examples are printing, painting
and coating processes where a wet ink or paint is used.  The carrier
fluid (solvent) then evaporates and leaves all the originally dissolved
non-volatile material behind in various deposition patterns on the substrate. 

If the aim of these processes is to produce a large-scale homogeneous
layer, ideally one would instantaneously produce an extended stable
homogenous film of solution that then evaporates slowly and
homogeneously. Although, this may be achieved to some extent, e.g.\ by
spin coating, this situation is rather exceptional. In practice, 
what generally happens is that even while the wet film is still being
deposited onto one part of the substrate, it is already dry on other
parts. Many techniques exist for applying such films, that are relevant
for different surface types and surface geometries, such as painting,
blade coating, deposition
of individual drops, spray coating, and processes involving liquid menisci.

Historically, the main aim of these processes was to produce
homogeneous coatings, so defects like holes or patterned areas would
not have been desirable. However, this has changed as nowadays wet
deposition may also be used to produce patterned functional layers via
the self-organisation of the solute. Although a large variety of
different apparatus geometries exists, the basic physical question
always is \textit{What types of deposition patterns exist and how can
  the self-organisation processes involved in forming the patterns be
  controlled?} A close inspection of the individual systems shows that
in most of them, the formation of deposition patterns occurs in a
localised zone at receding contact lines that move under the coupled
influence of solvent evaporation, convective motion of the solution,
capillarity and wettability.

In this paper we present a long-wave hydrodynamic thin film
model that focuses on such a moving contact line. Some of the main features
of our model were briefly presented by \citet{FAT11}. However, before we discuss the
model and the wide range of detailed results that we have obtained, we first briefly review
some of the relevant experimental systems and theoretical approaches that exist
in the literature.

Interest in deposition patterns has markedly increased over the last
decade, since Deegan and co-workers' detailed investigations of the
``coffee-stain effect'', i.e., the solute deposition patterns that are
left behind by a receding three-phase contact line of an evaporating
drop of a suspension upon a smooth solid
substrate.\cite{Deeg97,Deeg00,Deeg00b} In particular \citet{Deeg00} describes
and analyses a wide range of patterns: cellular and lamellar
structures, single and multiple rings, and Sierpinski gaskets. The
creation of multiple rings through a stick-slip front motion of
colloidal liquids was also observed.\cite{ADN95,SSS02} Since then,
interest widened and now encompasses all phenomena that accompany
evaporative and convective dewetting of colloidal suspensions and
polymer or macromolecular solutions in a number of different
geometries. Other early examples are the investigation by Parisse and
Allain of the shape changes that drops of colloidal suspension undergo
when they dry \cite{PaAl96,PaAl97} and the creation of semiconductor
nanoparticle rings through a similar deposition process.\cite{MDSY99}
Other observed drying structures include crack patterns,\cite{Dufr03}
chevron patterns,\cite{Bert10} and branched
patterns.\cite{GRDK02,Paul08} In fact, crack patterns in sol-gel
processes had already been studied somewhat
earlier,\cite{Brin92,AlLi95} however, we do not consider them here.
The related concept of using evaporation at contact lines to assemble
colloidal particles or proteins into crystals actually has a longer
history -- see for example the discussions and reviews by 
\citet{DTR1930,Denk92,AdNa97,MOY03,KCR08}. 

Generally, the evaporation of a macroscopic drop does not usually
result in the deposition of regular concentric ring patterns that
could potentially be employed to fabricate devices, but rather results
in irregular patterns of rugged rings and lines (see, e.g.,
\citet{ADN95,Deeg00}).  This changes when the experiments are
performed in a controlled way on smaller scales: Recently, both
polymer solutions \cite{YaSh05,Xu06,HXL07} and (nano)particle
suspensions \cite{RDLL06,XXL07,BDG10} have been employed in various
small-scale geometries where one is able to exercise greater control
over the contact line as it recedes due to evaporation. As a result,
strikingly regular line patterns are created, where the deposited
structures show typical distances ranging from 10--100$\mu$m.  Line
patterns can be parallel or perpendicular to the receding contact line
and are produced in a robust repeatable manner in extended regions of
parameter space. Besides the lines, a variety of other patterns may
also be found, including undulated stripes, interconnected
stripes;\cite{XXL07} ladder structures, i.e.\ superpositions of
perpendicular and parallel stripes;\cite{YaSh05} regular arrays of
drops;\cite{KIS96,YaSh05} and irregularly branched
structures.\cite{KGMS99,GRDK02,LZZZ08,Paul08} The occurrence of these more
complicated patterns is highly sensitive to the particular
experimental setup and the system parameters.

Several groups employ this type of wet evaporative deposition as a
non-lithographic technique for covering large areas with regular
arrays of small-scale structures, such as, e.g., concentric gold rings
with potential uses as resonators in advanced optical communications
systems \cite{HXL06} or ordered arrays of cyanine dye complex
micro-domes employed in photo-functional surfaces.\cite{HaKa07} Often
the patterns are robust and can be post-processed, e.g., to create
double-mesh structures by crossing and stacking two ladder
films.\cite{YaSh05}
A number of investigations focus on deposition patterns resulting from
more complex fluids, such as phase separating polymer
mixtures;\cite{Byun08} solutions of DNA,\cite{Shim99,MZZC08} collagen,
\cite{Maed00} liquid crystals,\cite{YAS09} dye
molecules,\cite{DTR1930,vanH06,HaKa07} dendrimers,\cite{LTLB06} carbon
nanotubes,\cite{Hong08,Zeng11} and graphene;\cite{Kim11} and biofluids
like blood.\cite{YaYa09,BSLS11} The latter has medical implications as
it is thought that one may learn how to detect some illnesses by
simple evaporation experiments on small samples.\cite{Tara04}

Surveying the literature, one finds that deposition patterns
consisting of regular stripes are a rather generic phenomenon. They
occur for many different combinations of substances \cite{FAT12_materials}
and in a range of different experimental setups that allow for
slow evaporation. To better control the contact line motion, various
techniques are employed. One may distinguish between \textit{passive} and
\textit{active} experimental set-ups. In the passive set-up, the solution or
suspension is brought onto the substrate and left to evaporate.
Examples include (i) the so-called ``meniscus technique'' (where a
meniscus with a contact line is created), e.g., in a sphere-on-flat
\cite{Xu06,XXL07,HXL07} or ring-on-flat \cite{Denk92,PaMo07} geometry,
(ii) the deposition of a single large drop onto a substrate
\cite{ADN95,Deeg00} and (iii) the deposition of flat films onto a
substrate using spin-coating.\cite{GeBr00,MTB02} These
`passive' set-ups are mainly controlled via the temperature, the partial
pressure of the solvent, and the solute concentration.

The active set-ups involve an additional control parameter that can
often be better adjusted than those in the passive set-ups.  An
example is a set-up similar to blade coating where a controlled
continuous supply of solution is provided between two glass
plates. The upper plate slides backwards with a controlled velocity
and in this way maintains a meniscus-like liquid surface where the
evaporation takes place and the patterns are deposited.\cite{YaSh05}
The deposition patterns are found to depend on the plate velocity.
Other examples are a receding meniscus between two glass plates whose
receding velocity is controlled by an imposed pressure
gradient,\cite{BDG10} an evaporating drop that is pushed over a
substrate at controlled velocity,\cite{RDLL06} or a solution that is
spread on a substrate by a roller that moves at a defined
speed.\cite{HaKa07}

Up to this point, we have only mentioned experiments involving
evaporating solutions or suspensions on solid substrates. There are
two systems that are closely related: On the one hand there exist
studies of evaporating films on a fluid substrate, in particular,
films of a nanocrystal dispersion in alkanes that spread and evaporate
simultaneously on the free surface of a polar organic fluid that is
immiscible with the alkane.\cite{Dong11} The defect-free liquid
substrate allows for highly regular periodic stripe patterns that
persists over a large area.  On the other hand there are many
experiments related to the transfer of Langmuir-Blodgett monolayers,
i.e., high density surfactant layers, from a trough filled with water
onto a solid plate that is withdrawn with a controlled velocity from
the bath. Depending, e.g., on the velocity of the plate and the
monolayer density on the trough, the transferred monolayer may exhibit
stripe patterns parallel or perpendicular to the direction of
withdrawal.\cite{RiSp92,SCR94,Lenh04} We will return to these
experiments in the conclusion and discuss in which way our results may
be related to these rather different systems.


Despite the large variety of experiments that produce regular line
patterns from polymer solutions and colloidal suspensions, a
theoretical description of their formation has been rather
elusive. Most authors agree that the patterns result from a stick-slip
motion of the contact line that is caused by pinning/depinning
events.\cite{Deeg00,HXL06,Xu06,BFA09} Various reduced models have been
developed that: relate the interaction between the contact line and
the deposit that is formed, in terms of a pinning force and derive how
this force depends on and scales with the experimental
parameters;\cite{Xu06,BDG10} develop evolution equations for the shape
of an individual deposited ring;\cite{Deeg00} study the time evolution
assuming a permanently pinned contact
line.\cite{Fisc02,OKD09,Witt09,TVI11} \citet{HuLa06} analytically
obtain a flow field that is combined with Brownian dynamics simulations to
study particle deposition.  \citet{WCM03} employs a thin film model
similar to the one we present below to describe the dewetting of a
film of a nanoparticle suspension. They study the regime where drop
arrays form via directed convective dewetting of the solvent before they
subsequently dry out.  A thin film model that produces rings is
presented by \citet{KBM10}, however, there the contact line is shifted
`by hand' if a certain condition is met.
A thin film description of an evaporating drop of a suspension may
also be coupled to a full description of the diffusion of vapour in
the gas phase.\cite{DoGu10} This model is used to predict the
dependence of the mean deposit thickness on the substrate velocity,
but is not used to describe the formation of deposition patterns.

In another approach, the system is described using a complex non-isothermal
Navier--Stokes model, i.e., with the complete set of transport equations
for momentum, energy, and solute and vapour concentration, thereby
incorporating evaporation, thermal Marangoni forces and rules for
contact line motion.\cite{BFA09} \citet{BFSA10} further
incorporate Derjaguin--Landau--Verwey--Overbeek (DLVO) interactions
between solute particles and the solid substrate in the form of
effective forces in the advection-diffusion equation for the solute
concentration. Simulations show the formation of and depinning from a
single deposit line, but for the parameter values used in this study,
they do not observe periodic deposits.\cite{BFA09}

To our knowledge, there exists no efficient model that is able to
capture the dynamics of the periodic deposition process, i.e., the
stick-slip character of the process. By `efficient' we mean a model
that allows for a numerical exploration of the parameter space.  The
full transport equations contain all or most of the physics, however,
they are tedious to use and tricky when it comes to incorporating
details like the wettability and contact line motion. Models
that assume a pinned contact line are only able to describe how a
deposit forms for a fixed drop base, even if they are fully dynamic
thin film models.  For instance, the model by \citet{OKD09}
distinguishes a fluid and a gel-like part of the drop and allows one
to follow the time evolution of the drop and concentration profiles. It
can distinguish between final deposits of basin-, crater-, and
mound-type. The crater-type deposits might be seen as resulting from
the deposition of a single ring. However, models like this one that fix
the drop base are not able to capture the deposition of multiple
rings or of a regular line pattern in planar geometry.
Note also that a first fully dynamic model exists for the related
phenomenon of the transfer of Langmuir--Blodgett surfactant monolayers
from a bath onto a solid substrate.\cite{KGFC10} There, the stripe
formation is related to a phase transition in the surfactant layer
that results from a substrate-mediated condensation effect. We discuss
similarities and differences of the model by \citet{KGFC10} and our
thin film model for a receding contact line of an evaporating solution
in section~\ref{sec:model}.

The aim of the present work is to analyse in detail the results from a
dynamic close-to-equilibrium thin film model that allows for a generic
description of a receding contact line of a solution in the passive
geometry, i.e., without an externally imposed velocity.  We show
that the model is able to capture the dynamics of the stick-slip motion of the receding
three-phase contact line of a colloidal suspension or polymer solution
on a solid substrate. For a range of parameter values, as the solvent
evaporates, the model predicts the deposition of regular and
irregular patterns of lines that are parallel to the receding contact line.
The model allows us to elucidate generic mechanisms that play an important
role in many of the experimental systems that are studied. Furthermore, we are able
to determine the influence of varying the basic model parameters on the
deposition patterns, in particular, on the spatial period, amplitude,
and morphology of the deposited lines. In the present work we
restrict our attention to one-dimensional deposition patterns, i.e.,
regular and irregular line patterns. The limitations of this approach
are discussed in the conclusion.

%
%
This paper is laid out as follows:
In Section~\ref{sec:model} the thin film model is introduced in the
form of evolution equations for the film thickness and concentration
profile, that are obtained from making a long-wave approximation. The model is
discussed in the context of related models in the literature.  In
Section~\ref{sec:results}, results from time simulations are
presented. It is shown that solely having a viscosity that diverges at
a critical solute concentration is sufficient to trigger a
self-organised periodic pinning-depinning process that results in the
deposition of regular line patterns in a well defined region of the
parameter space spanned by our non-dimensional control parameters,
i.e., evaporation rate, mean solute concentration, P\'eclet number,
parameters related to wettability, solvent vapour chemical potential, and the
intrinsic viscosity.  The final Section~\ref{sec:conc} gives our
conclusions and an outlook on future work. Two appendices discuss
details of our numerical scheme (Appendix~\ref{sec:numerics}) and the
measures we use to quantify the obtained line patterns
(Appendix~\ref{sec:measures}), respectively.

%
\section{The thin film model}
\label{sec:model}

\begin{figure}[tbh]
\includegraphics[width=0.95\hsize]{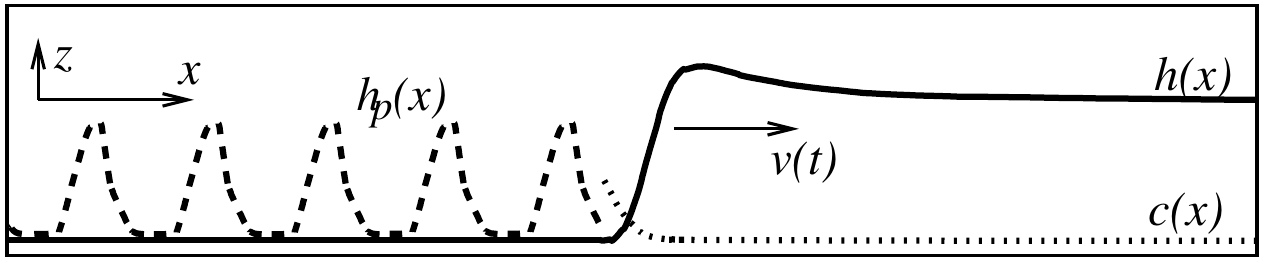}
\centering
\caption{Diagram showing a liquid front that recedes due to evaporation and
  convection. The front has varying velocity $v(t)$ and the liquid film thickness profile is
  given by the function $h(x,t)$. In the bulk film the profile $c(x,t)$ is the
  vertically averaged solute concentration profile and outside the liquid film, the
  solute layer thickness is given by $\hpfield(x,t)=c(x,t) h(x,t)$.}
\mylab{f:sketch}
\end{figure}

We consider a thin film of an evaporating partially wetting
nanoparticle suspension (or polymer solution) on a flat solid substrate
in contact with its vapour (see Fig.~\ref{f:sketch}).
We assume that the densities of the solvent and solute are matched, so there is little
or no sedimentation of the solute within the solvent, and slow evaporation
so that we may neglect the dependence on the vertical coordinate
of the solute concentration within the liquid film.
Assuming that all surface slopes are small, one may employ a long-wave
approximation \cite{ODB97,Thie07} and derive two coupled evolution equations
for the film thickness profile $h(x,t)$ and the
concentration of the solute $c(x,t)$, which can be written in compact
form as a pair of continuity
equations for the solution and the solute, respectively:
\begin{align}
\partial_th & = -\partial_xj_\mathrm{c}(h,c) -
j_\mathrm{e}(h),\label{e:tfeqhj}\\[2ex]
\partial_t(c h) &= -\partial_x\left[c j_\mathrm{c}(h,c) +
j_\mathrm{d}(h,c)\right],\label{e:tfeqhpj}
\end{align}
The equation for the solution, i.e., the film thickness evolution equation
(\ref{e:tfeqhj}), contains a non-conserved evaporative source term and 
a conserved convective transport term. The equation for
the solute i.e., the evolution
equation (\ref{e:tfeqhpj}) for the effective solute layer thickness
$\hpfield= c h$ only has a conserved dynamics, which is made up of two
terms: the first is a convective transport term and the second describes
transport due to diffusion. These coupled convective, evaporative, and
diffusive fluxes are given by
\begin{align}
j_\mathrm{c}(h,c) &= -Q_\mathrm{c}(h,c)\partial_x p(h)
 = -\frac{h^3}{3\eta(c)}\partial_x p(h),\label{e:j_c}\\
j_\mathrm{e}(h) &= \frac{\beta}{\rho}\left[p(h)-\mu\rho\right],\label{e:j_e}\\
j_\mathrm{d}(h,c) &= -Q_\mathrm{d}(h,c)\partial_xc
 = -D(c)h\partial_xc.\label{e:j_d}
\end{align}

\begin{figure}[tbh]
\includegraphics[width=0.7\hsize]{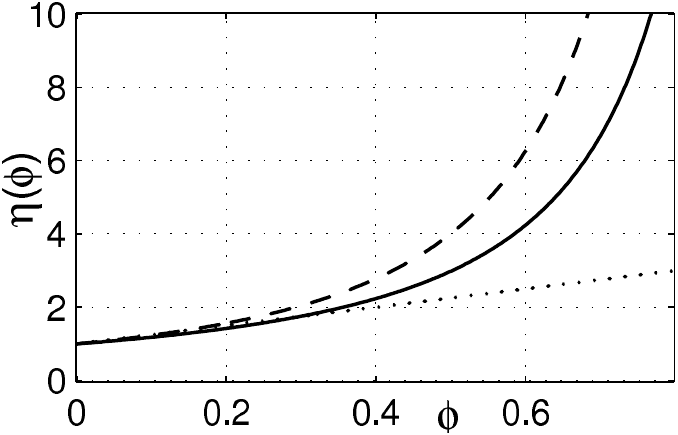}
\centering
\caption{Dependence of the viscosity on the scaled solute
  concentration $\phi=c/c_\mathrm{c}$, as described by the
  dimensionless form of the Krieger--Dougherty law in
  Eq.~(\ref{e:eta(phi)}).  The solid line is for the exponent
  $\nu=1.575$, whereas the dashed one is for $\nu=2$. The linear
  Einstein relation is given as dotted line.}
\mylab{f:viscosity}
\end{figure}
The first term on the right hand side of Eq.~(\ref{e:tfeqhj}) is the
conserved part of the dynamics, where $j_{\mathrm c}$ [given in
Eq.~(\ref{e:j_c})] is the total horizontal volume flux (integrated
over film thickness).  This flux is driven by the pressure
gradient. The pressure $p$ is discussed below. The convective mobility
function $Q_\mathrm{c}(h,c)=h^3/3\eta(c)$ in $j_{\mathrm c}$
[Eq.~(\ref{e:j_c})] results from the Poiseuille flow of the liquid in
the case of a no-slip boundary condition at the substrate and the free
upper surface.  The dynamic viscosity, $\eta(c)$, exhibits a strong
nonlinear dependence on the local solute concentration and obeys the
Krieger--Dougherty law \cite{Lars98,Quem77}
\begin{equation}
\eta(c)= \eta_0\,\left(1-\frac{c}{c_\mathrm{c}}\right)^{-\nu},
\mylab{e:visc}
\end{equation}
where $\eta_0$ is the dynamic viscosity of the pure solvent.  Note
that the solute bulk concentration field $c(x,t)$ is a dimensionless
volume fraction concentration and $c_\mathrm{c}$ is the critical
value at random close packing, where the viscosity diverges,
i.e.\ $\eta(c)\to\infty$ when $c \to c_\mathrm{c}^-$. For hard-spheres
$c_\mathrm{c}=0.63$. The viscosity concentration dependency
Eq.~(\ref{e:visc}) is illustrated in Fig.~\ref{f:viscosity}.

In general, the precise value of the exponent $\nu$ depends on the
type of suspension employed. Here we mainly consider particles that
have no net attractive forces between them and only have excluded
volume interactions, i.e., they interact via a hard-core repulsion
when coming into direct contact.  For such particles, values for $\nu$
between 1.4 and 3 are discussed in the literature, depending on their
shape.\cite{Lars98}
The exponent $\nu$ is sometimes written as $\nu =
[\eta]c_\mathrm{c}$, where $[\eta]$ is the intrinsic viscosity,
see,\citet{Lars98} defined by
\begin{equation}
[\eta]= \lim_{c\rightarrow 0}\frac{\eta-\eta_0}{\eta_0c}.
\mylab{e:intrvisc}
\end{equation}
For spherical particles $[\eta]=2.5$, resulting in $\nu =
1.575$. Other thin film models use $\nu=2$.\cite{CBH08,WCM03} For
solute particles with net attractive forces between them, values for
$\nu$ as low as 0.13 are reported.\cite{Trap01} Depending on the
particular system, the transition at $c_\mathrm{c}$ is either referred
to as jamming or gelation.\cite{Trap01,OKD09} Here we fix $\nu=1.575$,
corresponding to particles that only interact via excluded volume
interactions, in order to gain an understanding of these types of
systems and also to determine the regions of qualitatively different
behaviour in the phase plane spanned by the evaporation rate and
concentration, defined below. Additional investigations show (see
Section~\ref{sec:rheology}) that the formation of periodic deposits is
somewhat more pronounced for smaller $\nu$, which corresponds to more
strongly attracting solute particles.

\begin{figure}[tbh]
\includegraphics[width=0.8\hsize]{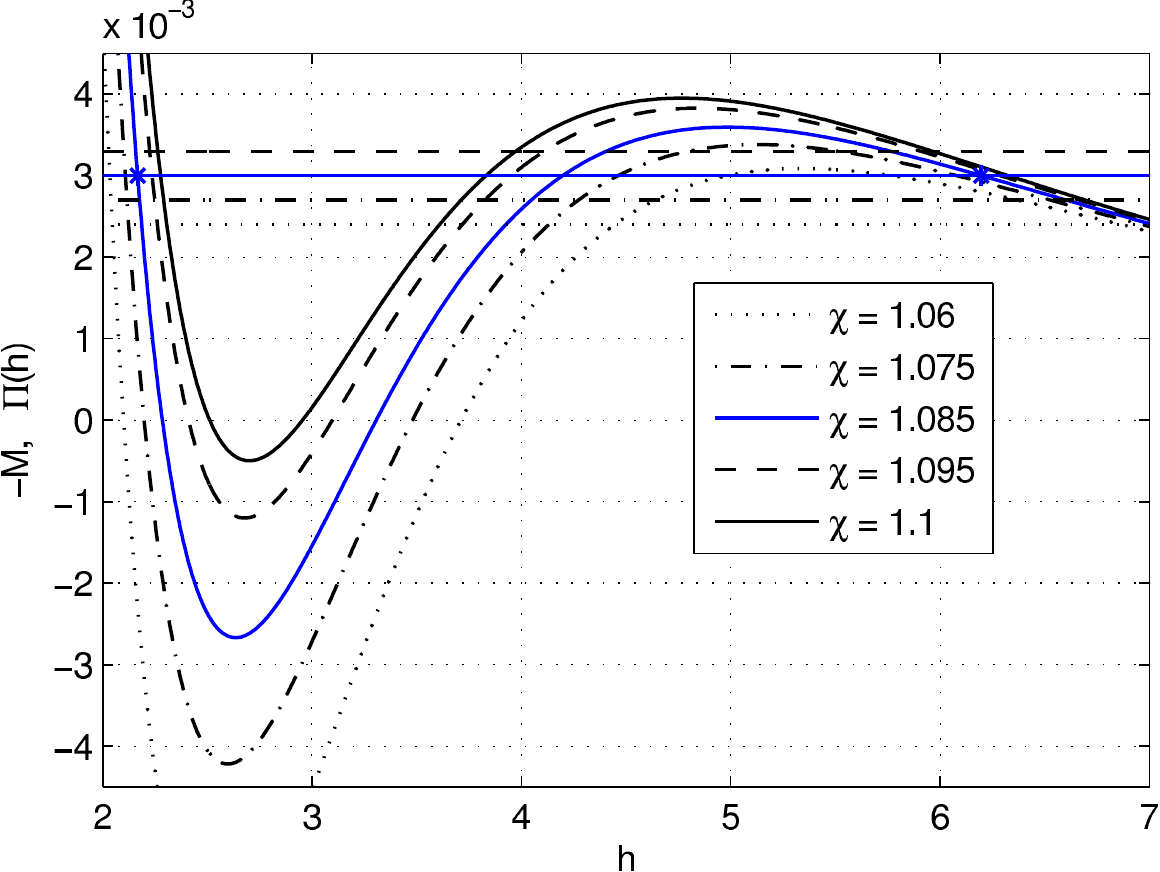}
\centering
\caption{(Color online) The non-dimensional disjoining
  pressure $\varPi(h)$, given in Eq.~(\ref{e:p(h)}), for various values of
  $\chi$, as
  given in the legend. The horizontal straight lines indicate various
  values of the non-dimensional chemical potential $M$,
  given in Eq.~(\ref{e:M}). The blue solid lines
  for the disjoining pressure and chemical potential correspond to our
  `standard' values of $\chi=1.085$ and $M=-0.003$, respectively (see
  the end of Section~\ref{sec:model}). The
  other lines correspond to parameter values employed in Section~\ref{sec:wett}
  below. The blue stars denote the stable film heights: $h_1 < h_2$
  for the standard values.}
\mylab{f:disjp}
\end{figure}
The convective flow is driven by the gradient of the pressure
\begin{equation}
 p(h)= - \gamma\partial_{xx}h -\varPi(h),
\mylab{e:press}
\end{equation}
where the first term is the Laplace pressure ($\gamma$ is the surface
tension) and the second term is the disjoining (or Derjaguin) pressure 
\begin{equation}
\varPi(h) = \frac{2S^\mathrm{LW}d_0^2}{h^3}
 + \frac{S^\mathrm{P}}{l_0}\exp\frac{d_0-h}{l_0}
\end{equation}
that models wettability effects for a
partially wetting fluid;\cite{deGe85,Shar93,Thie07} see
Fig.~\ref{f:disjp} for a graphical representation of its
non-dimensional form.
$l_0$ is the Debye
length, $d_0$ is a molecular interaction length, $S^\mathrm{LW} = -A/12\pi
d_0^2$ and $S^\mathrm{P}<0$ are the apolar and polar spreading coefficient,
respectively, and $A<0$ is the Hamaker constant.  We expect
qualitatively similar behaviour for other combinations of stabilising
and destabilizing terms in $\varPi(h)$.\cite{TVN01,TVNP01,Thie10}

The second term on the r.h.s.\ of Eq~(\ref{e:tfeqhj}) represents the non-conserved
part of the dynamics and models evaporation, where $j_{\mathrm e}$
[given in Eq.~(\ref{e:j_e})] is the evaporative
flux density at the free surface of the film.
Here, we assume that the system is close to equilibrium and that the
vapour is near to saturation and so evaporation is slow.  In this
limit evaporation with a rate $\beta$ is driven by the difference of
the scaled pressure $p/\rho$ and the chemical potential of the ambient
vapour $\mu$.\cite{LGP02,Pism04} Latent heat effects may be neglected,
and the density $\rho$ is assumed to be equal for the solute particles
and the solvent.

The first and second terms on the right hand side of
Eq.~(\ref{e:tfeqhpj}) model convective and diffusive transport of the
solute, respectively. Since the solute is passively advected by the
fluid, the convective flux is given by $cj_\mathrm{c}$.  For the
diffusive flux [given in Eq.~(\ref{e:j_d})] we assume Fick's law and we
set the diffusive mobility to be $Q_\mathrm{d}(h,c)=D(c)h$,
where $D(c)$ is the concentration dependent diffusion coefficient,
and we employ the Einstein--Stokes relation
\begin{equation}
D(c) =\frac{k_BT}{6\pi r_0\eta(c)},
\end{equation}
where $k_B$ is the Boltzmann constant, $T$ the temperature, and $r_0$
the solute particle radius.

To facilitate a comparison of our results with the literature we employ
a scaling identical to that employed by \citet{LGP02} and,\citet{FAT11} namely:
time, horizontal $x$-coordinate, and film thickness scales are
\begin{align}
\tau   &= \frac{3\eta_0\gamma}{\delta |\widetilde{S}^\mathrm{P}|^2},
\qquad
\alpha =\left(\frac{\delta\gamma}{|\widetilde{S}^\mathrm{P}|}\right)^{1/2},
\nonumber\\
\mbox{and}\qquad\delta &=
\left(\frac{A}{6\pi|\widetilde{S}^\mathrm{P}|}\right)^{1/3},\label{e:scales}
\end{align}
respectively, where
\begin{equation}
\widetilde{S}^\mathrm{P} = \frac{S^\mathrm{P}}{l_0}\exp\frac{d_0}{l_0}.
\end{equation}

Using this scaling, Eqs.~(\ref{e:tfeqhj}) and (\ref{e:tfeqhpj}) are
brought into the following non-dimensional form
\begin{align}
\partial_th
 &= \partial_x\left[\frac{h^3}{\eta(\phi)}\partial_x p(h)\right]
 - \varOmega_0\left[p(h)-M\right],\mylab{e:tfeqh}\\
\partial_t(\phi h)
 &= \partial_x\left[\frac{\phi h^3}{\eta(\phi)}\partial_x p(h)\right]
 + \partial_x\left[\frac{h}{\mathrm{Pe}\,\eta(\phi)}\partial_x\phi\right],
\mylab{e:tfeqhp}
\end{align}
where
\begin{equation}
p(h) = -\partial_{xx}h - \varPi(h) = -\partial_{xx}h - h^{-3}+\exp(-\chi h).
\label{e:p(h)}
\end{equation}
Note that starting with Eqs.~(\ref{e:tfeqh}) and (\ref{e:tfeqhp}) in
the remainder of the paper, the symbols $h$, $\hpfield$, $t$, $x$,
$p$, $\eta$, and $\varPi(h)$ stand for the non-dimensional quantities
whereas up to this point they denoted the dimensional
quantities. However, for the scaled concentration we introduce $\phi =
c/c_\mathrm{c}$.  The diffusion coefficient is expressed as
$\left[\mathrm{Pe}\,\eta(\phi)\right]^{-1}$, where $\mathrm{Pe}$ is
the P\'eclet number and the dimensionless viscosity function is
\begin{equation}
\eta(\phi) = (1-\phi)^{-\nu}.\label{e:eta(phi)}
\end{equation}
The dimensionless numbers in Eqs.~(\ref{e:tfeqh})--(\ref{e:p(h)})
are defined as
\begin{align}
\chi &= \left(\frac{|A|}{6\pi|\widetilde{S}^\mathrm{P}|l_0^3}\right)^{1/3},
 \label{e:chi}\\
\varOmega_0 &= \frac{18\pi\beta\eta_0\gamma}{\rho\left[6\pi A^2
 |\widetilde{S}^\mathrm{P}|\right]^{1/3}},\label{e:Omega_0}\\
M &= \frac{\rho\mu}{|\widetilde{S}^\mathrm{P}|},\label{e:M}\\
\mathrm{Pe}^{-1} &= \frac{3k_BT}{r_0\left[6\pi A^2
|\widetilde{S}^\mathrm{P}|\right]^{1/3}},\label{e:Pe^-1}
\end{align}
where the evaporation number, $\varOmega_0$, is the ratio of the time scales of
convection and evaporation of a film without solute and the reciprocal
P\'eclet number, $\mathrm{Pe}^{-1}$, is the ratio of the time scales of
convection and diffusion.

There exist a number of models in the literature that are similar to
that defined in Eqs.~(\ref{e:tfeqh}) and (\ref{e:tfeqhp}). One is used
to study macroscopic particle-laden film flow down an
incline.\cite{CBH08} There, the solvent is non-volatile and changes in
the particle concentration result from the settling of particles due
to gravity.  As the particles in the model of \citet{CBH08} are
assumed to be large, no diffusion is included in the model, nor is
wettability; the advancing of the contact line is facilitated by a
precursor film of imposed height.
Another example is the case of a surface-passive solute studied by
\citet{WCM03} [their Eqs.~(54) and (55)]. They use a different
Derjaguin pressure that, however, also models partially wetting
liquids. The main difference is in the evaporation model that they
use, which incorporates a vapour recoil effect. They also model
particle diffusion in a manner that is independent of solute
concentration.  Both, \citet{CBH08} and \citet{WCM03} use a
Krieger--Dougherty law to model the dependence of the viscosity on the
concentration. This is also done by \citet{CMS09} where the
spreading and retraction of evaporating droplets containing nanoparticles
is studied with a thin film model that involves a structural
disjoining pressure.
There are some similarities to other models
\cite{Fisc02,OKD09,TVI11} where, however, the contact lines are kept
pinned. Since moving contact lines are an intrinsic part of the solute deposition
process that we study here, we do not refer any more to these other
approaches.

The evaporation model that we use is valid close to equilibrium where
the film may be considered isothermal. The evaporation is limited by
the kinetics of the phase transition (or by the boundary layer
transfer, but not by the diffusion of solvent vapour in the gas phase)
in the contact line region, which is influenced by the effective
molecular interactions, i.e., the saturated vapour pressure depends on
the disjoining pressure and curvature in the manner employed in
studies of evaporating films and
drops.\cite{LGP02,PKS99,Pism04,Thie10,TTP11} Our evolution equation
(\ref{e:tfeqh}) reduces to the model by,\citet{LGP02} in the limit
$\phi_0 \to 0$.  Note, that one may also obtain our evaporation model
by taking the isothermal limit of the models by.\citet{Ajae05,ReCo10}
For a further discussion see.\citet{TTP11}

In the present work, we restrict our attention to line patterns
deposited through evaporative dewetting, i.e., we assume a
one-dimensional geometry, as sketched in Fig.~\ref{f:sketch}. It is
known from the experiments that line patterns are not always transversally
stable. We discuss this point in relation to our results in the
conclusion.  We start our numerical computations with an initial
condition that corresponds to a spatially one-dimensional
semi-infinitely extended film of constant thickness that is connected
to a thin precursor layer. The thickness of the precursor film corresponds
to the equilibrium height where the Derjaguin pressure and evaporation
balance. The liquid in the film is a suspension with a constant solute
concentration. We discretize the non-dimensional equations
(\ref{e:tfeqh}) and (\ref{e:tfeqhp}) over a finite domain $x\in[0,L]$
where the dewetting front is located close to the boundary of the
domain at $x=0$, see Fig.~\ref{f:sketch}. The details of the
implementation including the procedure we use for shifting the spatial
computational domain at certain times are given in
Appendix~\ref{sec:numerics}.

Initially, we employ a set of values of our dimensionless parameters
that we will refer to as the {\em standard configuration}: We set the
dimensionless chemical potential, $M=-0.003$, and the dimensionless
disjoining pressure parameter $\chi=1.085$, which are the same values
as used by.\citet{LGP02} The other parameters are modified to
accommodate the effect of the solute.  In particular, for the initial
results presented here, we use the evaporation rate
$\varOmega_0=4.64\times10^{-7}$; the scaled constant bulk
concentration, $\phi_0=0.41$, that enters the simulation as a given
initial value; the reciprocal P\'eclet number,
$\mathrm{Pe}^{-1}=0.0003$; and the exponent of the Krieger--Dougherty
law, $\nu=1.575$, as discussed above after Eq.~(\ref{e:intrvisc}).

The main control parameters in our simulations are $\varOmega_0$ and
$\phi_0$. These two parameters span the plane within which we
determine regions where periodic patterns and other characteristic
deposits occur. Then, for selected fixed parameter combinations
$(\varOmega_0, \phi_0)$, we vary one of the parameters
$\mathrm{Pe}^{-1}$, $M$, $\chi$, and $\nu$ while the remaining
parameters are kept fixed, in order to determine how sensitive the behaviour
of the system is to variations in the value of these.
%
\section{Results}
\mylab{sec:results}
\subsection{Dynamics of the evaporative dewetting front}
\mylab{sec:front-dyn}

%
\begin{figure}[tbh]
\centering
\includegraphics[width=0.8\hsize]{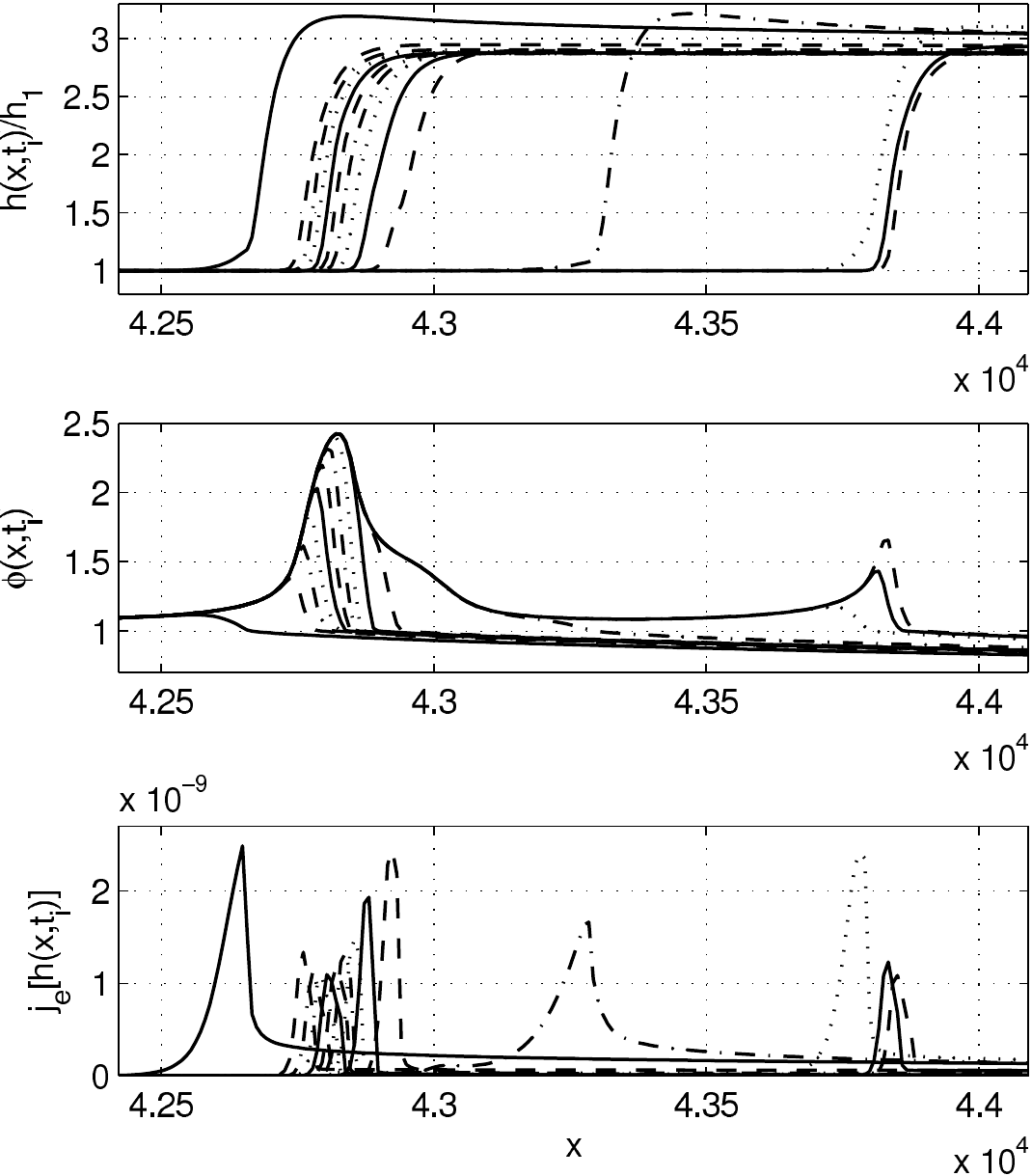}
\caption{Sequences of snapshots of film thickness profile $h(x,t)$ (top panel),
  mean concentration $\phi(x,t)$ (middle), and evaporation flux
  $j_\mathrm{e}(x,t)$ (bottom), which illustrate the pinning-depinning dynamics.
  The individual profiles in each sequence are plotted starting at the same
  time, with a time-increment $\Delta t = 10^9$ between each successive curve,
  and are displayed with
  periodically repeated line styles (solid, dashed, dash-dotted,
  dotted, solid, ...). All the model parameters are set to the `standard
  values', which are defined at the end of Sec.~\ref{sec:model}.}
  \mylab{f:regimes}
\end{figure}
\begin{figure}[tbh]
\includegraphics[width=0.9\hsize]{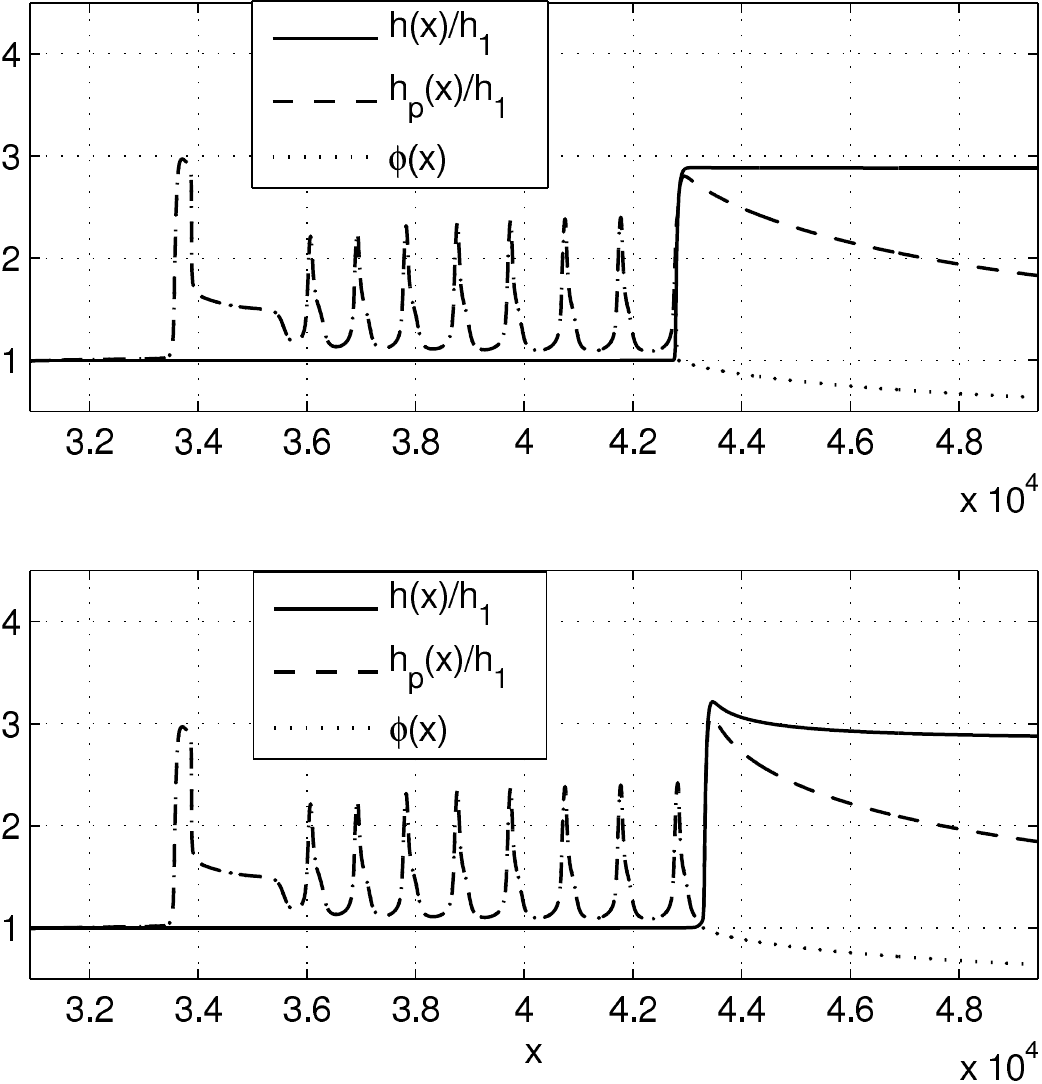}
\caption{Snapshots from the deposition process at two characteristic
  stages: evaporation-dominant (top) and
  convection-dominant (bottom). The system parameters are as in Fig.~\ref{f:regimes}.
  The profiles in the top and bottom panel correspond to the 4th
  and the 11th profile in time in Fig.~\ref{f:regimes}, respectively.
  Note the capillary ridge behind the front in the lower plot, which is absent
  in the upper evaporation-dominated stage of the cycle.}
\mylab{f:timesnaps}
\end{figure}
\begin{figure}[tbh]
\includegraphics[width=0.8\hsize]{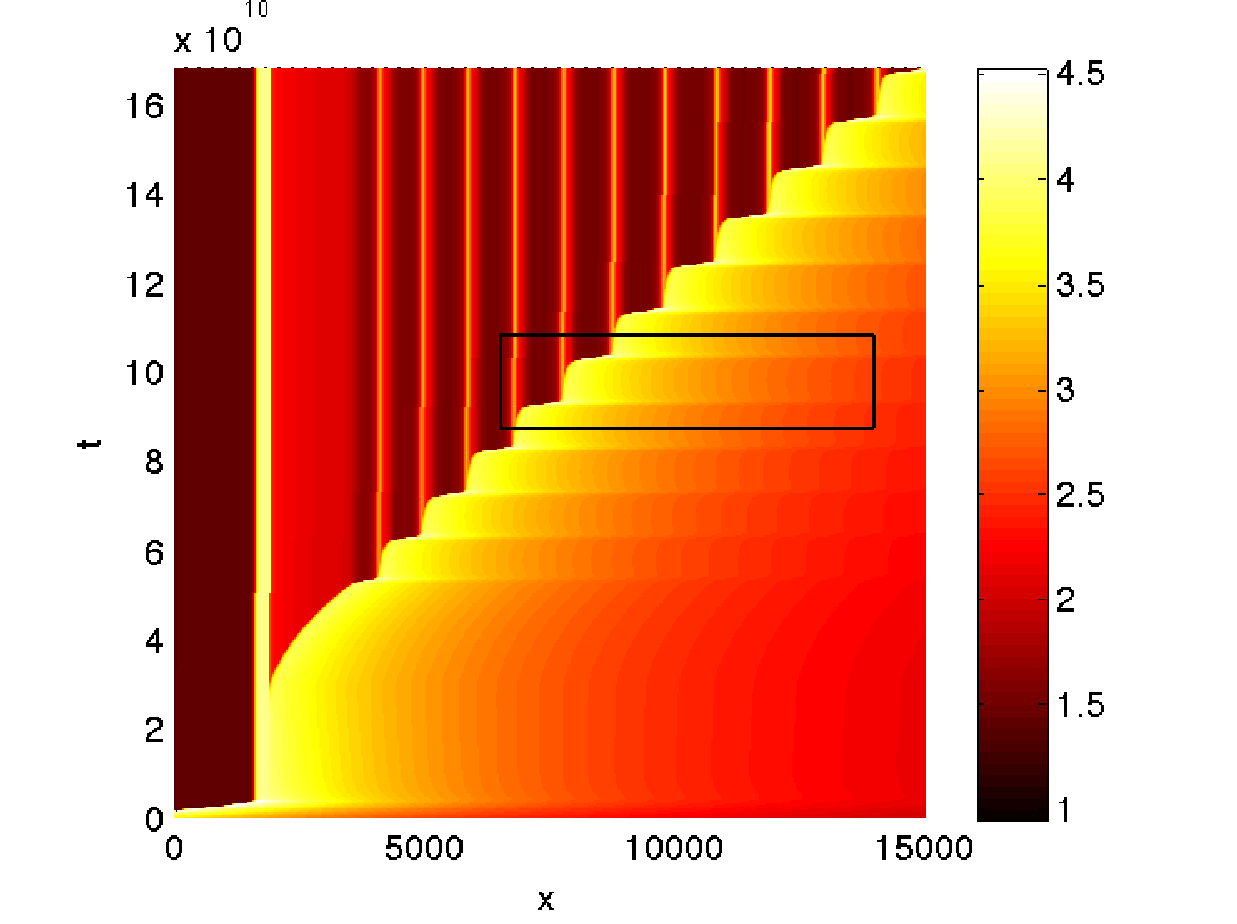}\\
\includegraphics[width=0.7\hsize]{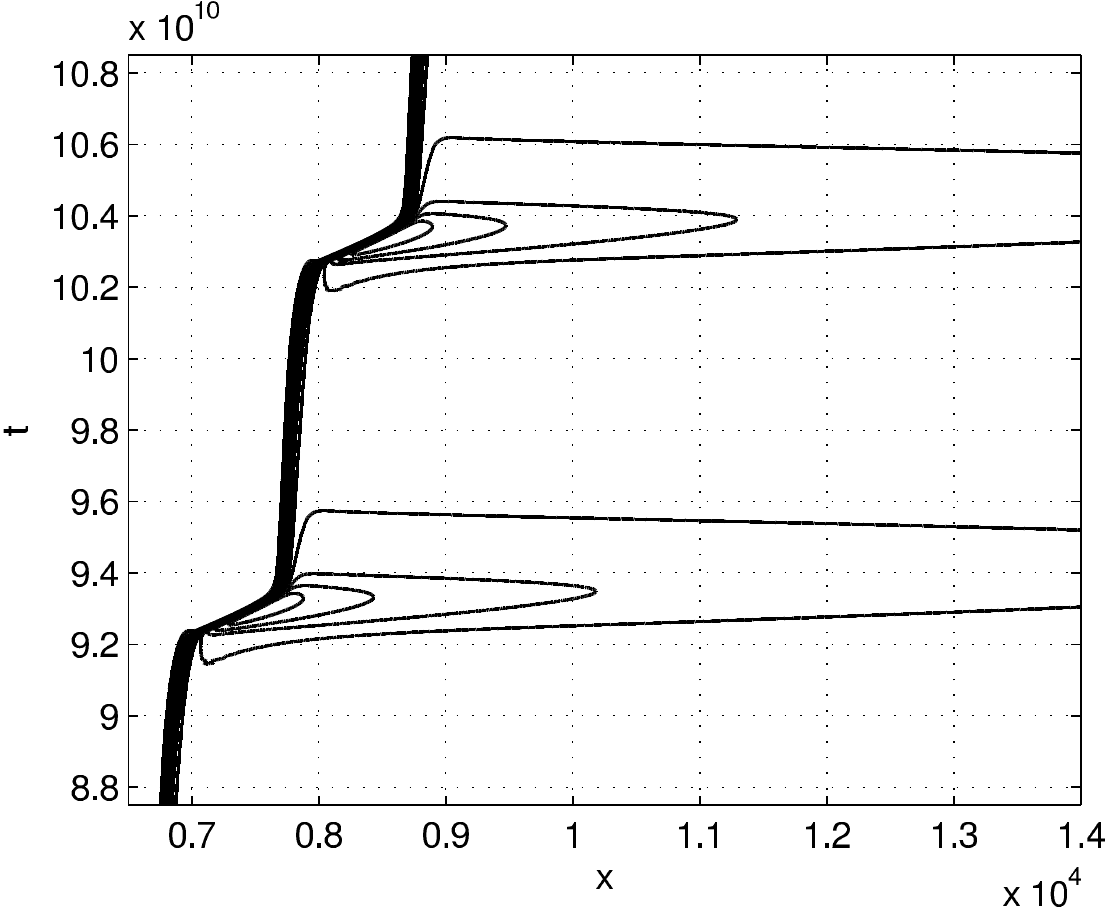}
\caption{(Color online) Space-time plots of the nanoparticle deposit
thickness $\hpfield(x,t)$ (top) for the standard values of the system parameters,
where the deposited lines have characteristic heavy right tail (see
Fig.~\ref{f:vertcutprof}) and a contour plot of the film height $h(x,t)$
(bottom) in the rectangular
region marked in the upper plot. There are
24 equally spaced level contours in the interval
of heights $[1.2,3.25]$ which allows the steep dewetting front and capillary
ridges to be resolved.}
\mylab{f:xtph}
\end{figure}
\begin{figure}[tbh]
\includegraphics[width=0.8\hsize]{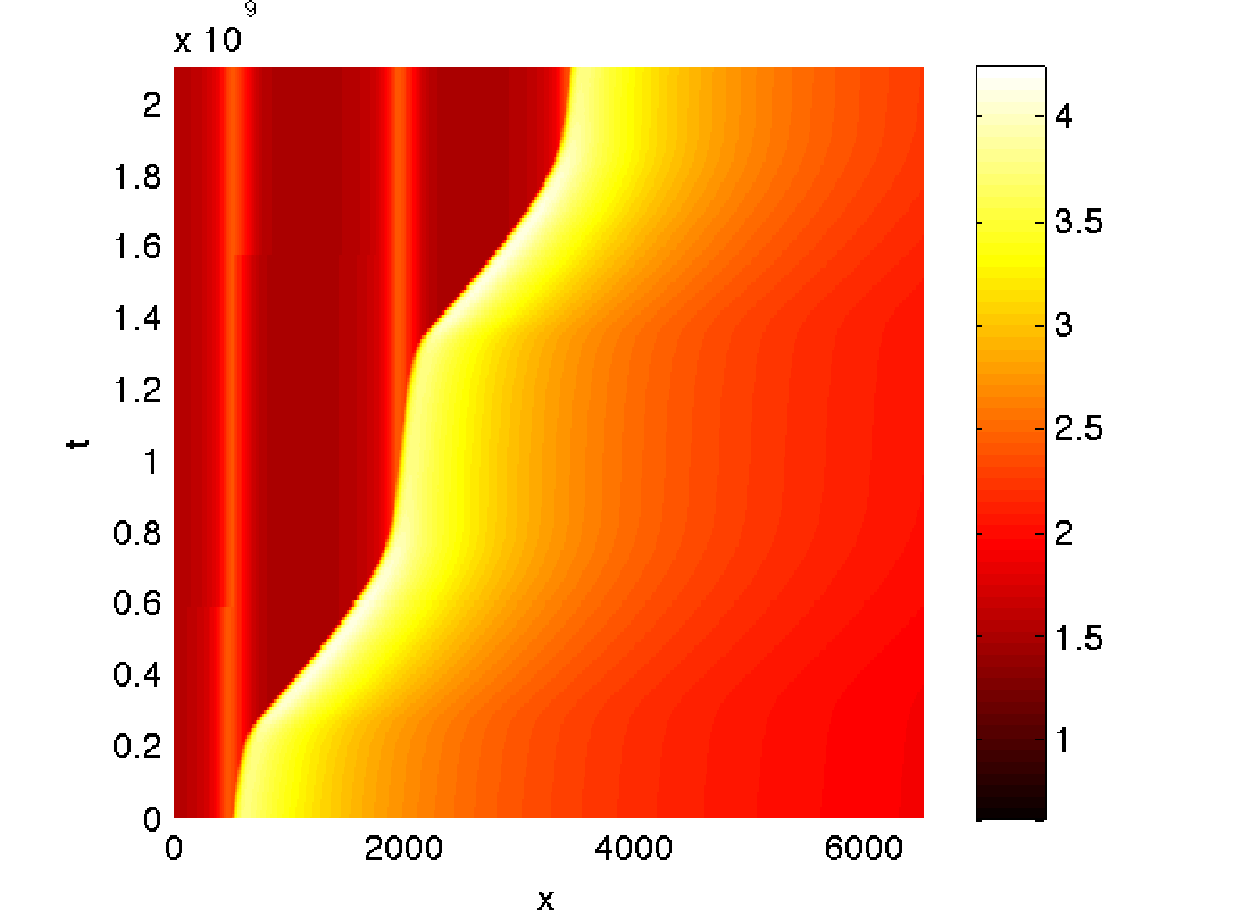}\\
\includegraphics[width=0.8\hsize]{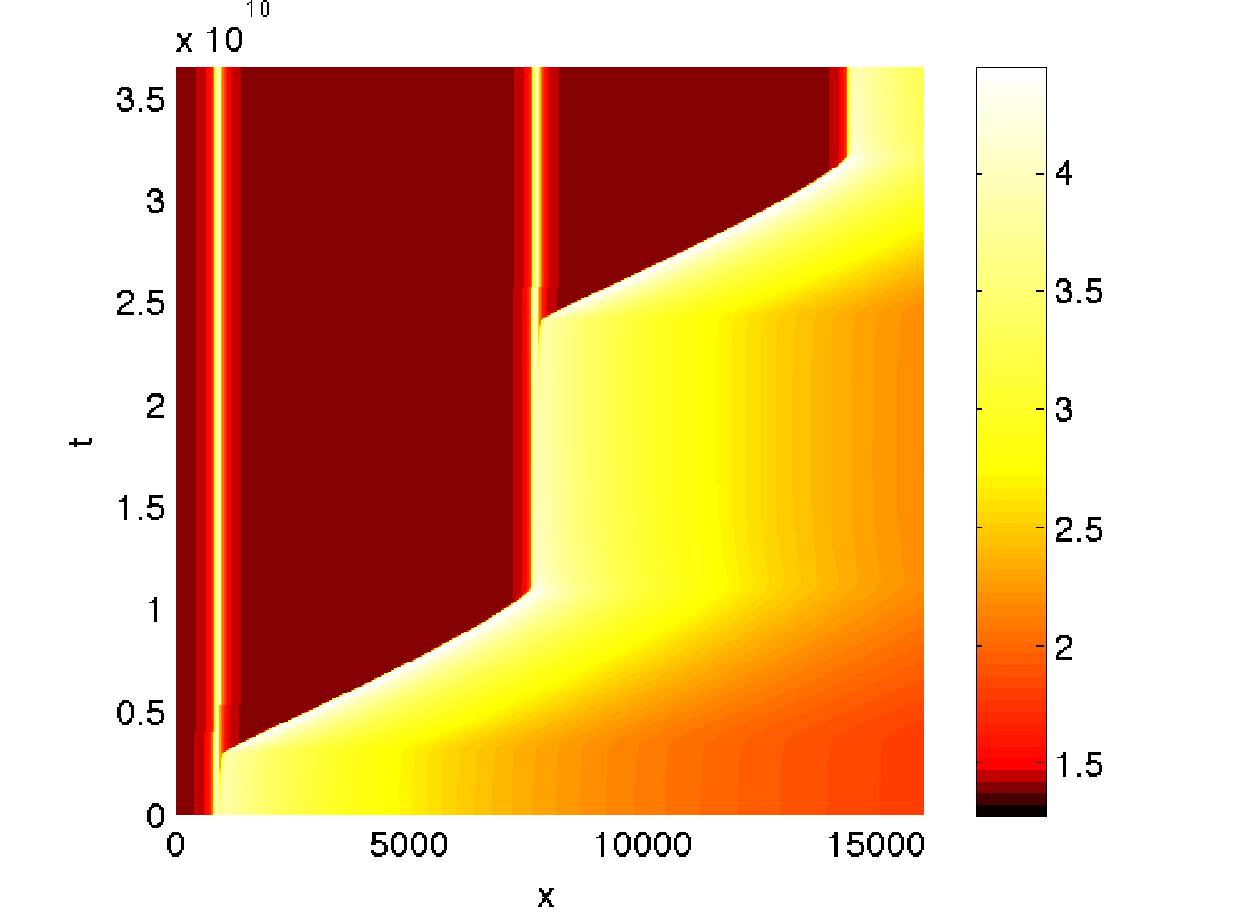}
\caption{(Color online) Space-time plots of the nanoparticle film height
$\hpfield(x,t)$ for 2 characteristic cases:
the upper panel shows deposit lines with a
dominant left `tail' ($\varOmega_0=4.64\times10^{-6}$, other
parameters are standard). The regular line pattern corresponding to
this situation is shown in Fig.~\ref{f:vertcutprof}.
The lower panel shows a long-period large-amplitude pattern
($\phi_0=0.31$, other parameters are standard). A regular line pattern for the close,
but slightly different value $\phi_0=0.3016$ is shown in Fig.~\ref{f:vertcutprof}.}
\mylab{f:xtphp}
\end{figure}
Our investigations are mainly based on time simulations of the front
motion and the deposition process. We start with a semi-infinitely
extended liquid film that coexists with
an ultrathin precursor film. The initial dewetting front has a
step-wise profile, but is smoothed by capillarity during the very
first time steps. The concentration is initially set to a uniform
value $\phi(x,0) = \phi_0$. The chosen value of the vapour chemical potential
ensures that the front recedes by evaporation and/or convection. As
the front recedes, it deposits part of the solute in an initially
smooth layer. Then, in most cases, the front settles after some transient
into a different type of regular motion.

The time-dependent behaviour of the system is well known in the case
without solute $(\phi_0=0)$.\cite{LGP02} There, the front profile
always converges to a constant shape that moves with constant
velocity, i.e., the front motion is stationary. In this situation, one
may still distinguish between two qualitatively different limiting
cases: (i) \textit{convection-dominated} and (ii)
\textit{evaporation-dominated dewetting}. They are found for small and
large values of the evaporation number $\varOmega_0$, respectively.
In case (i) the convective motion maintains a capillary ridge despite
the ongoing evaporation whereas in case (ii) convection is much slower
than evaporation and there is no capillary ridge.

In the presence of a solute, the situation is more complex and there
exist extended regions in parameter space where no stationary front
motion is found. Instead, the receding front shows an unsteady motion
with periodically changing front velocity and shape.  An example of
such a dynamics is shown in Figs.~\ref{f:regimes}, \ref{f:timesnaps}
and \ref{f:xtph}. These figures illustrate various aspects of the
periodic pinning-depinning dynamics of the front that is related to a
periodic transition between evaporation- and convection-dominated
regimes of the front motion.  Fig.~\ref{f:regimes} shows snapshots of the film
height $h(x,t_i)$, concentration $\phi(x,t_i)$ and evaporation flux
$j_\mathrm{e}(x,t_i)$ profiles. They are equidistant in time and cover
just over one period in time.  The pinned (evaporation-dominated)
regime is characterised by densely spaced profiles when the front is
at about $x=4.275$. During this stage the front moves very slowly,
the height profile has no capillary ridge and the evaporative flux is
rather localised. In contrast, during the depinned
(convection-dominated) stage of the cycle, the profiles are sparsely distributed
(i.e., the front is fast), the convective motion supplies enough
solution to maintain a capillary ridge and therefore the evaporative
flux is spread over a wider $x$-region.  Note that the maximal
$j_\mathrm{e}(x,t)$ is actually larger during the depinned regime.

Transitions between the two regimes can be explained as follows: In
the early stage of the convection-dominant phase the front moves
relatively fast, although the evaporation flux, $j_{\mathrm e}(x,t)$,
is still greatest in the contact line region. This results in an
increase of the local concentration $\phi(x,t)$ and, in consequence,
in a strongly nonlinear increase of the viscosity in this region that
suppresses the convective motion of the front. As a result, the front
slows down and if the local solute concentration reaches random close
packing ($\phi(x,t) \to 1$) the convective motion in the contact
region stops completely. Thus, the suspension becomes locally
jammed. In the case of a polymer solution the transition may be
referred to as a local gelling transition (as employed, e.g., in the
piece-wise model by.\citet{OKD09}) With the convective motion
arrested, the front only moves by evaporation. As the resulting front
velocity can be orders of magnitude slower than during the convective
motion; the front seems to be pinned. This is clearly visible in the
space-time plots of Figs.~\ref{f:xtph} and \ref{f:xtphp}. The typical
front shape at this stage is then monotonic, and no capillary ridge
exists [see the solid line in Fig.~\ref{f:timesnaps}(a)]. During this
phase of slow evaporative motion, the front effectively leaves
deposits of the highly concentrated solute. As a result, the
concentration and therefore viscosity in the contact line region
decrease, allowing the convective motion to start again. The front
speed is then much larger than in the evaporation-dominated phase, and
the front seems to depin. The typical front shape at this stage is
non-monotonic, due to the presence of a capillary ridge [see the solid
line in Fig.~\ref{f:timesnaps}(b)]. Note that the width of the region
at the front where the concentration of the solute is locally
increased varies over time. The extent (along the $x$-axis) of this
region is visible in Fig.\ \ref{f:xtphp} as the bright (white) region
of the field $\hpfield(x,t)$ at the receding dewetting front. We see
that in the convective regime the width increases as the solute gets
concentrated and in the evaporative (pinned) regime the width
decreases as the solute is deposited and left behind the front.

\begin{figure}[!tbh]
\centering
\includegraphics[width=0.9\hsize]{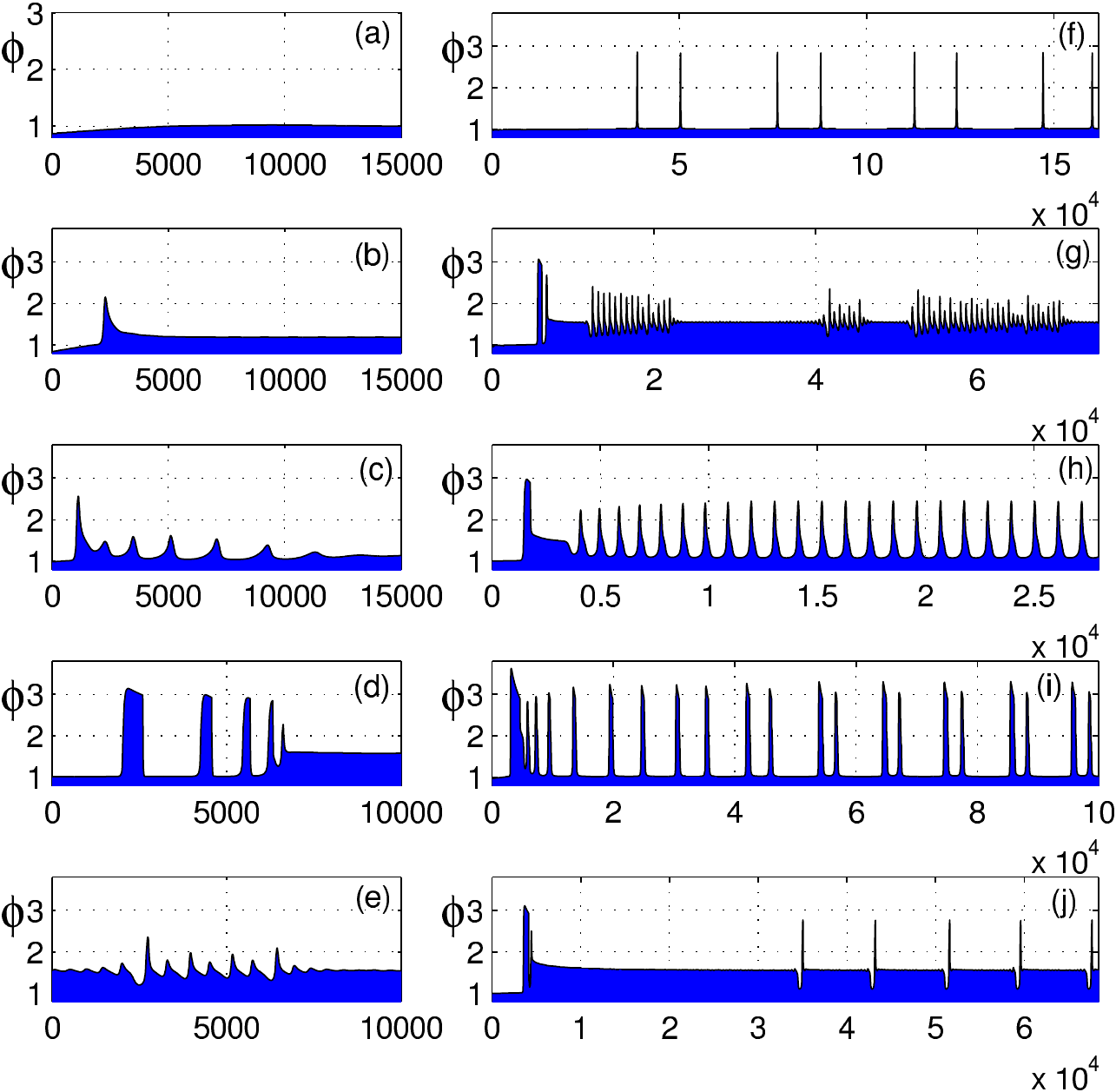}
\caption{(Color online) A selection of typical deposit profiles,
 including the spatial region where the
 transient to long-time behaviour occurs. The panels show deposition of
(a) no lines (but in the transient stage $\hpfield$ may change nonmonotonically);
(b) a single line followed by a flat layer;
(c) transient lines (whose amplitude decays first fast then slow) followed by a flat
layer;
(d) transient lines (whose amplitude decays first slow then fast) followed by a
flat layer (decreased $\nu=1.25$);
(e) an intermittent line pattern [which is a magnification of the pattern in panel (g)];
(f) transient double lines (converging to regular lines);
(g) an intermittent line pattern;
(h) transient lines followed by a regular line pattern;
(i) transient lines converging to a regular pattern of double lines
(decreased $\chi=1.065$);
(j) a long-period pattern switching between a flat layer and a single
line with a leading depression (decreased $\nu=1.25$).
The corresponding parameters for (a--j) are: $\varOmega_0=($4.64,
14.68, 4.64, 0.167, 0.147, 0.464, 0.147, 0.464, 0.7,
0.464)$\times10^{-6}$, $\phi_0=($0.3343, 0.41, 0.3588, 0.41, 0.41,
0.2983, 0.41, 0.41, 0.41, 0.498). The remaining parameters are the
standard values, defined in section~\ref{sec:model}. The letters
(a-d, g, h) are used to mark the corresponding regions in the phase
diagram displayed in Fig.~\ref{f:phasediag}.  }
\mylab{f:sampleprof}
\end{figure}

The resulting periodic pinning-depinning cycle -- that is perceived as
a stick-slip motion -- can be best appreciated in the space-time plot
displayed in Fig.~\ref{f:xtph}(b) which shows the film thickness profile.
The contours of $h(x,t)$ in the steep dewetting front region are
so closely bunched that they appear to be a single thick line whose
slope, $\diff x/\diff t$, corresponds to the velocity of the dewetting
front. It is clearly visible that the velocity periodically changes
between two rather different values. Further contour lines to the
right of the moving front show the periodic appearance of a capillary
ridge in the phases of convective motion. The pinning-depinning cycle
can also be seen in the space-time plots of the effective nanoparticle
layer thickness $\hpfield(x,t)$ presented in Fig.~\ref{f:xtph}(a) and
\ref{f:xtphp}. There, however, the capillary ridges are not as clearly
visible because $\hpfield(x,t)$ decreases faster when moving away from the
dewetting front into the wet region.

The significant changes in the relative importance of convective and
evaporative fluxes can be conveniently described by a local
evaporation number
$\varOmega_\mathrm{loc}=(\eta(\phi)/\eta_0)\varOmega_0$. When
$\varOmega_\mathrm{loc}$ diverges, the front seems pinned, but actually
still moves extremely slowly by evaporation alone, and deposits a line
or a thick layer of solute. When $\varOmega_\mathrm{loc}$ sufficiently
decreases, the front depins and convective motion resumes.

We use the simulation set up described above to investigate the
transient and long-time deposition behaviour. It is found that after
the initial transient (that may involve the deposition of some irregular
lines) various scenarios are possible, depending on the values of our
model control parameters. Most importantly, there exists an extended region
in parameter space, where after a transient, very regular line patterns
are deposited (see Figs.~\ref{f:timesnaps}, \ref{f:xtph} and
\ref{f:xtphp}). In these cases the pinning-depinning process is
repeated in a regular manner. Particular features of the deposited
pattern depend on the particular values of the model parameters and are discussed
below. The occurrence of periodic deposits for an extended parameter
range indicates that
this phenomenon is robust and explains why the deposition of line
patterns is found in a wide variety of experimental settings with
various suspensions and solutions, where it is often described as
resulting from a regular stick-slip motion of the contact line.\cite{HXL06,Xu06,BDG10}

Before we discuss the various types of deposition patterns that we
observe as the system parameters are varied, we return to the space-time plots in
Figs.~\ref{f:xtph} and \ref{f:xtphp} that already indicate how the
ratio of typical time scales for the convection- and evaporation-dominated
front motion changes with increasing evaporation number and mean
concentration. Fig.~\ref{f:xtph} shows (as do Figs.~\ref{f:regimes} and
\ref{f:timesnaps}) our standard case (described at the end of
Sec.~\ref{sec:model}). There, the ratio of the velocities of the
dewetting front in the depinned and pinned regime is $45.07$.
Increasing the evaporation rate coefficient to $\varOmega_0=4.64\times10^{-6}$
(keeping the other parameters at the standard values), the evaporative
flux becomes stronger and the convective motion is not fast enough to
create a large capillary ridge. As a result, the difference between the depinned and pinned
regimes is smaller and so the ratio of the velocities decreases to $8.16$
[see Fig.~\ref{f:xtphp}(a)]. Further increasing of $\varOmega_0$
eventually results in the deposition of a layer of constant thickness.
If instead one decreases the concentration to $\phi_0=0.31$
(keeping the other parameters at standard values), see
Fig.~\ref{f:xtphp}(b), the viscosity in the convective regime
decreases and the capillary ridges become larger. Because there is
less solute present, it takes longer to build up a high concentration
of solute in the contact line region and to pin the front. This results in an
increased length of the convective phase of the cycle (compare
Figs.~\ref{f:xtphp}(b) and \ref{f:xtph}) and to an increase of the
ratio of velocities to $80.36$. A further decrease in the
concentration prolongs the convective phase even more until eventually
the length of the convective phase diverges (at finite concentration value,
see below).

\begin{figure}[tbh]
\centering
\includegraphics[width=1.0\hsize]{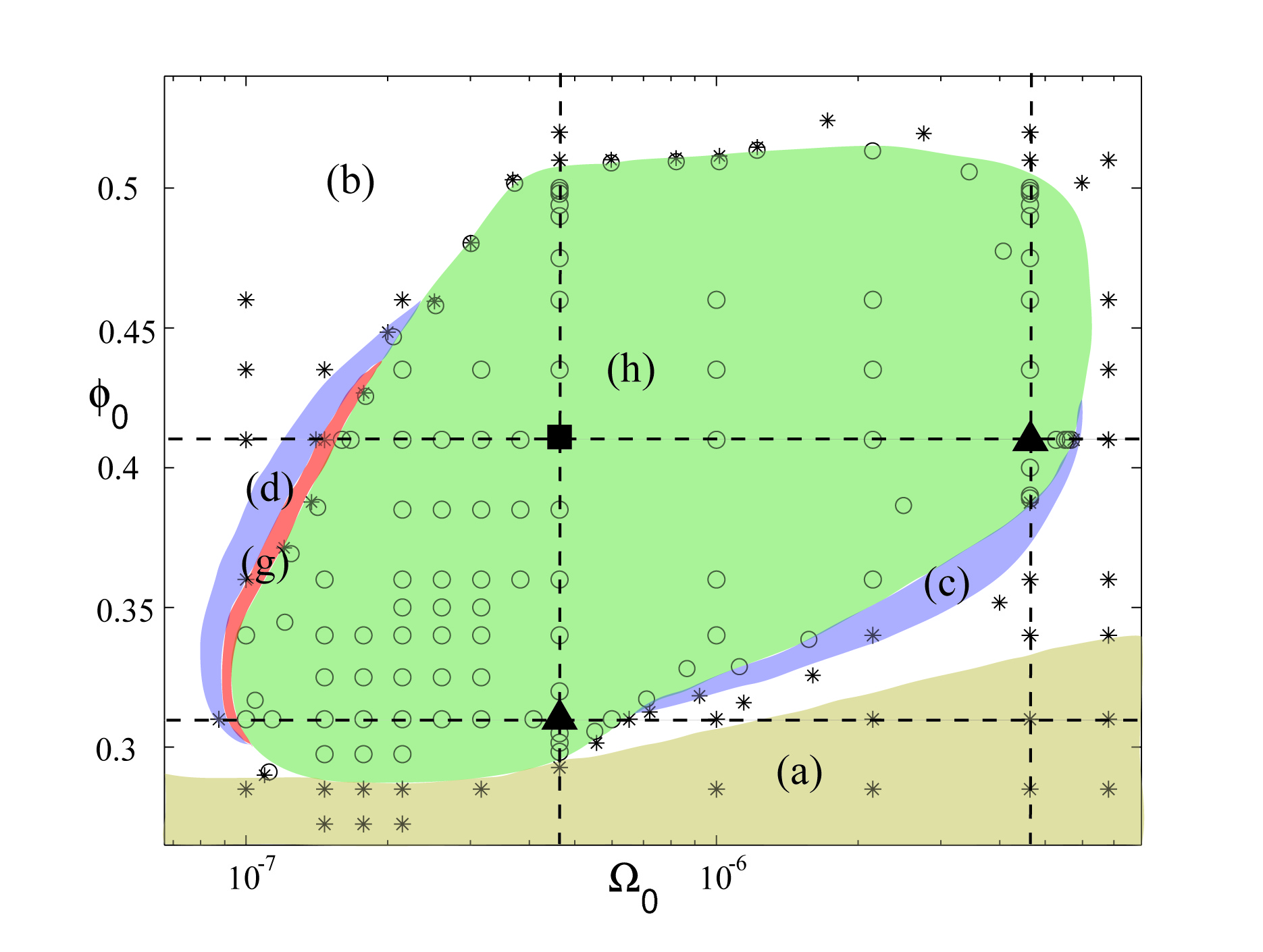}
\caption{(Color online) Morphological phase diagram of deposition
  patterns, in the plane spanned by the evaporation number
  $\varOmega_0$ and the bulk concentration $\phi_0$. Letters (a)-(d),
  (g) and (h) in the differently shaded areas (coloured online)
  indicate the type of pattern that is found there using the same
  letters as in the caption of Fig.~\ref{f:sampleprof}.  In
  particular, in the central grey area (green online) regular line
  patterns are found after some transient (simulations denoted by
  $\circ$), while outside this area (simulations denoted by $\ast$) a
  layer of constant height is deposited after a variety of transients
  indicated by subregions of different shading (colour online). As an
  exception, an intermittent pattern is found in the small (red
  online) region (g).  Results obtained along the dashed vertical and
  horizontal lines are presented in subsequent figures and discussed
  in detail in the main text. The standard configuration defined at
  the end of Section~\ref{sec:model} and shown in
  Figs.~\ref{f:regimes} to \ref{f:xtph}, is marked by the small filled
  square where the left vertical and upper horizontal dashed lines
  cross. Fig.~\ref{f:xtphp} gives results for the two other crossing
  points that are marked by filled triangles.}
\mylab{f:phasediag}
\end{figure}
%

\subsection{Types of deposition patterns}
\mylab{sec:patt-type}

An extensive parameter scan in the space spanned by our control
parameters -- $\varOmega_0$, $\phi_0$, $\mathrm{Pe}^{-1}$, $\chi$,
$M$, and $\nu$ -- reveals a zoo of various different deposition patterns and
allows one to study the dependence of the pattern morphology on
the control parameters.
An overview of typical deposit profiles, $\phi(x)$, is displayed in
Fig.~\ref{f:sampleprof}. The letters (a)--(j) that indicate the
individual panels are used in the following list that describes
important properties of the individual patterns. They are also used to
indicate the corresponding regions in the morphological phase diagram
in Fig.~\ref{f:phasediag}.
\begin{itemize}
\item[(a)]  After a short transient during which the height of the deposit can vary
 nonmonotonically (there can be a very small bump but the solution is not in
 jammed state), a layer of constant thickness is left behind the moving
 front. This is normally observed for very dilute solutions at any
 evaporation number. The solute concentration in the resulting film is everywhere
 below the jamming threshold, i.e.\ the solute may still diffuse within the precursor film.
 This is in contrast to all the other
 cases where at least parts of the deposit (normally the lines) are
 in the jammed state.
\item[(b)] A single line is deposited as the final part of a long initial
 transient before a flat layer is deposited. This represents the case
 closest to the original coffee stain effect where a
 single ring is deposited from an evaporating droplet. This behaviour
 is observed for a wide parameter range outside the region of
 periodic patterns [see (h) below] and for denser solutions
 than in case (a). Apart from in the region of the initial
 line deposit, the solute in the flat layer is not necessarily jammed.
\item[(c), (d)] A finite number of lines is deposited as part of a
  long initial transient before a flat layer of solute is
  deposited. This represents the experimental case where multiple
  irregular rings are deposited from an evaporating droplet. This
  behaviour occurs in bands around the region of periodic patterns.
  Panels (c) and (d) in Fig.~\ref{f:sampleprof} present deposit
  profiles from the regions with higher and lower $\varOmega_0$ than
  in the periodic line region (cf.~Fig.~\ref{f:phasediag}),
  respectively.
\item[(f)]
 A very long transient of a periodic pattern of double lines. The
 pattern slowly evolves towards one with regularly spaced lines by slowly
 changing the distance between the two lines in the respective pairs. The
 initial transient deposit is a flat layer and the size and shape of the
 first line is not significantly different from the following lines.
 This pattern is observed for very low solute concentrations and for somewhat small
 evaporation rates in the region of periodic lines close to its boundary.
\item[(g)] An intermittent pattern (a magnification of a portion of
 this is displayed in Fig.~\ref{f:sampleprof}(e)).
 A rather irregular line pattern is deposited in an
 intermittent manner, i.e., there are times when a nearly flat layer is
 left behind the front which alternate in a non-periodic way with episodes where 
 line patterns are deposited. These line patterns are not periodic
 but exhibit a typical timescale; see the magnification in Fig.~\ref{f:sampleprof}(e).
 This behaviour is found in a very narrow band
 between the region of periodic patterns and the band where single or
 multiple lines form (cf.\ Fig.~\ref{f:phasediag}). 
\item[(h)] After a short monotonous transient, the deposition of lines
 starts with a large first line followed by a small number of
 transient lines. The transient lines in Fig.~\ref{f:sampleprof}(h)
 are of increasing period, however, the case of decreasing period is
 also observed. The line deposition rapidly converges to a regular
 periodic state. For instance, Fig.~\ref{f:sampleprof}(h) shows
 a case where a rather high initial line terminates in a shoulder
 that is then followed by lower lines that slowly increase in height
 and converge to the truly periodic line pattern. This profile is at
 the standard values of our control parameters; see also the results
 in Figs.~\ref{f:timesnaps} and \ref{f:xtph}. This
 behaviour is found in a large region of the space spanned by our
 main control parameters (Fig.~\ref{f:phasediag}). The region may
 shrink and disappear if other parameters are strongly changed away
 from their standard values, such as, e.g.\ increasing the diffusion
 coefficient leads to the pattern
 formation being suppressed (see section~\ref{sec:diff}).
\item[(i)] Stable double lines can be observed for values of the parameter
 $\chi$ in the disjoining pressure that are less than our standard value.
 For the case displayed in Fig.\ \ref{f:sampleprof}(i), we observe that
 a large initial line is deposited followed by some irregular lines
 with monotonically increasing line distance. However, unlike the
 transient double lines in case (f), here the distances adjust till a
 regular pattern of double lines is deposited in a stable manner.
\item[(j)] After the initial transient deposition of a finite number of lines it
 at first seems that a region of flat film follows. However, the flat
 film actually becomes a long-period pattern that switches between
 a flat layer and single
 lines with a leading depression. This pattern was observed for a lower
 value of the viscosity exponent, $\nu=1.25$, that corresponds, e.g., to
 a suspension of non-spherical particles. A similar pattern was
 observed at the boundary of region (h), where $\phi_0=0.31$ and
 $\varOmega_0$ is small. In this case, switching between a flat layer and
 multiple lines occur. 
\end{itemize}
Note, that in some cases where the concentration is small, the concentration
in the `valleys' between the periodic lines is sufficiently low that the solution
is not jammed in these regions; an example is given in Fig.~\ref{f:sampleprof}(f).

As our parameter space is 6 dimensional and the equations of the model
(\ref{e:tfeqh}) and (\ref{e:tfeqhp}) are highly nonlinear, our list of
typical patterns must certainly be incomplete. As our investigation is
numerical in nature, we are unable to determine to which state some of
the very long initial transients converge. Below, we discuss the
various transitions between the observed patterns on changing a
control parameter. This allows us to speculate with more confidence
which further types of patterns might be expected.

Although one is easily able to qualitatively classify the transient
patterns, a quantitative analysis is cumbersome and therefore is not
pursued here. Discarding the initial transients, we can nevertheless
distinguish several types of deposits: (a)--(d) flat layers; (f), (h)
regular periodic lines; (g) intermittent line patterns; (i) periodic
arrays of double or multiple lines; (j) periodic switching between a
depression-line combination and a flat layer. Our main aim here is to
analyse the periodic line patterns that are observed in region (h) of
Fig.~\ref{f:phasediag}. This is done in the following sections, where
we analyse the dependence of the line morphology on selected control
parameters, while keeping the other parameters fixed. We have
carefully checked that the patterns are robust by using various values
of the parameters of our numerical solvers; see
Appendix~\ref{subs:robustness}.

\begin{figure}[tbh]
\centering
\includegraphics[width=0.9\hsize]{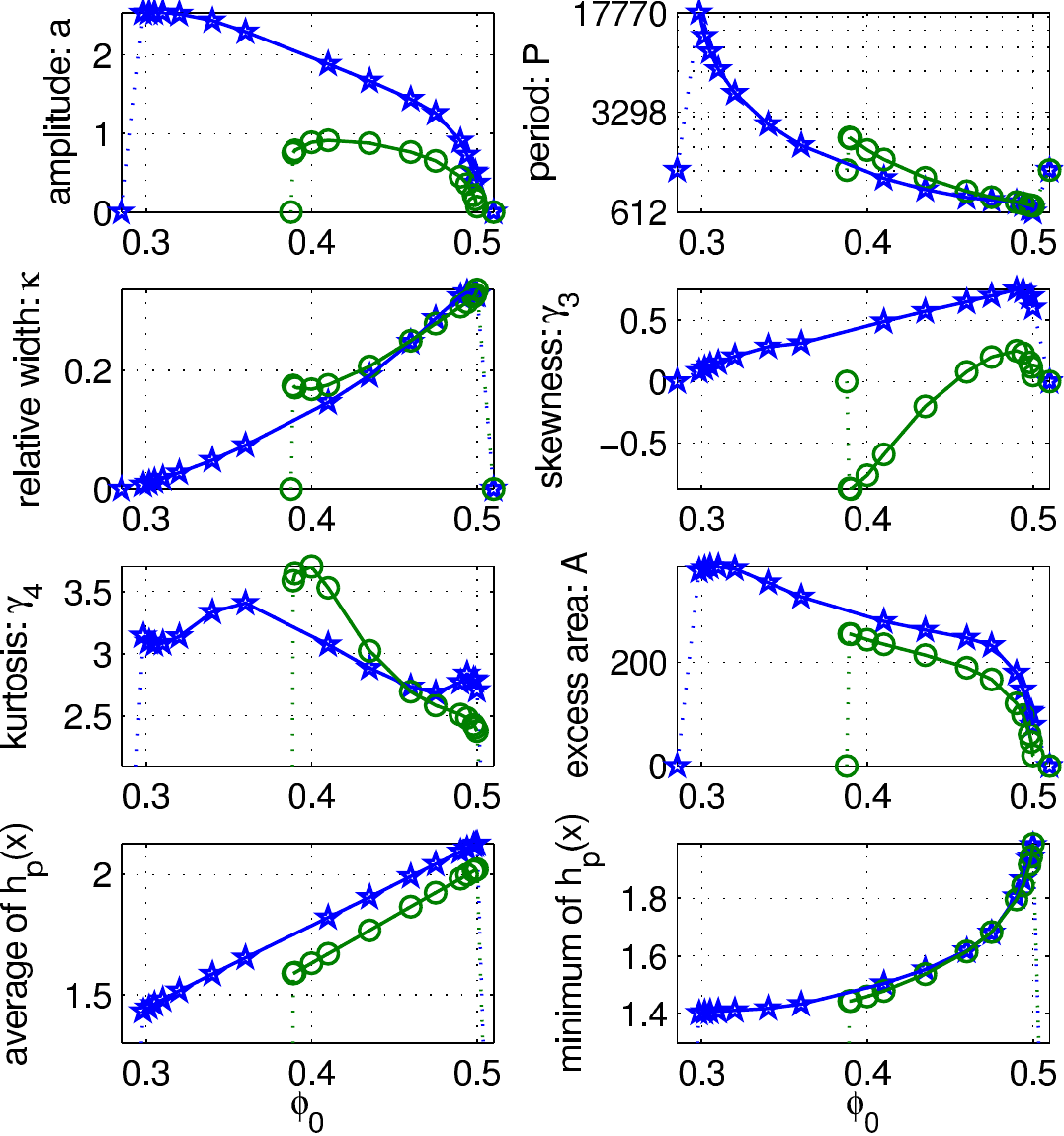}
\caption{ (Color online) Measures characterising the regular lines as
  a function of $\phi_0$, for fixed standard
  $\varOmega_0=4.64\times10^{-7}$ (blue line with $\star$) and for
  fixed $\varOmega_0=4.64\times10^{-6}$ (green line with
  $\circ$).\cite{FAT12_noteplots} The other parameters are fixed at
  their standard values. Selected corresponding line patterns are
  displayed in in Fig.~\ref{f:vertcutprof}.  }
\mylab{f:vertcut}
\end{figure}

\begin{figure}[tbh]
\centering
\includegraphics[width=0.9\hsize]{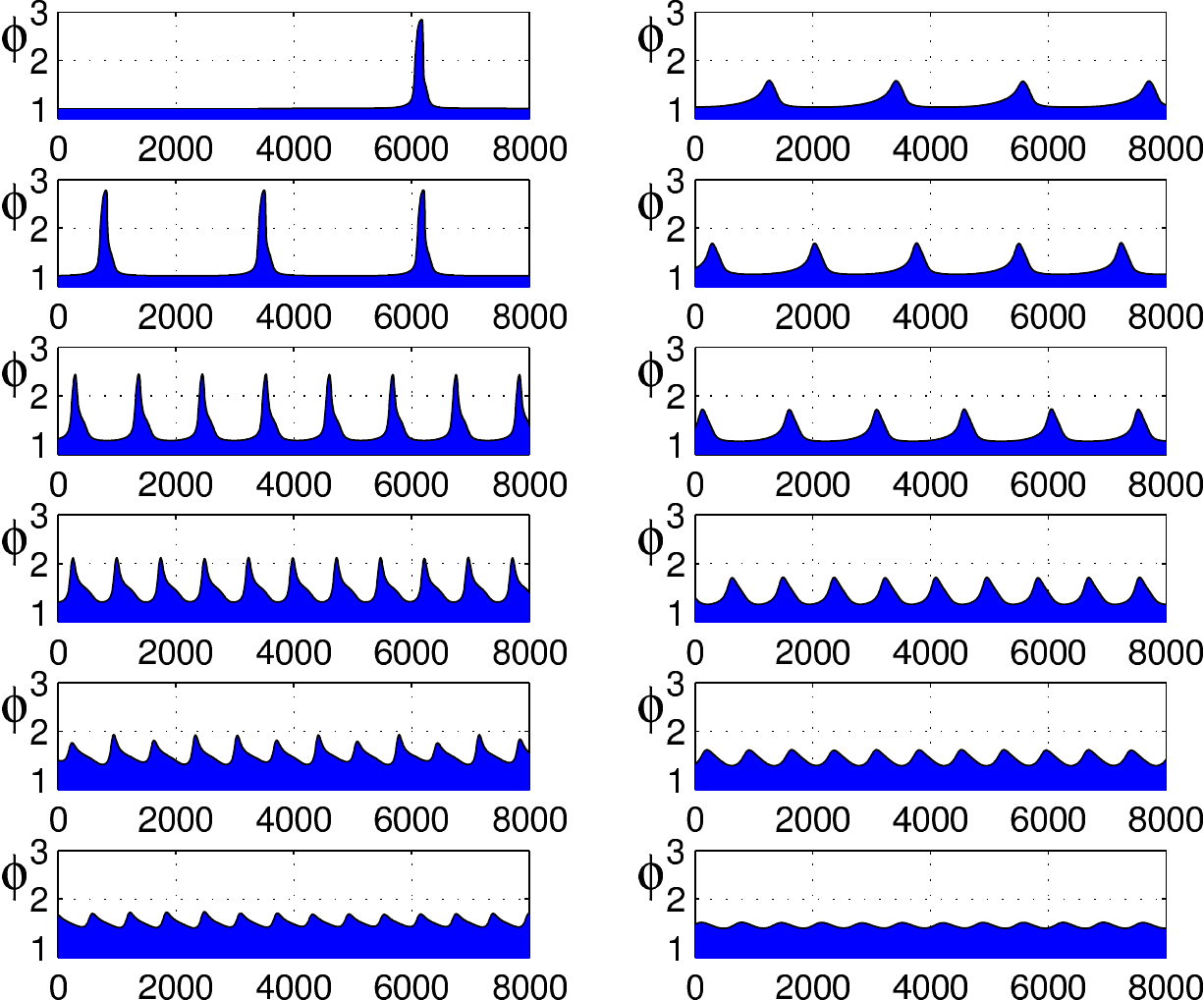}
\caption{(Color online) Morphology changes of the line pattern as
$\phi_0$ is varied, corresponding to Fig.~\ref{f:vertcut}.
Left column: For fixed standard $\varOmega_0=4.64\times10^{-7}$ and,
from the top to the bottom, $\phi_0
= 0.3016, 0.34, 0.41, 0.475, 0.494, 0.5$, and other parameters are standard.
Right column: For fixed $\varOmega_0=4.64\times10^{-6}$ and, from the top
to the bottom, $\phi_0
= 0.389, 0.4, 0.41, 0.46, 0.49, 0.499$, and other parameters are standard.}
\mylab{f:vertcutprof}
\end{figure}

\subsection{Regular patterns in the $(\varOmega_0,\phi_0)$ plane}
\mylab{sec:line-prop}

%
Having discussed the dynamics of the deposition
process (section~\ref{sec:front-dyn}) and the main types of transient
and long-time deposition patterns (section~\ref{sec:patt-type}), we
now embark on a more detailed analysis of the morphology of
the regular line patterns and its dependence on
the location in the parameter plane $(\varOmega_0,\phi_0)$. 
On the basis of our exploration of this parameter plane we detected regions
of characteristic deposition patterns -- see Fig.~\ref{f:phasediag}.
The sub-regions marked by letters in Fig.~\ref{f:phasediag} relate to
the patterns under the same letters in Fig.~\ref{f:sampleprof} and the
list of typical patterns in Sec.~\ref{sec:patt-type}.
The following analysis is based on a large number of long-time simulations (for
details see Appendix~\ref{sec:numerics}) in the region (h) of periodic
lines, cf.\ Fig.~\ref{f:phasediag}. Excluding the initial
transient, we take a sequence of $N$ regular deposition periods
(lines), where $10 \lessapprox N \lessapprox 100$, depending on the
spatial period of the deposit and the required CPU time. We process
the profiles and extract measures that characterise the individual
lines and their arrangement. In particular, we obtain the amplitude
$a$, spatial period $P$, relative width $\kappa=2\sigma/P$ (where
$\sigma$ is the standard deviation), skewness $\gamma_3$, kurtosis
$\gamma_4$, excess cross-sectional area of the lines $A$, the mean
deposit height $\overline{\hpfield}$, and the minimal height of
the deposit between the lines $\tilde\hpfield$. Definitions of
all these measures are given in Appendix~\ref{sec:measures}. Our
investigation shows that these quantities strongly depend on both the
evaporation number $\varOmega_0$ and the concentration $\phi_0$. We
focus on two vertical and two horizontal cuts through the parameter
plane $(\varOmega_0,\phi_0)$ that are indicated by dashed straight
lines in Fig.~\ref{f:phasediag}.

First, we vary the concentration $\phi_0$ for two evaporation numbers,
$\varOmega_0=4.64\times10^{-7}$ and $\varOmega_0=4.64\times10^{-6}$,
while the remaining system parameters are fixed at our standard values (see
end of section~\ref{sec:model}). Fig.~\ref{f:vertcut} presents the
line characteristics for both cases and Fig.~\ref{f:vertcutprof} shows
a number of corresponding deposit profiles.

Most of the line characteristics behave qualitatively similar for the
two cases. We first focus on $\varOmega_0=4.64\times10^{-7}$ and
then point out the differences. On increasing $\phi_0$ from a
region where there is no deposition of periodic lines, one first encounters large
amplitude almost solitary peaks that are separated by very large distances. On
further increasing $\phi_0$, the amplitude first hardly changes and
later decreases. The period rapidly decreases while the
relative line width and skewness increase almost linearly with a
slight drop in skewness at very high concentrations
$\phi_0\gtrapprox0.49$. For higher
$\phi_0$, the deposit pattern almost turns into a constant thickness layer,
with a small amplitude harmonic modulation. Finally, the amplitude goes
to zero (at finite period) at the upper border of region (h).
For $\varOmega_0=4.64\times10^{-6}$, the evaporation is stronger and one
must go to a higher $\phi_0$ than when
$\varOmega_0=4.64\times10^{-7}$, to see periodic deposition of lines. The
period of the lines does not seem to diverge at this border and the
amplitude of the lines is much smaller. Actually, the amplitude first
increases with increasing $\phi_0$ and has a maximum well inside the
region of regular lines before it decreases towards zero at the other
border of the region. Interestingly, in contrast to the former case,
the skewness changes from negative to positive values, i.e., the
individual lines change their morphology from having a `tail' away
from the receding front to having a tail in the direction towards the
front, as is always the case for $\varOmega_0=4.64\times10^{-7}$.
In both cases the excess area greatly decreases, the minimum height of
the deposited valleys between lines greatly increases, and the period and
amplitude decrease as the
boundary of region (h) at higher $\phi_0$ is approached. The average
height of the deposit linearly increases with the concentration as
expected but, interestingly, the two linear dependencies for the two
evaporation rates are shifted with respect to each other. This aspect
is further discussed in Appendix~\ref{subs:bc}.

\begin{figure}[tbh]
\centering
\includegraphics[width=0.9\hsize]{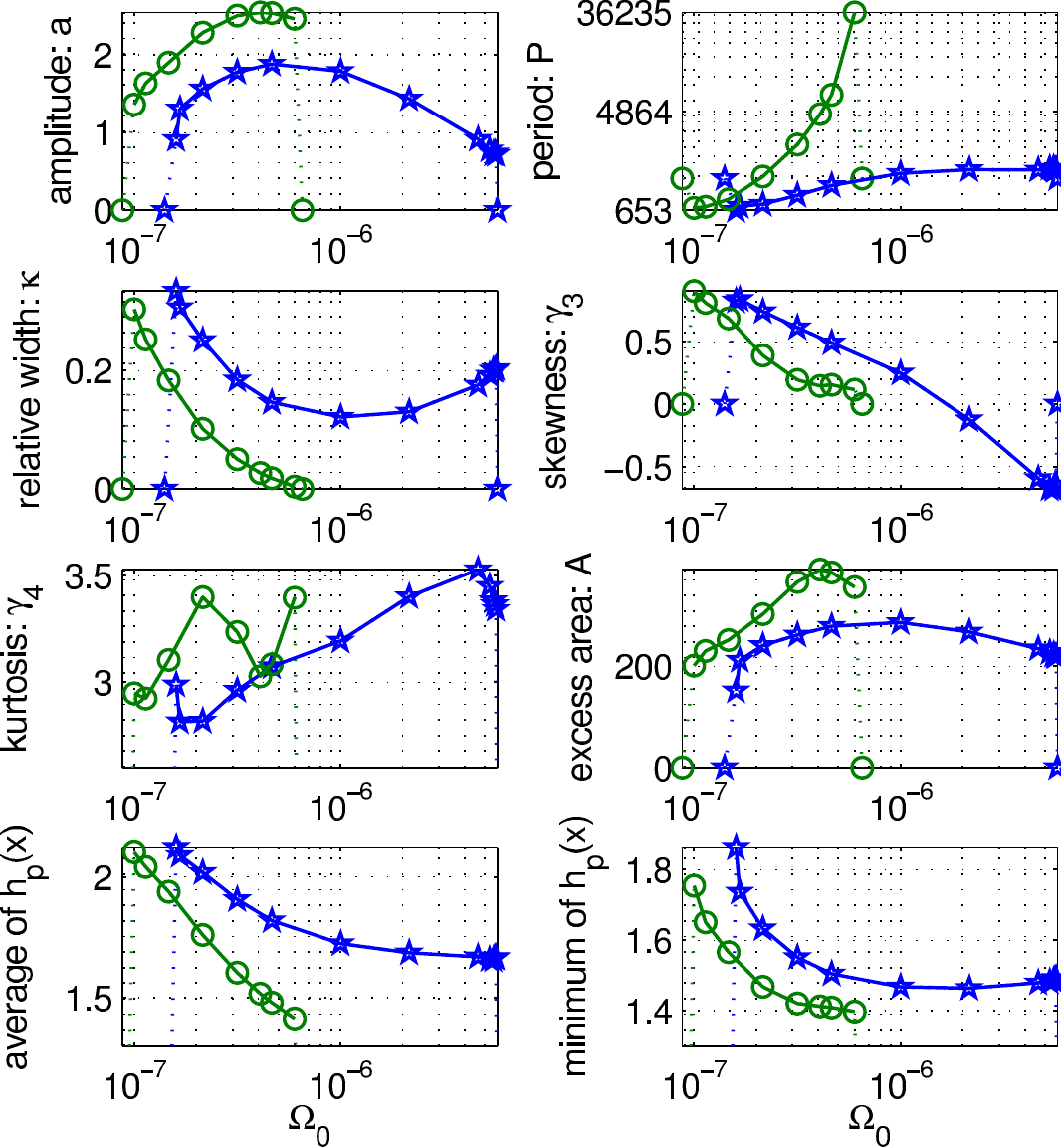}
\caption{(Color online) Measures characterising the regular lines as a
  function of $\varOmega_0$, for fixed standard $\phi_0=0.41$ (blue
  line with $\star$) and for fixed $\phi_0=0.31$ (green line with
  $\circ$).\cite{FAT12_noteplots} The rest of the parameters are fixed
  at the standard values.  Selected corresponding line patterns are
  displayed in Fig.~\ref{f:horizcutprof}.  }
\mylab{f:horizcut}
\end{figure}
\begin{figure}[tbh]
\centering
\includegraphics[width=0.9\hsize]{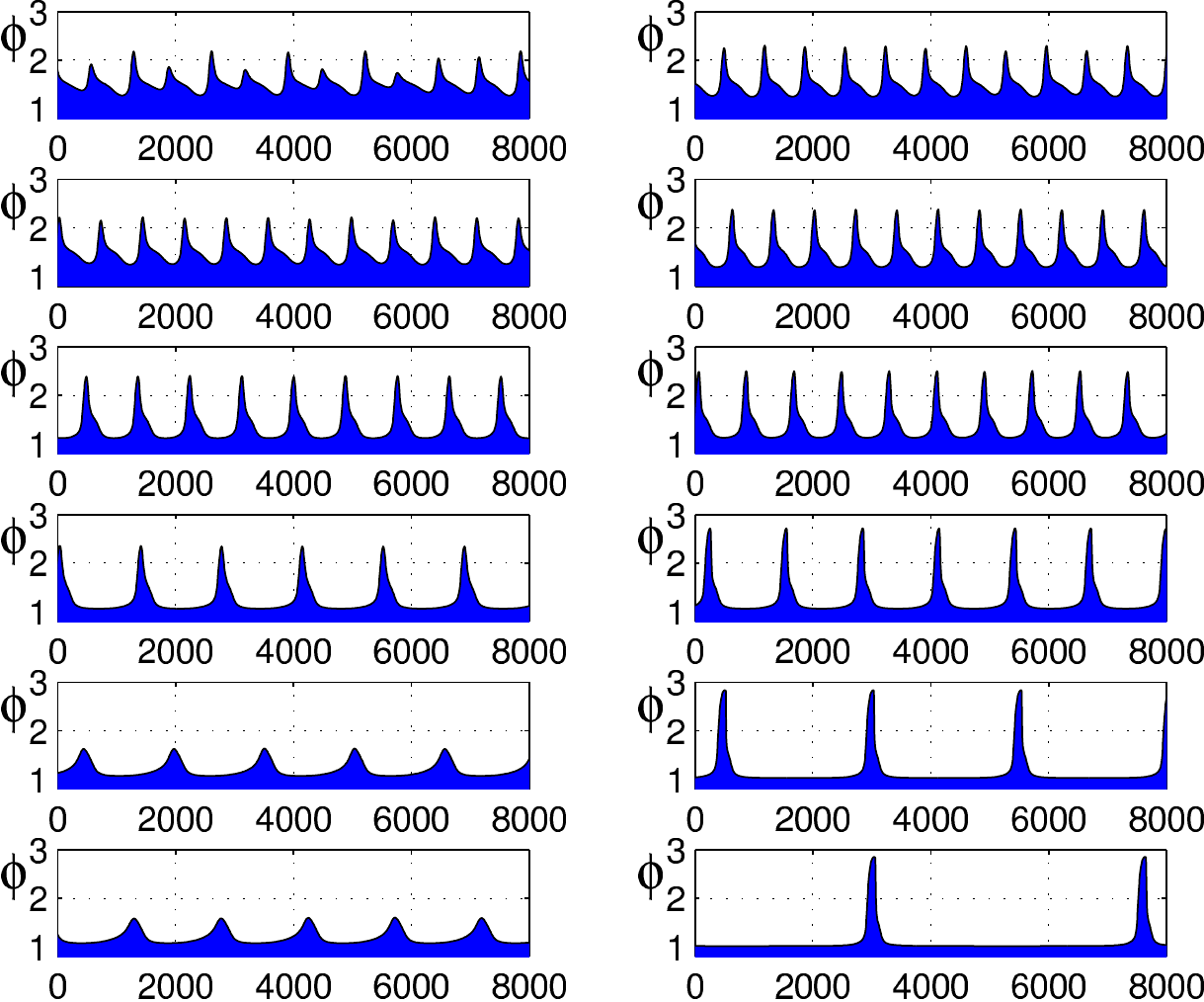}
\caption{(Color online) Morphology changes of the line pattern as
$\varOmega_0$ is varied, corresponding to the results in Fig.~\ref{f:horizcut}.
Left column: Fixed standard $\phi_0=0.41$ and, from the top to the bottom,
$\varOmega_0=(0.1598,0.1667,0.3162,1,5.274,5.663)\times10^{-6}$.
Right column: Fixed $\phi_0=0.31$ and, from the top to the bottom,
$\varOmega_0=(1,1.137,1.468,2.15,3.162,4.083)\times10^{-7}$.}
\mylab{f:horizcutprof}
\end{figure}
Second, we vary the evaporation number $\varOmega_0$ for two
concentrations $\phi_0=0.41$ and $\phi_0=0.31$, while the remaning
parameters are fixed at our standard values (see end of
section~\ref{sec:model}). Fig.~\ref{f:horizcut} presents the line
characteristics for both cases and Fig.~\ref{f:horizcutprof} displays a
number of the corresponding deposit profiles.

We first focus on the case $\phi_0=0.41$ and then point out the differences
found for the lower concentration value. Increasing $\varOmega_0$,
one moves from the narrow region (d) of transient multiple lines
(see Fig.~\ref{f:phasediag}), passes through a very narrow band of
intermittent line patterns (g) followed by the region (h) in which we observe the regular
line patterns. For the lowest values of $\varOmega_0$ in (h), the
patterns have a relatively small period and a small but non-zero
amplitude. The strongly anharmonic peaks are skewed to the right with
their tail pointing towards the wet side. On increasing
$\varOmega_0$ the period increases. The amplitude, however, first
increases and then decreases, until at a certain threshold, the pattern
ceases to be periodic and we arrive in the narrow border region (c),
where only a finite number of lines are deposited. Correspondingly,
the relative width of the lines decreases as one moves deep into
the region of periodic lines, where the lines become more
peaked. Remarkably, the skewness changes sign here, i.e.\ as $\varOmega_0$
increases the
tail of the lines shifts from pointing towards the receding wet film, to
pointing away.

The second cut with varied $\varOmega_0$ is at fixed $\phi_0=0.31$
and the standard values for the remaining parameters.
For small $\varOmega_0$, the onset is similar as for the previous cut at
$\phi_0=0.41$. However, on increasing $\varOmega_0$,
one finds large-amplitude long-period line patterns. The
period seems to diverge in a manner similar to that found when
increasing $\phi_0$ for
fixed $\varOmega_0=4.64\times10^{-7}$ (see Fig.~\ref{f:vertcut}).
Note that in Fig.~\ref{f:horizcut} the average deposit thickness
changes as $\varOmega_0$ is varied, even though the initial concentration
$\phi_0$ is fixed. The deposited average thickness for more dilute solute
concentrations can be higher than when a denser solution is used,
as long as the evaporation rate is higher. This effect stems from the influence
the moving evaporation front has on the concentration field in the
thick liquid film. This is related to our boundary conditions and is
further discussed at the end of Appendix~\ref{subs:bc}.

Note, that a change in the sign of the skewness as described above, was
also observed in experiments on nanoparticle suspensions \cite{philnote}
and can be explained as follows: For smaller values of $\varOmega_0$
and/or higher $\phi_0$, the capillary ridge is large and accommodates a
large amount of solute as it recedes. When the front pins, the
capillary ridge evaporates and the solute contained in the ridge is deposited in a
thick tail pointing towards the receding front. When the liquid front
depins, it carries much of the solute within the liquid front away with it, which results in a
further drop in the deposition thickness (seen as the final shoulder
of the tail). For higher values of $\varOmega_0$ the capillary ridge
is smaller and so the tail pointing towards the receding front is
smaller. On the other hand a negative skewness is typical for higher
$\varOmega_0$ and (possibly) low $\phi_0$. There, one encounters smaller tails
towards the receding front and also a pronounced tail away from the
front. In this situation, it takes some time (during which the front travels some distance)
for the solute concentration to build up at the front. During this time the front
gradually slows down and deposits the solute with growing thickness until
local jamming occurs. Then the front pins and starts to deposit a line.
The resulting asymmetry is seen as negative skewness.

\subsection{Onset of formation of periodic deposits}
\mylab{sec:limit-period}

To further clarify how the onset of pattern formation occurs we next
focus on the small band in the parameter plane $(\varOmega_0,\phi_0)$
that bounds the region of periodic deposits. From inspecting
Fig.~\ref{f:phasediag} and the line deposition descriptions outlined in the previous section,
one clearly sees that the onset of the formation of periodic deposits
may occur through a number of different transitions.

The most intricate transition is the one that involves the
intermittent deposits shown in Fig.~\ref{f:sampleprof}(g) and (e),
and occurs in the narrow region (g) of Fig.~\ref{f:phasediag}. We find
that this behaviour is very sensitive to computational details (which
is not the case for the regular line patterns), a fact which bolsters our
opinion that the intermittent line patterns represent a `chaotic
deposition'. Taken in this context, the occurrence of periodic deposits of
double lines (Fig.~\ref{f:sampleprof}(f),(i) would be related to the periodic
deposition of single lines through a period doubling
bifurcation that occurs when changing the relevant control parameter.
Both effects point towards the presence of chaos in the system, as
they are elements of the intermittency and period-doubling route to
chaos, respectively.\cite{Stro94} However, to investigate these
effects further, simpler asymptotic models for the moving
material-depositing front need to be constructed that are numerically
less challenging.

\begin{table*}
\centering
\begin{tabular}{| l ||c|c|c|c|c|}
\hline
fixed & $\varOmega_0=4.64{\cdot}10^{-7}$
 & \multicolumn{2}{c|}{$\phi_0=0.41$}
 & \multicolumn{2}{c|}{$\varOmega_0=4.64{\cdot}10^{-6}$}\\
\hline
varied
 & \ high $\phi_0$ \ 
 & \ low $\varOmega_0$\  & \ high $\varOmega_0$\ 
 & \ low $\phi_0$\      & \ high $\phi_0$\ \\
\hline\hline
$\varDelta_\mathrm{abs}$ & 0.0004 & 0.0359 & 0.0546 & 0.0134 & 0.0023 \\\hline
$\varDelta_\mathrm{rel}$ & 0.0018 & 0.0197 & 0.0299 & 0.0596 & 0.0102 \\\hline
\end{tabular}
\caption{The measured hysteresis intervals, $\varDelta_\mathrm{abs}$ and
$\varDelta_\mathrm{rel}$, are listed for the various transitions 
between the long-time deposition of periodic patterns and flat layers
that occur along the straight lines in Fig.~\ref{f:phasediag}. In these intervals
different deposition patterns may be obtained, depending on initial
conditions. The first row indicates which parameter is held fixed
whilst the second row states which parameter is varied and which
border of the region of periodic deposits is being considered. The third row
gives the absolute hysteresis interval $\varDelta_\mathrm{abs}$ in
terms of the parameter that is being varied, whereas the fourth row gives the
relative interval size $\varDelta_\mathrm{rel}=\varDelta_\mathrm{abs}/\varDelta$,
where $\varDelta$ is the corresponding size of the entire region of periodic deposits
in the $(\log_{10}\varOmega_0,\phi_0)$ plane ($\varDelta= 1.8239$ in $\log_{10}\varOmega_0$ coordinate and
$\varDelta=0.225$ in $\phi_0$ coordinate). Note that
$\varDelta_\mathrm{abs}$ is measured in terms of the particular parameter
that is being varied (either $\log_{10}\varOmega_0$ or $\phi_0$),
i.e., it represents a length in the semilogarithmic phase plane Fig.~\ref{f:phasediag}.
Note, that we do not include results for fixed $\varOmega_0=4.64{\times}10^{-7}$ at low $\phi_0$.
There the line period
becomes very large and long pieces of nearly flat layers occur between the lines.
This results in rather expensive computations. However, our limited results
indicate that there is no hysteresis in this case.
}
\mylab{tab:hyst}
\end{table*}
In general, it has proved to be difficult to detect the exact location
of the boundary of the region of periodic line deposits. A detailed
study reveals that close to the boundary different patterns may often
emerge for identical parameter values, depending on the initial
condition. In other words, many of the transitions are hysteretic: If one starts
inside the region (h) of Fig.~\ref{f:phasediag} with a simulation that
gives regular lines and then moves by small parameter increments
closer to the boundary one can detect at which parameter values the
deposit turns into a layer of constant thickness, and in this way
define a boundary of region (h). However, alternatively one may start
in the region outside (h), where after some initial transient one
obtains a flat deposit layer, and then slowly move towards the
boundary of (h). Detecting at which parameter value the flat deposit turns
into a regular line pattern, one finds that this transition occurs
at a point inside the region (h), i.e.\ there is a small
hysteretic region where both patterns may occur. Using this technique
we numerically obtain the width of the hysteresis region on the cuts
through the $(\varOmega_0,\phi_0)$ parameter plane that were discussed
in section~\ref{sec:line-prop} and indicated in
Fig.~\ref{f:phasediag}. The results for the width of the hysteresis region
are listed in Table~\ref{tab:hyst}.

From our investigations of the transitions in the
$(\varOmega_0,\phi_0)$ plane, we find
four typical scenarios for the onset of the formation of periodic
line patterns:
\begin{itemize}
\item[(i)] At low $\phi_0$ and small or intermediate $\varOmega_0$ the
  spatial period of the lines diverges while the line amplitude first
  slowly increases and then converges to a finite value (see
  Fig.~\ref{f:vertcut}). This indicates the occurrence of an infinite
  period bifurcation, that could be either a SNIPER (Saddle Node
  Infinite PERiod) bifurcation or a homoclinic
  bifurcation.\cite{Stro94} The fact that for
  $\varOmega_0=4.64{\times}10^{-7}$ we do not see any hysteresis (see
  Table~\ref{tab:hyst}) points towards a SNIPER bifurcation, as a
  homoclinic bifurcation often involves some hysteresis between a
  stable steady or stationary state (in the present case, the
  deposition of a flat layer) and a stable time-periodic state (here,
  the deposition of line patterns). Note, however, that we are not
  able to come sufficiently close to the bifurcation point to test
  whether the typical power law relation between the period and the
  distance to the bifurcation point also holds here.  A similar
  behaviour is found at high $\varOmega_0$ and fixed $\phi_0=0.31$
  (Fig.~\ref{f:horizcut}).  We discuss further at the end of this
  section how this finding compares to results for depinning
  transitions in other soft matter systems.
\item[(ii)] For high evaporation rates $\varOmega_0$ and high $\phi_0$ the 
  line amplitude decreases with increasing $\varOmega_0$ as
  the boundary of region (h) is approached, before suddenly jumping to zero
  (see Fig.~\ref{f:horizcut}).  At the same time, the line period tends to a finite value.
  Outside region (h), the pattern ceases to be periodic and the
  simulations show an initial transient deposition of a finite number of lines,
  followed by a flat layer. Since the hysteresis here is rather large
  (see Table~\ref{tab:hyst}), we conclude that the transition most likely
  corresponds to a subcritical Hopf bifurcation.
\item[(iii)] For low $\varOmega_0$ and all but the very small values of $\phi_0$, as
  the boundary is approached
  the line amplitude first decreases (with increasing rate)
  with decreasing $\varOmega_0$, before suddenly jumping to zero
  (see Fig.~\ref{f:horizcut}).  At the same time, the line period tends to a
  finite value. Just outside this boundary to region (h) the
  simulations exhibit a transient deposition of a finite number of lines,
  followed by a flat layer. Since there is some hysteresis
  (see Table~\ref{tab:hyst}), at first sight the transition seems to correspond to a subcritical Hopf
  bifurcation, and therefore seems to be rather similar to the case
  described in the previous point (ii).
  However, at intermediate $\phi_0$, close to the boundary of (but
  within) region (h) one sees signs of a period doubling and there is
  also the band of intermittent patterns of type (g) just outside of
  region (h) [cf.~Fig.~\ref{f:horizcutprof} (left column, first panel)
  and Fig.~\ref{f:sampleprof}(g) and (e), respectively]. This
  indicates that the transition might involve several complex
  eigenvalues and be related to the intermittency and period-doubling
  route to chaos.\cite{Stro94} Whatever is happening at the boundary,
  it is certainly more complex than case (ii).
\item[(iv)] The last case we mention is a hypothetical transition
  scenario that we did not see but that we believe is likely to occur. Our
  study is based on a model that can be computationally expensive,
  especially for values of $(\varOmega_0,\phi_0)$ that are close to the onset of the
  deposition of periodic lines. Table~\ref{tab:hyst} shows that the
  width of the hysteresis region $\varDelta_\mathrm{abs}$ varies along
  the boundary. In particular, at large concentrations
  $\varDelta_\mathrm{abs}$ becomes very small. We believe it is likely that 
  there may exist a small interval along the boundary where
  there is no hysteresis, i.e., $\varDelta_\mathrm{abs}=0$, and the
  line amplitude gradually decreases to zero while the line period
  approaches some non-zero value, and the deposition pattern becomes a
  small amplitude harmonic modulation. Such a transition would
  correspond to a supercritical Hopf bifurcation.
\end{itemize}

Note, that the present study only considers one-dimensional deposition
patterns. Two dimensional deposition patterns are beyond our scope. We
expect the full two-dimensional behaviour to be very rich. In
particular, at small evaporation rates one should expect the
evaporative dewetting front to be transversally unstable, even in the
situation without solute.\cite{LGP02} We expect this to also occur in
the case with solute, close to the transition discussed in points (i)
and (iii). This argument is bolstered by the observation that
  the particular transition described in (i) involves deposits that
  are nearly homoclinic in space. There exist generic results
  \cite{ACR01} that show that patterns near homoclinic solutions are
  prone to instabilities.

Before we move on to discuss in the next section the influence of
system control parameters besides the solute concentration $\phi_0$
and the evaporation number $\varOmega_0$, we make a few comments to 
put our findings into a wider context. It is important to
understand that the observed transitions from stationary front motion
to the deposition of lines may be seen as depinning transitions
in the frame moving with the mean front speed: When a front of constant
speed deposits a flat layer, in the comoving frame the concentration
profile is steady. Then, one may say that the concentration
profile is pinned to the moving front as it does not move relative to
it. However, at the transition to depositing a periodic
pattern, the concentration profile starts to stay behind the moving
front, and one may say it depins from the front.
Note, however, that after depinning, the concentration profile does not
move relative to the front as a whole. Instead, only a part of it
(the jammed part) starts to move relative to the front, resulting in
the deposition of a line. This process then repeats periodically.

From this observation it becomes clear why the transition scenarios
(i) to (iv) described above are analogous to similar scenarios found in
studies of depinning in other driven soft matter systems. To
illustrate how universal such transitions are, we mention three
systems: First, drops of simple nonvolatile liquids that sit on
heterogeneous substrates and depin from the heterogeneities under the
influence of external driving forces. Depending on the particular
setting and parameter regime, one may observe SNIPER, homoclinic and
super- or subcritical Hopf bifurcations.\cite{ThKn06,BHT09,BKHT11}
There, however, the entire depinning drop or ridge slides along the
heterogeneous substrate in a periodic manner.
A second system consists of clusters of interacting colloidal
particles that shuttle under the influence of external forces through
a heterogeneous nanopore.\cite{Poto10,poto11} Under weak dc driving, the peak
in the particle density distribution is pinned by the heterogeneities of the
pores. However, depending on driving force and the attraction between
the colloids, depinning transitions via Hopf and homoclinic
bifurcations occur resulting in time periodic fluxes. 
The third system is much closer related to the one studied here, as it
also concerns patterns that are produced at a three-phase contact
line: If a Langmuir--Blodgett surfactant monolayer is transferred from
a bath onto a solid substrate, one may observe stripe patterns in the
deposit resulting from substrate-mediated
condensation.\cite{RiSp92,SCR94,Lenh04} Time simulations of a
dynamical model for this system show the occurrence of patterns of
stripes parallel to the contact line (and other patterns
too).\cite{KGFC10} In a reduced model the depinning transitions from a
steady (pinned) concentration profile to the time periodic (depinned)
state may occur via a homoclinic bifurcation or a subcritical Hopf
bifurcation.\cite{KGFT12}

The comparison with these different soft matter systems that show
depinning, corroborates the picture we have given above in points (i)
to (iv). However, a systematic analysis of the various transitions
related to the onset of the deposition of line patterns in the present
system requires a simplified model to be developed.

In the following sections we briefly describe how varying other
physical aspects of the system influence the behaviour of the
system. In Sec.\ \ref{sec:diff} we examine the effect of varying the
diffusivity of the solute in the solvent. In Sec.\ \ref{sec:wett} we
discuss the influence of varying the wettability and also of
varying the chemical potential of the solvent in the ambient vapour.
In Sec.\ \ref{sec:rheology} we discuss effects of solvent rheology.

%
\subsection{Influence of diffusion}
\mylab{sec:diff}
%
\begin{figure}[tbh]
\centering
\includegraphics[width=0.9\hsize]{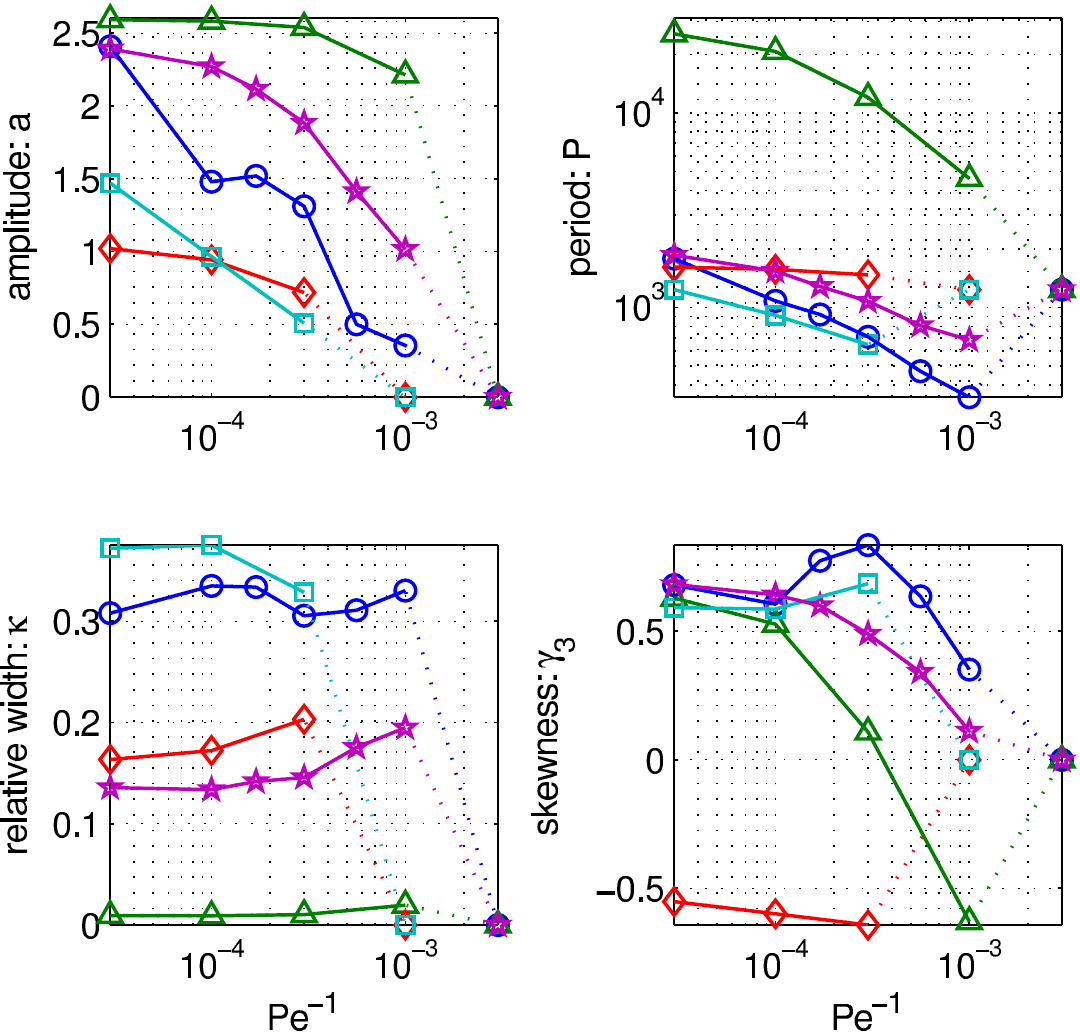}
\caption{(Color online) Measures characterising the regular line
  patterns as a function of $\mathrm{Pe}^{-1}$ for fixed
  parameters:\cite{FAT12_noteplots} $\star$ (purple line), the
  standard parameter values; $\circ$ (blue line),
  $\varOmega_0=1.667\times10^{-7}, \phi_0=0.41$;
  $\scriptstyle\triangle$ (green line),
  $\varOmega_0=4.64\times10^{-7}, \phi_0=0.3016$;
  $\scriptstyle\diamondsuit$ (red line),
  $\varOmega_0=5.663\times10^{-6}, \phi_0=0.41$; and
  $\scriptstyle\square$ (light blue line),
  $\varOmega_0=4.64\times10^{-7}, \phi_0=0.498$.  The purple line
  corresponds to states in the central part of the region (h) on
  Fig.~\ref{f:phasediag}, whereas the other lines correspond to points
  close to its boundary. Selected corresponding line patterns are
  shown in Figs.~\ref{f:diffcutprof} and \ref{f:diffcutprof2}.  Note,
  that for $\scriptstyle\triangle$ (green) the period $P$
  monotonically increases to very large values with decreasing
  $\mathrm{Pe}^{-1}$ (up to $P\approx30000$ at $\mathrm{Pe}^{-1}= 0$).}
  \mylab{f:diffcut}
\end{figure}
\begin{figure}[tbh]
\centering
\includegraphics[width=0.9\hsize]{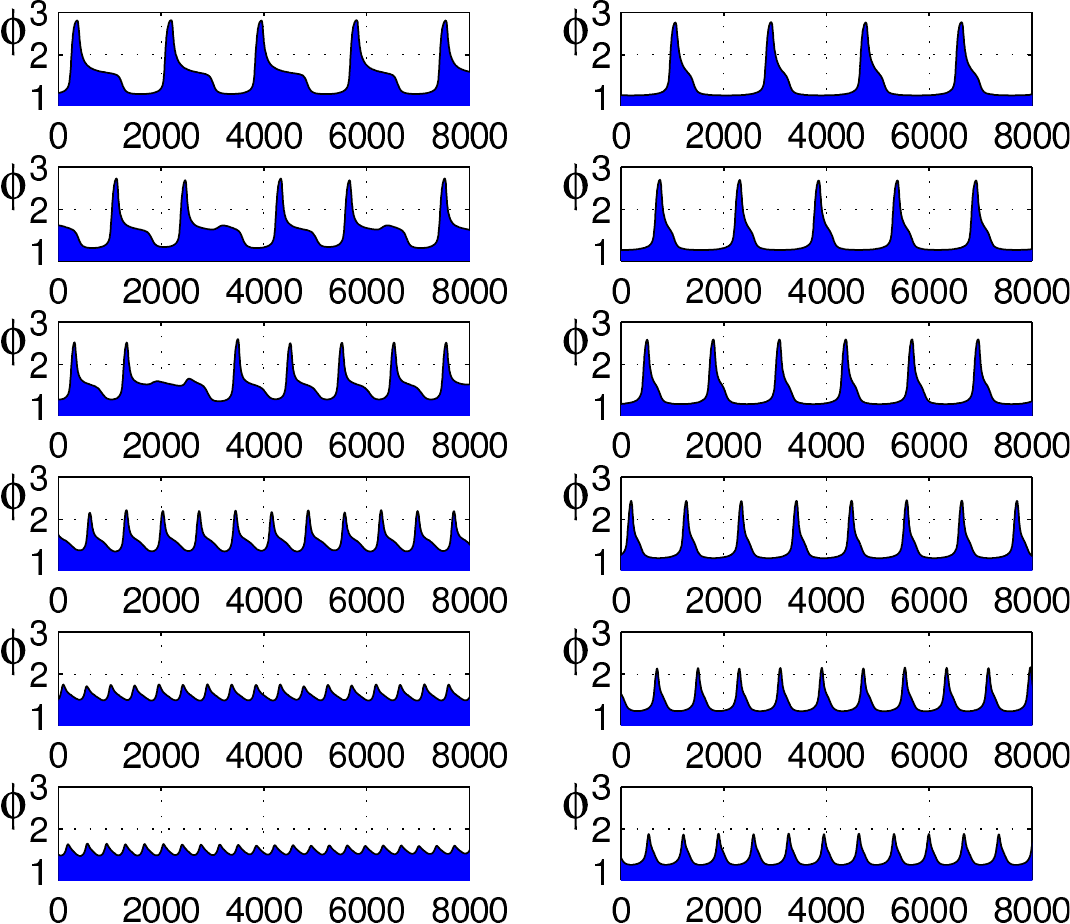}
\caption{(Color online) Morphology changes of the line pattern as
  $\mathrm{Pe}^{-1}$ is varied, corresponding to
  Fig.~\ref{f:diffcut}. The left column shows profiles corresponding to
  the blue line with $\circ$ symbols and the right
  column to the purple line with $\star$ symbols in Fig.~\ref{f:diffcut}.
  From top to bottom, the panels in both columns
  correspond to $\mathrm{Pe}^{-1}=3\times10^{-5}, 10^{-4}, 1.7\times10^{-4}, 3\times10^{-4},
  5.6\times10^{-4}$ and 0.001.}
\mylab{f:diffcutprof}
\end{figure}
\begin{figure}[tbh]
\centering
\includegraphics[width=0.9\hsize]{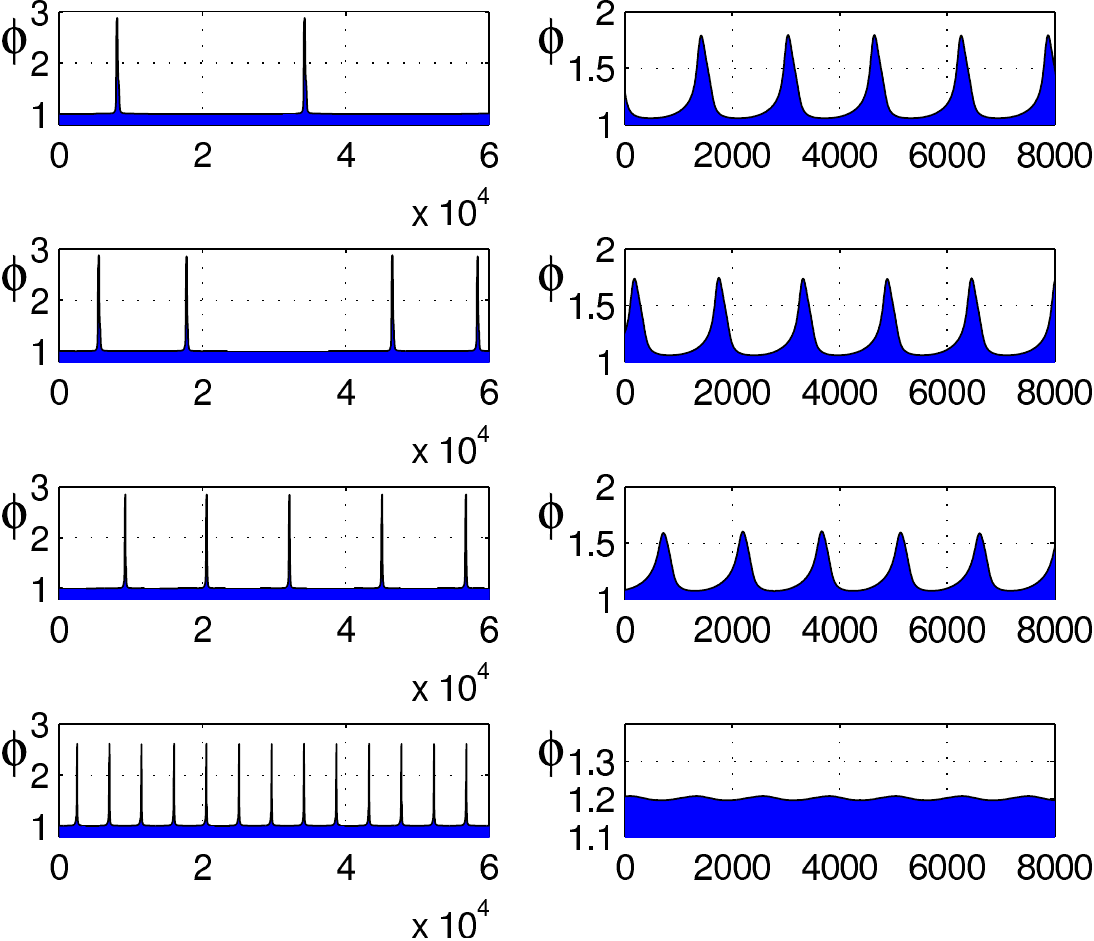}
\caption{(Color online) Morphology changes of the line pattern as
  $\mathrm{Pe}^{-1}$ is varied, corresponding to
  Fig.~\ref{f:diffcut}. The left column shows profiles corresponding to
  the green line with $\scriptstyle\triangle$ symbols and the right column to the
  red line with $\scriptstyle\diamondsuit$ symbols in Fig.~\ref{f:diffcut}.
  From top to bottom, the panels in both columns
  correspond to $\mathrm{Pe}^{-1}=3\times10^{-5}, 10^{-4}, 3\times10^{-4}$ and
  0.001.}
\mylab{f:diffcutprof2}
\end{figure}

The influence of diffusion is quantified in our model by the inverse P\'eclet
number $\mathrm{Pe}^{-1}$. Up to this point we have discussed the basic
mechanism as being based on a subtle balance of convective and
evaporative motion. In the results presented so far,
the solute diffusivity was low
and contributed little to the overall transport.

Fig.~\ref{f:diffcut} presents results for several measures
 which characterise the periodic
line patterns, as a function of the inverse P\'eclet number
$\mathrm{Pe}^{-1}$, which is a ratio of the time scales for
convection and diffusion. Results are shown for five sets
of values of the parameters $(\varOmega_0,\phi_0)$,
corresponding to locations within the region of periodic lines [region
(h) in Fig.~\ref{f:phasediag}]: symbols $\star$ indicate our standard
values of $(\varOmega_0,\phi_0)$ and the other sets correspond to
locations close to the boundary of region (h). Examples of the
corresponding line patterns are displayed in Figs.~\ref{f:diffcutprof} and
\ref{f:diffcutprof2}. All the other parameters, $\chi$, $M$, and $\nu$ are
equal to the standard values (see the end of Section~\ref{sec:model}).

Around and below the standard value of $\mathrm{Pe}^{-1}=0.0003$, the
deposition pattern is almost independent of the value of
$\mathrm{Pe}^{-1}$. Decreasing $\mathrm{Pe}^{-1}$ to zero has almost
no effect on the size of region (h), and the only effect is that the
deposit patterns becomes slightly sharper.  However, on increasing
$\mathrm{Pe}^{-1}$ above $0.0003$, the size of the region (h)
(Fig.~\ref{f:phasediag}) starts to shrink considerably, until it
vanishes entirely as the effects of solute diffusion increase
(roughly, when $\mathrm{Pe}^{-1}>0.003$).  The shape of the lines
changes monotonically and they become more sinusoidal in shape: the
amplitude, period and skewness all become smaller with increasing
$\mathrm{Pe}^{-1}$, while the relative width, $\kappa=2\sigma/P$
increases (see Fig.~\ref{f:diffcut}).  We remind the reader that
highly peaked or separated lines have small $\kappa$ values, whereas a
deposit profile that resembles a harmonic wave has a large $\kappa$
value. When the diffusive mobility of the solute is large the effect
of diffusion counteracts the solute build up due to evaporation. The
transition between the deposition of periodic lines and a flat layer
occurs in between $\mathrm{Pe}^{-1}=0.001$ and
$\mathrm{Pe}^{-1}=0.003$ for the standard values of
$(\varOmega_0,\phi_0)$ and at somewhat smaller $\mathrm{Pe}^{-1}$ for
parameter values close to the boundary of region (h), as the region
itself is shrinking. Such a large diffusivity is unlikely for large
nanoparticles in suspension but might occur for very small particles
or for small molecules in solution. Since the line amplitude
approaches zero at this transition point, whilst the period remains
finite and the line profiles become nearly harmonic, we suspect that
this transition corresponds to a Hopf bifurcation
(cf.~section~\ref{sec:limit-period}) but we did not study this
transition in as much detail as the transitions discussed in
Sec.~\ref{sec:limit-period}.

In the lower right panel of Fig.\ \ref{f:diffcut} we display
results obtained for the skewness, $\gamma_3$.
As $\mathrm{Pe}^{-1}$ is increased, the curves
marked by symbols $\circ$ and $\scriptstyle\square$ show at first an
increase of $\gamma_3$ before it decreases again just before the
pattern vanishes. The reason for this
non-monotonic behaviour can be seen in the left column of
Fig.~\ref{f:diffcutprof} that displays deposit profiles which correspond
to the $\circ$ curve in Fig.~\ref{f:diffcut}. There, going from the
first to the fourth panel from the top we see new secondary lines emerge out
of the tail of the primary lines, i.e., decreasing the effects of diffusion, leads to a
period doubling. The period doubling is not visible in the dependence
of the period on $\mathrm{Pe}^{-1}$ (see Fig.~\ref{f:diffcut}) because
around these parameter values the pattern is not very regular and the
secondary lines appear through a smooth transition.
These irregular patterns seem to be analogous to the intermittent
patterns [Fig.~\ref{f:sampleprof}(g)] or the depression--line patterns
[Fig.~\ref{f:sampleprof}(j)] but here these intermediate patterns
appear in a somewhat wider band of parameter values close to the
border of region (h) in the phase plane shown on
Fig.~\ref{f:phasediag}. The period of these irregular patterns
$P\approx P_\mathrm{shift}/2$, which indicates that the shift length
(see Appendix~\ref{sec:numerics}) interacts with the period of the
pattern. In general, all such irregular patterns appear only close to
the boundary of region (h) (Fig.~\ref{f:phasediag}) of periodic lines,
where $\varOmega_0$ is small and $\phi_0$ is moderate to high. In
other words, where patterns have large positive $\gamma_3$ (heavy
right tail), we observe the `chaotic deposition' described in
Sec.~\ref{sec:limit-period}, and we expect transversal instability
effects. In Appendix~\ref{subs:robustness} we discuss how robust the
results we obtain are, in relation to the numerical methods we use to
solve our model equations.  The profiles in Figs.~\ref{f:diffcutprof}
and \ref{f:diffcutprof2} also show that as the value of
$\mathrm{Pe}^{-1}$ is increased, the parameter regions where lines of
large positive skewness (long tail towards front) are found (smaller
$\varOmega_0$ and moderate and high $\phi_0$) become replaced by lines
that are nearly symmetric (small skewness) or point the tail away from
the front (negative skewness). In addition, the period of the lines
decreases.

Finally, we mention the importance of diffusion in experiments and
applications of such evaporation driven self-organisation processes.
The relative importance of diffusion depends on both the mobility of
the solute and the thickness of the solution layer. For instance, for
very thin films \cite{MOY03,Paul08} one sees from Eqs.~(\ref{e:j_c})
and (\ref{e:j_d}), that the diffusive mobility $Q_\mathrm{d}\propto h$
dominates the convective mobility $Q_\mathrm{c}\propto h^3$; see also
the discussion by \citet{Vanc08}, where
the dynamics in ultrathin postcursor films left behind mesoscopic
dewetting fronts of nanoparticle suspensions \cite{Paul08} is modelled by either
kinetic Monte Carlo models \cite{RRGB03,Vanc08,Stan08} or dynamical
density functional theory \cite{ART10,RAT11} that only incorporate
diffusive transport because in this situation diffusion is the dominant
process. The present thin film model captures the
competition of convective and diffusive transport relevant for larger film
thicknesses where the key influence is the mobility of the 
solute itself. \citet{XXL07} found that as
the size of nanoparticles is decreased, one obtains spoke-like structures
and irregular patterns instead of the regular rings seen for larger
nanoparticles. This indicates that an increase of the influence of
diffusion brings the system closer to the border of the region of
periodic line deposition, where patterns become irregular and unstable
in accordance with our present findings.

In summary, a low value for the solute diffusivity means diffusion
does not influence the deposition
patterns and diffusion may actually be neglected in the model. However, fast
diffusion is able to counteract evaporation and effectively suppresses
the occurrence of deposition patterns.

\subsection{Influence of wettability and chemical potential}
\mylab{sec:wett}
%

%
\begin{figure}[tbh]
\centering
\includegraphics[width=0.9\hsize]{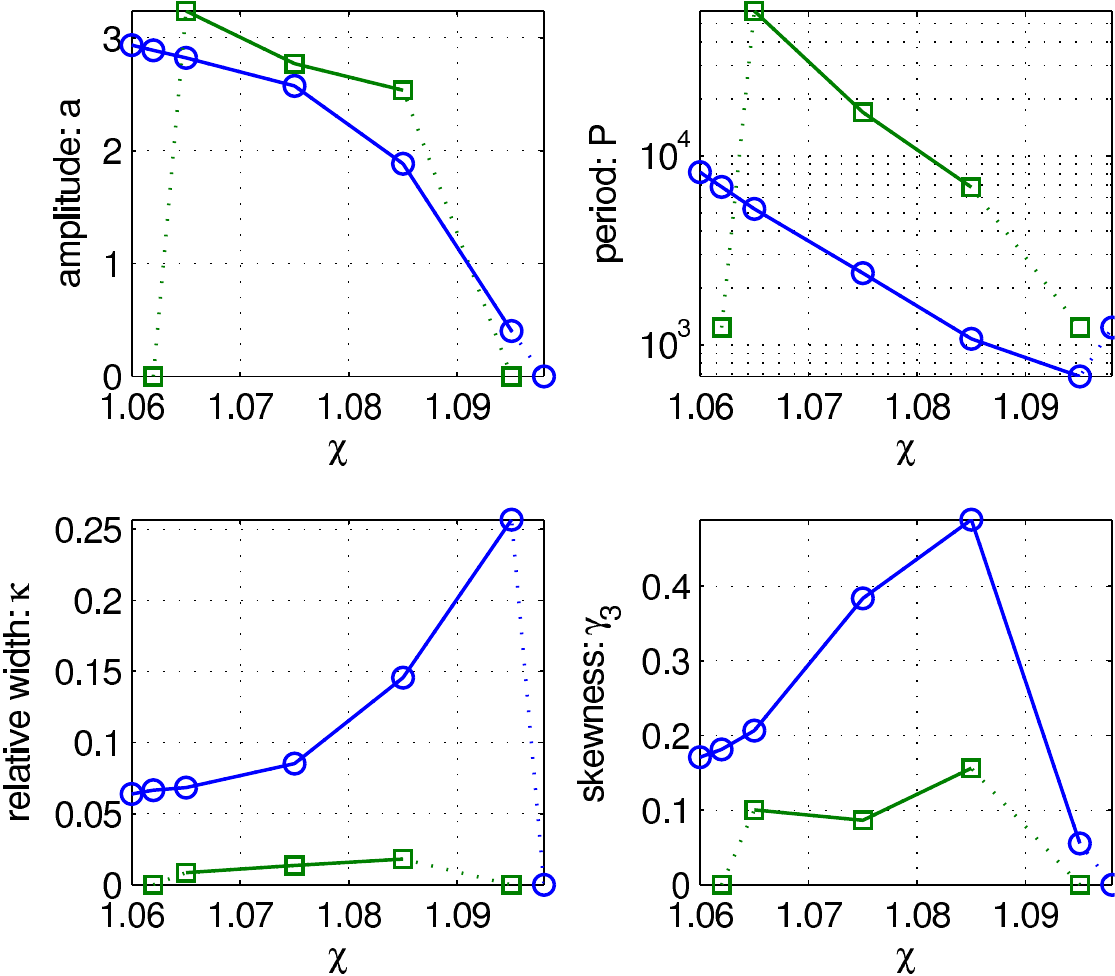}
\caption{(Color online) Measures characterising the regular lines as a function of
$\chi$.\cite{FAT12_noteplots} The blue lines with
$\circ$ symbols have fixed $\varOmega_0=7\times10^{-7}$, and the standard
concentration $\phi_0=0.41$, whilst the green lines with
$\scriptstyle\square$ symbols have the (standard) evaporation rate
$\varOmega_0=4.64\times10^{-7}$ and concentration
$\phi_0=0.31$. All other parameter values are the standard ones, defined at the end of
Sec.~\ref{sec:model}.
The blue lines corresponds to configurations in the central part of the
region (h) on Fig.~\ref{f:phasediag}.
Selected line patterns corresponding to these are shown in Fig.~\ref{f:chicutsprof}.
Note that the period $P$ seems to decrease exponentially with increasing
$\chi$ but we also note that for $\chi$ smaller than the standard value, we observe
stable double lines.}
\mylab{f:chicuts}
\end{figure}
\begin{figure}[tbh]
\centering
\includegraphics[width=0.9\hsize]{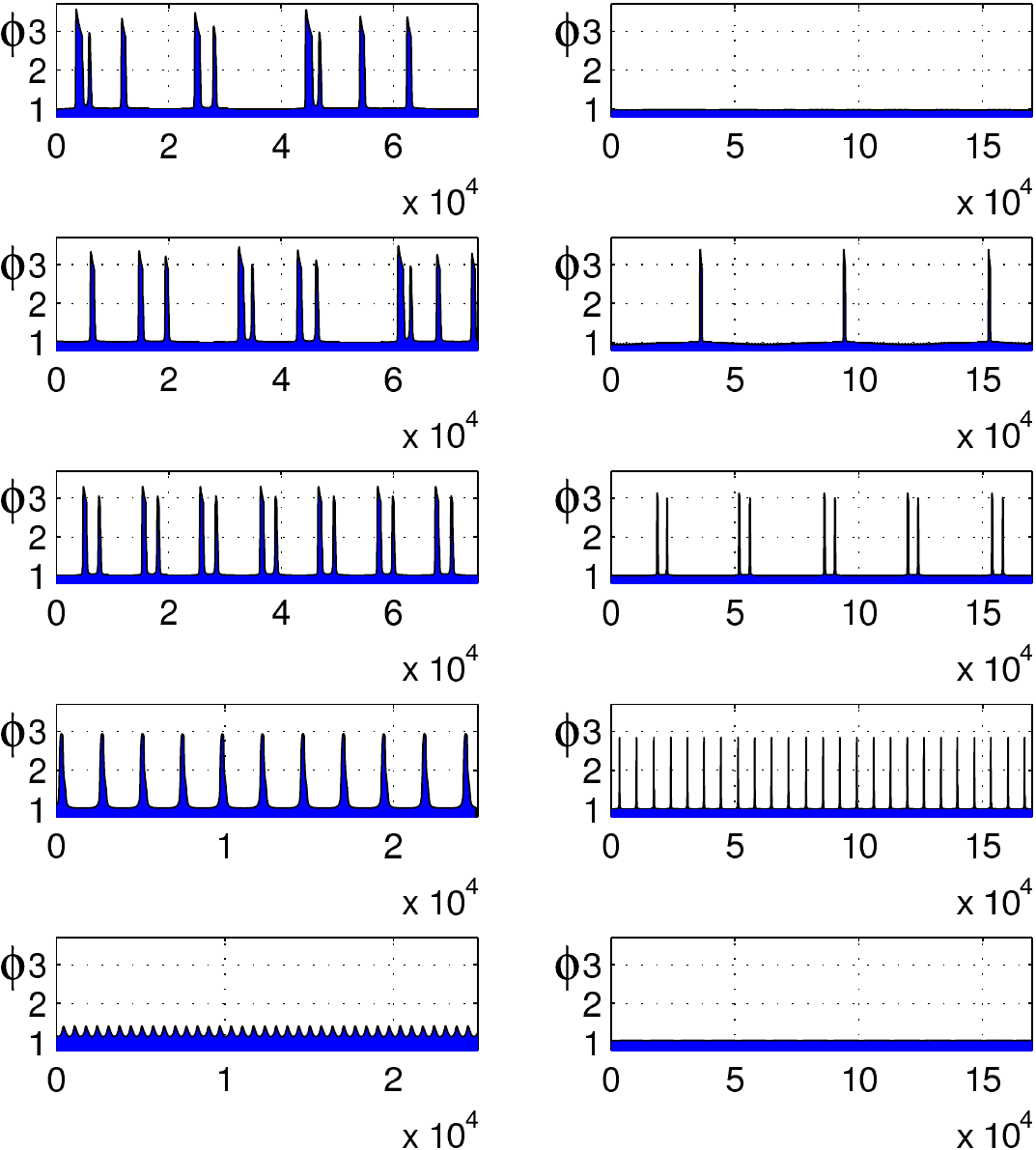}
\caption{(Color online) Morphology changes as $\chi$ is varied,
  corresponding to the results in Fig.~\ref{f:chicuts}.  Left column:
  the profiles correspond to the blue lines with $\circ$ symbols in
  Fig.~\ref{f:chicuts}; from top to bottom the results are for
  $\chi=1.06, 1.062, 1.065, 1.075$ and $1.095$.  Right column:
  corresponding to the green lines with $\scriptstyle\square$ symbols
  in Fig.~\ref{f:chicuts}; from top to bottom the results are for
  $\chi=1.062, 1.065, 1.075, 1.085$ and $1.095$.  The other parameters
  have the standard values.}
\mylab{f:chicutsprof}
\end{figure}
\begin{figure}[tbh]
\centering
\includegraphics[width=0.9\hsize]{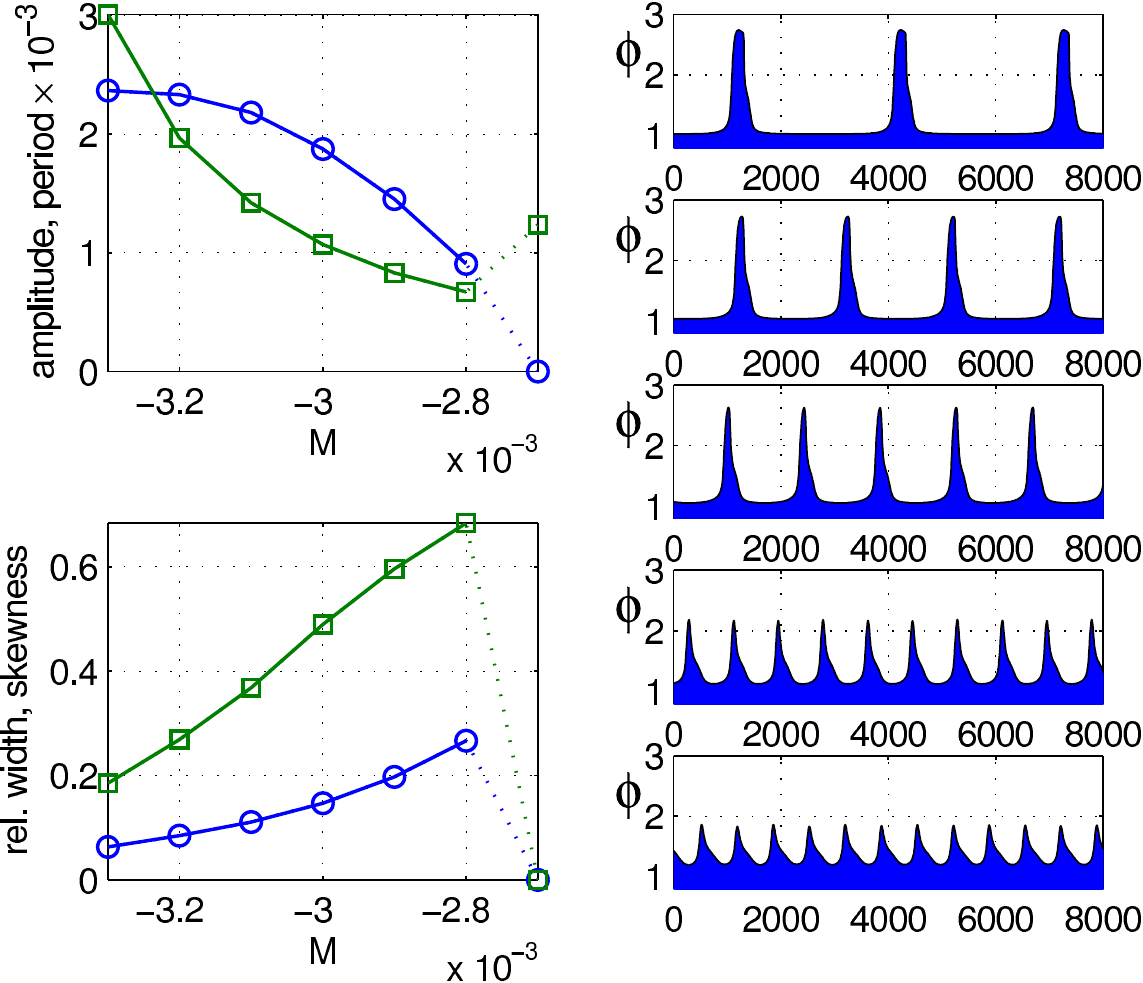}
\caption{(Color online) Left column: Measures characterising the
  regular line pattern as $M$ is varied.\cite{FAT12_noteplots} The
  upper panel shows both the amplitude (blue line with $\circ$
  symbols) and the period (green line with $\scriptstyle\square$).
  The Lower panel shows the relative width (blue line with $\circ$)
  and skewness (green line with $\scriptstyle\square$).  All the
  parameters except $M$ have the standard values as defined at the end
  of Sec.~\ref{sec:model}.  Right column: Corresponding profiles for
  $M=-(3.3, 3.2, 3.1, 2.9, 2.8)\times10^{-3}$, from top to bottom.}
\mylab{f:mcut}
\end{figure}
After having discussed the influence of solute diffusion in the
previous section, here we briefly consider the influence of wettability
and the chemical potential of the solvent
vapour. We start with the influence of the wettability
that is quantified in our model by the parameter $\chi$, contained in the polar
contribution to the disjoining pressure $\varPi=1/h^3-\exp(-\chi h)$.
The influence of changes in $\chi$ on $\varPi$ can be appreciated by
inspecting Fig.~\ref{f:disjp}. A decrease [increase] from our standard value 
$\chi=1.085$ results (at fixed chemical potential $M$)
in a decrease [increase] of the precursor film
thickness and of the thickness of the bulk film; an increase [decrease] of the
thickness where the front profile has its inflection point; 
a decrease [increase] of the wetting energy at the precursor film height, and in
consequence an increase [decrease] of the slope at the inflection
point, that is our measure of a `nonequilibrium contact angle'. It also leads
to larger [smaller] energy difference between the two stable heights, $h_1$
and $h_2$, so the evaporation will be stronger [weaker].

Corresponding results for the deposition patterns are displayed in
Fig.~\ref{f:chicuts} (line characteristics) and
Fig.~\ref{f:chicutsprof} (selected deposition profiles). The line
amplitude and period decrease monotonically with increasing $\chi$,
while the skewness first increases and then drops in value. The
period, amplitude and relative width behave in a similar manner as
when increasing $\phi_0$, see Sec.~\ref{sec:line-prop}. This indicates
that the transition towards a flat deposit at large $\chi$ is most
likely via a Hopf bifurcation, i.e.\ via scenario (ii) or (iv)
discussed in Section~\ref{sec:limit-period}. On decreasing $\chi$, we
find that the lines become more anharmonic and sharper and after a
period doubling become deposited in pairs. A further decrease results
in groups of multiple lines and in general to the deposition of a more
irregular pattern, i.e.\ the transition towards a flat deposit at
small $\chi$ is most likely via scenario (iii) introduced in
Section~\ref{sec:limit-period}.  Some corresponding profiles are
displayed in the left hand column of Fig.~\ref{f:chicutsprof}.

The main features of an evaporative dewetting front are influenced not
only by $\chi$, but also by the dimensionless chemical potential $M$
(cf.~Fig.~\ref{f:disjp}). A decrease in $M$ at fixed $\chi$ results in
a decrease of the precursor film thickness and also of the upper film thickness;
an increase of the thickness where the front profile has its
inflection point; a small increase of the wetting energy at precursor
film height but also in an increase of the energy difference between
the two stable heights, $h_1$ and $h_2$; and, as a consequence an
increase of the contact angle. A further decrease in $M$ leads to
faster evaporation in the front region, i.e., a larger front velocity,
and so the capillary ridge decreases and the 'nonequilibrium
contact angle' decreases. Many of these features can already be seen
in the case without solute.\cite{LGP02}

Results obtained for varying $M$ are presented in Fig.~\ref{f:mcut}.
A decrease in $M$, i.e.\ increase in $|M|$, results in an increase of
the line amplitude and period. We do
not consider $M<-0.0033$ as then no second stable film height exists, a
case not covered by our numerical set-up. Increasing $M$, the line
amplitude decreases towards zero, the period approaches a constant
value and the relative width and skewness of the lines
both increase. This is similar to the case of
increasing $\chi$ (compare the two columns of Fig.~\ref{f:chicuts}). 
We see this as an indication of a Hopf bifurcation as the most likely 
transition mechanism, i.e.\ scenario (ii) or (iv) of
Section~\ref{sec:limit-period}. If this is correct, the skewness should decrease
strongly over a small range of values of $M$.
Note finally, that it is difficult to separate the influences of the
parameters $\chi$ and $M$ as they both influence the stable film
heights, the contact angle, and whether a capillary ridge exists or
not.
%
\subsection{Influence of solution rheology}
\mylab{sec:rheology}

%
\begin{figure}[tbh]
\centering
\includegraphics[width=0.9\hsize]{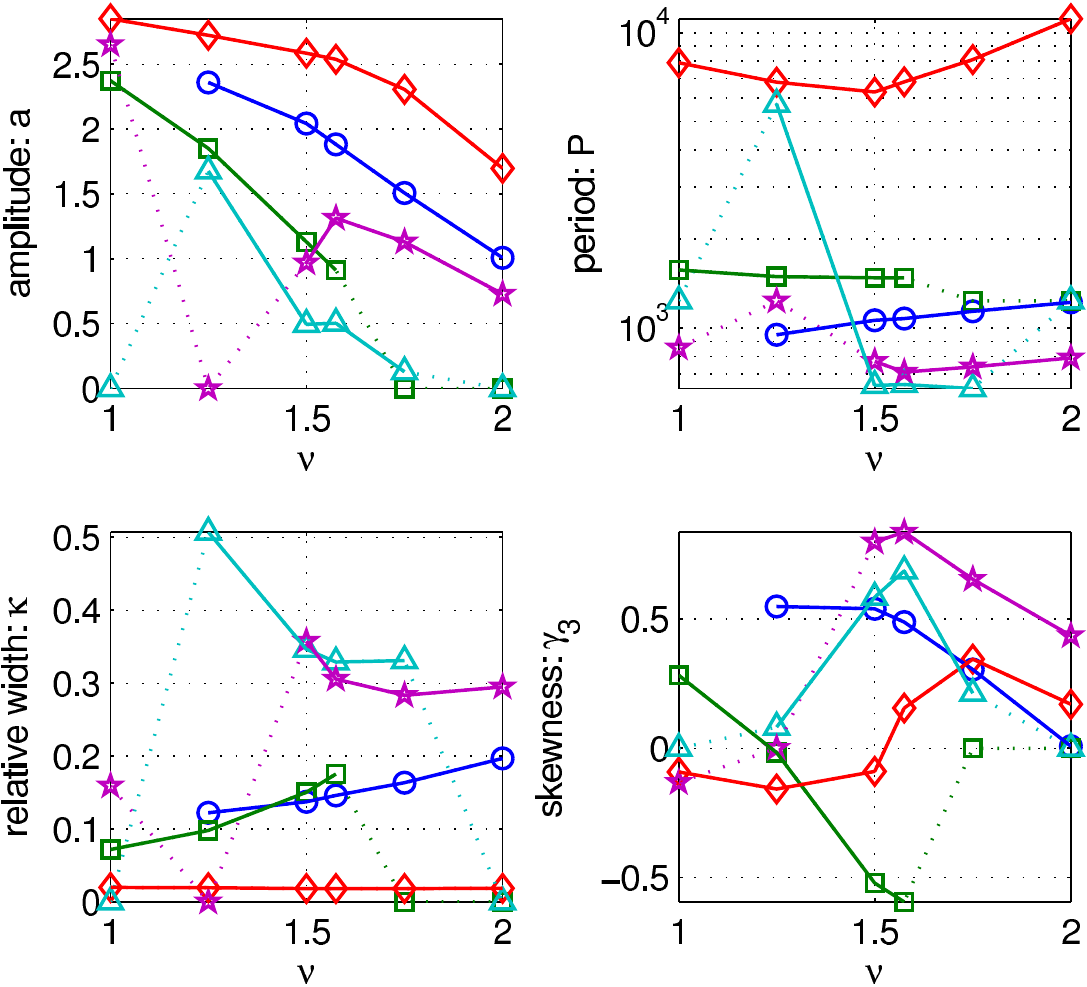}
\caption{(Color online) Measures characterising the line pattern as a
  function of the viscosity exponent $\nu$.\cite{FAT12_noteplots} The
  blue curve with $\circ$ symbols is for fixed standard
  $\varOmega_0=4.64\times10^{-7}$ and standard $\phi_0=0.41$; The
  green curve with $\scriptstyle\square$ is for
  $\varOmega_0=4.64\times10^{-6}$ and standard $\phi_0$; The red curve
  with $\scriptstyle\diamondsuit$ is for the standard value of
  $\varOmega_0$ and for $\phi_0=0.31$; The light blue line with
  $\scriptstyle\triangle$ is for the standard value of $\varOmega_0$
  and for $\phi_0=0.498$; The purple line with $\star$ is for
  $\varOmega_0=1.667\times10^{-7}$ and standard $\phi_0$. All the
  other parameter values are standard, as defined at the end of
  Sec.~\ref{sec:model}.  The blue line corresponds to configurations
  in the central part of the region (h) on Fig.~\ref{f:phasediag}
  whereas the other lines correspond to points close to its boundary.
  The corresponding line patterns are shown in
  Figs.~\ref{f:nucuts045-015-prof} and \ref{f:nucuts048-340-prof}.  }
\mylab{f:viscocut}
\end{figure}
\begin{figure}[tbh]
\centering
\includegraphics[width=0.9\hsize]{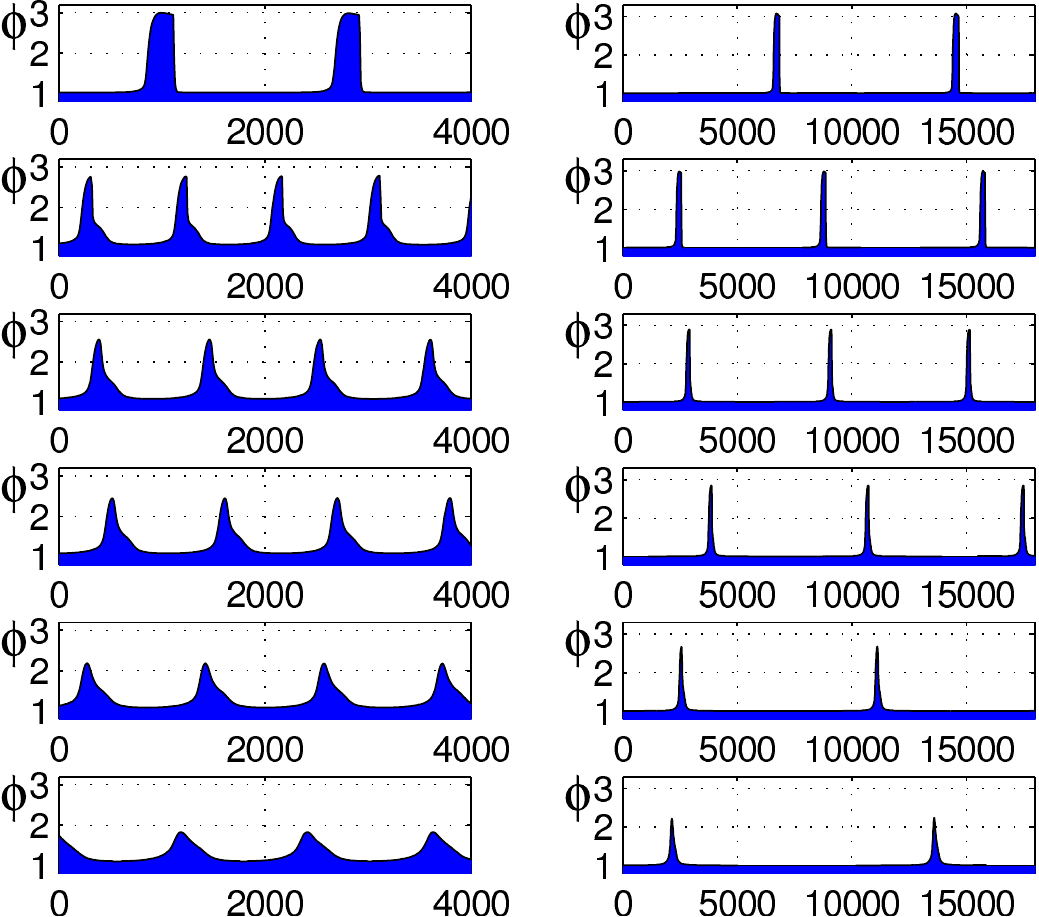}
\caption{(Color online) Morphology changes for various values of
the exponent $\nu$, corresponding to Fig.~\ref{f:viscocut}.
The left column corresponds to the blue lines with $\circ$ symbols
and the right column to the red lines with $\scriptstyle\diamondsuit$
in Fig.~\ref{f:viscocut}.
From top to bottom panels in each column correspond to
$\nu=1, 1.25, 1.5, 1.575, 1.75$ and  $2$.}
\mylab{f:nucuts045-015-prof}
\end{figure}
\begin{figure}[tbh]
\centering
\includegraphics[width=0.9\hsize]{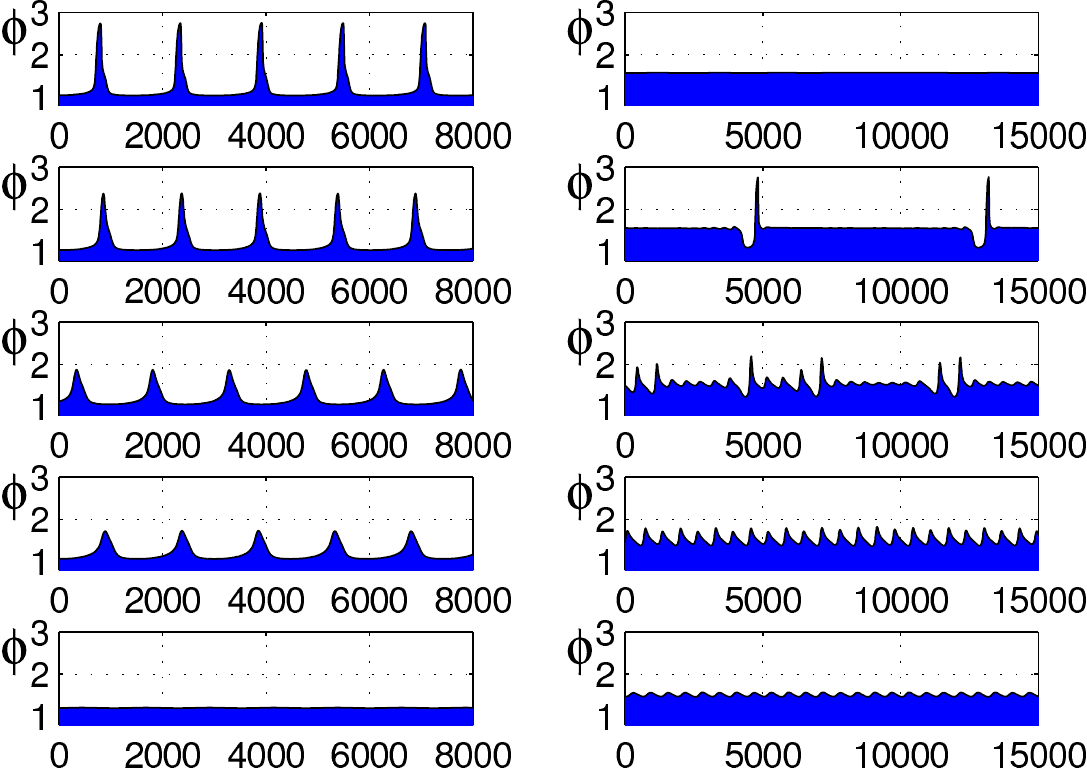}
\caption{(Color online) Morphology changes for various values of
the exponent $\nu$, corresponding to Fig.~\ref{f:viscocut}.
The left column corresponds to the green line with $\scriptstyle\square$ symbols
and the right column to the light blue line with $\scriptstyle\triangle$
in Fig.~\ref{f:viscocut}.
From top to bottom panels in each column correspond to
$\nu=1, 1.25, 1.5, 1.575$ and  $1.75$.}
\mylab{f:nucuts048-340-prof}
\end{figure}
In this final results section we discuss the influence of solution
rheology -- i.e.\ we present results for varying the exponent $\nu$ in
the Krieger--Dougherty law (\ref{e:eta(phi)}). In the thin film
literature the value $\nu=2$ is often used,\cite{CBH08,WCM03} and our
standard configuration employs the value $\nu = 1.575$ corresponding
to spherical colloidal particles as discussed above following
Eq.~(\ref{e:intrvisc}). However, some authors suggest that for some
systems, much lower values are required, in particular, in the context
of jamming transitions for attractive colloidal particles. There,
values as low as $\nu=0.13$ are mentioned.\cite{Trap01}

Here we consider moderate variations in the interval
$1\le\nu\le2$, and
investigate how the deposition patterns change for five selected
points in the parameter plane $(\varOmega_0,\phi_0)$. Three of the
configurations correspond to the crossing points of the four straight dashed
lines in Fig.~\ref{f:phasediag} that lie within the region (h) of
periodic lines [one is the standard case described at the end of
Sec.~\ref{sec:model} and the other two lie close to the boundary of
region (h)]. The other two selected points also lie close to the
boundary of the region (h). Results for measures characterising
the deposition
patterns are displayed in Fig.~\ref{f:viscocut} and selected corresponding
deposition profiles are displayed in
Figs.~\ref{f:nucuts045-015-prof} and \ref{f:nucuts048-340-prof}.

First, we consider the case of the standard set of parameter values,
that is in the central
part of the region (h) in Fig.~\ref{f:phasediag}. The measures are
shown with the symbols $\circ$ (blue curve) in Fig.~\ref{f:viscocut} and
selected corresponding profiles are displayed in the left column of
Fig.~\ref{f:nucuts045-015-prof}.
As $\nu$ decreases from the value $\nu =2$, the amplitude of the deposition lines
increases, the relative width decreases and the skewness departs from
zero indicating that the lines become more peaked and anharmonic at
lower $\nu$. Note that simulations tend to be more computationally
demanding at lower $\nu$ and e.g., the point at $\nu=1.0$ has a larger
error, because effects of the initial
transient are still present in the analysed data.

The other selected points, located close to the boundary of region (h)
in Fig.~\ref{f:phasediag} behave somewhat similarly, but with some
differences.  The symbols $\scriptstyle\diamondsuit$ (red line) in
Fig.~\ref{f:viscocut} correspond to lines with a large amplitude and a
long period; a selection of profiles for this case are displayed in
the right column of Fig.~\ref{f:nucuts045-015-prof}. Interestingly,
the period changes nonmonotonically, while the skewness decreases as
$\nu$ is decreased. Profiles that correspond to the symbols
$\scriptstyle\square$ (green line) in Fig.~\ref{f:viscocut} are
displayed in the left column of Fig.~\ref{f:nucuts048-340-prof}. The
lines have negative skewness and vary with $\nu$ in a manner similar
to the results for the standard set of parameter values. The symbols
$\scriptstyle\triangle$ (light blue line) and $\star$ (purple line)
correspond to the points close to the boundary of region (h) where
period doubling transitions are observed and results become quite
sensitive to the details of the numerical solution method.  The
computations are very sensitive for decreased $\nu$ values for the
case when $\phi_0=0.498$, with standard $\varOmega_0$, which
correspond to the symbols $\scriptstyle\triangle$ (light blue line) in
Fig.~\ref{f:viscocut}, with corresponding profiles displayed in the
right column of Fig.~\ref{f:nucuts048-340-prof}. There we see patterns
that switch between a depression--line pair and a flat layer, and
irregular intermittent patterns.

In general, the deposition of regular lines is robust with respect to a change in
$\nu$ and we observe them over a broad interval of parameter values. This is to
be expected from the experimental results, where periodic lines were seen for various
(non)spherical nanoparticles,\cite{XXL07} colloids \cite{ADN95} and
polymeric solutes.\cite{Byun08}  The general trend is for the regular line
patterns to become more sharply pronounced as the value
of $\nu$ becomes smaller,
i.e., the line amplitude increases and their relative width decreases.

We do not study in detail the onset of periodic deposits when varying
$\nu$. For very small values of $\nu$ the simulations are computationally too
demanding, since the lines are more sharply peaked. For some
configurations close to the boundary of region (h) in
Fig.~\ref{f:phasediag}, we observe irregular deposits and a
transition to a flat deposit indicating that region (h) may shrink
as $\nu$ is decreased -- cf.\ Fig.~\ref{f:nucuts048-340-prof}.  As the
value of $\nu$ is increased, the line amplitude decreases, and we expect that
for values of $\nu$ larger than 2, the region of periodic deposits will also
vanish.

\section{Discussion and conclusion}\label{sec:conc}

We have presented a thin film model for the close-to-equilibrium
self-organised deposition of material onto a smooth flat solid surface
from a receding three-phase contact line of a polymer solution or
nanoparticle suspension. The model consists of a pair of highly
non-linear coupled long-wave evolution equations for the film
thickness $h$ and the effective solute layer height $\hpfield=h\phi$,
where $\phi$ is the height-averaged scaled solute concentration. The
evolution of the film is driven by a number of terms in the equations
which describe several physical effects, including: (i) capillarity
through a Laplace (or curvature) pressure, (ii) wettability through a
Derjaguin (or disjoining) pressure, (iii) evaporation due to a
difference in the local solvent chemical potential and that of the
vapour, and (iv) forces due to gradients in the solute
concentration. The transport processes that are involved in the
dynamics correspond to convective and diffusive transport (both give a
conserved dynamics), and evaporation (a non-conserved dynamics). An
important ingredient in the model is the rheological property of the
solution/suspension, that leads to an arrest of the convective motion
at some critical solute concentration, e.g., at random close packing,
for a suspension of spheres that have no net attractive forces between
them and only interact via excluded volume repulsive interactions.  To
model these effects, we have employed the Krieger--Dougherty power law
\eqref{e:visc} for the viscosity, but we expect similar behaviour to
occur for other such laws. The model we use is related to some other
models used in the literature (e.g.,\cite{CBH08,WCM03}, as explained
in Section~\ref{sec:model}).

Numerically solving our model equations, we have investigated the
deposition of regular and irregular line patterns that has been
observed in numerous experiments utilising a wide range of materials
and experimental set-ups.  Assuming the front profiles only vary in
one spatial direction, we have found that regular line patterns are
formed over an extended region of the parameter space. Other
one-dimensional patterns that we have encountered, include the
transient deposition of a single or a finite number of lines, periodic
arrays of double lines, a periodic switching between depression--line
pairs and a flat layer, and irregular intermittent line patterns.

We believe that the model explains one of the basic mechanisms for the
formation of regular line patterns. They result from a self-organised
cycle of deposition-caused pinning-depinning events that is
experimentally often described as a `stick-slip'
motion.\cite{HXL06,Xu06,BDG10} The stick-slip motion is caused by the
highly nonlinear rheology: Evaporation leads to a rapid increase of
the solute concentration in the contact line region resulting in a
strong increase of the viscosity. This in turn eventually leads to
arrest of the convective motion of the receding front. However, the
front is not entirely stuck, it still moves due to evaporation, albeit
at a much slower speed. During this phase more material is deposited
resulting in a line deposit. After a sufficient amount of material is
deposited, the concentration in the contact line region decreases, the
front depins from the line deposit, and starts moving with much
greater speed. This appears to be a stick-slip motion. Thus, the
self-organisation of the patterned deposit results from a subtle
interplay of all three of the transport processes (convection,
diffusion and evaporation) and the stick-slip motion results from the
large difference in the timescales of convective and evaporative
dewetting.

After establishing this basic mechanism, we have performed a detailed
study to investigate the influence of solute concentration,
evaporation rate, diffusion, wettability, chemical potential, and
solution rheology on the patterns that are formed. In general, line
patterns emerge at intermediate values of both the evaporation rate
and the solute concentration, indicating that both are important
quantities in the deposition.  Furthermore, line patterns are
suppressed when the solute diffusivity is sufficiently high, because
when this occurs, the moving front is not able to collect the
solute. In the opposite limit, when solute diffusion is set to zero,
we find that this does not significantly affect the line patterns. In
this paper, we have probed the influence of wettability solely through
the parameter $\chi$ in the short-range part of the Derjaguin
pressure. Our results indicate that an increase in the wettability may
lead to a suppression of the patterns. Note, however, that our
investigations of this aspect are not exhaustive. We have not fully
studied this aspect here, because our present numerical set-up does
not allow for the study of a completely wetting solution. Such a study
will be pursued in the future, employing a modified set-up for the
case of an active geometry. The influence of the solute on the system
rheology is quite notable but not easy to categorise: decreasing the
exponent $\nu$ in the power law relation between the viscosity (and
diffusivity) and the solute concentration, increases the amplitude of
the lines for all the values of the other parameters that we have
tested. However, the influence on the period of the line pattern
depends on the other parameters: it can increase, decrease or even
behave in a non-monotonic manner with decreasing $\nu$.

In general, these findings are in agreement with experimental
results.\cite{LiGr05,XXL07,BDG09,BDG10,KYY11} However, to our
knowledge, no experimental study has systematically mapped out the
regions where the various line patterns exist (analogue to our
Fig.~\ref{f:phasediag}), or presented a detailed (quantitative)
analysis of how the line pattern characteristics depend on the system
control parameters. Therefore, before we make a comparison of the
typical dependencies found in the experiments and our model, we first
discuss the typical length and time scales in the deposition process.

The scales we have used [cf.~Eq.~(\ref{e:scales})] for the time,
$x$-coordinate, and film thickness are $\tau=3\eta_0\gamma/\delta
|\tilde{S}^P|^2$, $\alpha=(\delta\gamma/|\tilde{S}^P|)^{1/2}$, and
$\delta=(A/6\pi|\tilde{S}^P|)^{1/3}$, respectively. For our standard
parameter values (see Sec.~\ref{sec:model}), we find
$\tau=2.3\times10^{-8}$s, $\alpha=2.9$nm, and
$\delta=0.5$nm,\cite{LGP02} i.e., the line pattern in
Fig.~\ref{f:timesnaps} has a period of $\lambda\approx3\mu$m and the
mean deposition velocity is $v\approx1\mu$m/s. For Fig.~\ref{f:xtph}
we have $\lambda\approx3\mu$m and $v\approx0.1\mu$m/s, whereas the
periods of the patterns displayed in Fig.~\ref{f:vertcut} range from
about $1.7\mu$m to about $52\mu$m. These values are similar to those
of the experiments by \citet{XXL07}. Their Fig.~1 shows line patterns
with $3$--$5\mu$m periods.  Fig.~2 of \citet{Xu06} gives line
  periods between $4$ and $20\mu$m, a range similar to results by
  \citet{Hong08}.
Some characteristics that we find are
rather similar to experimental findings by \citet{BDG09}: The contact
line velocity is similar to the $0.8\mu$m/s imposed in their Fig.~1;
the spatial periodicity of the pattern in their Fig.~2 is $210\mu$m
and the contact line speed is $6.2\mu$m/s; In general, their typical
imposed velocities are $1$--$10\mu$m/s.

It is hard to compare with experimental results in the literature how
pattern properties typically depend on the control parameters, as most
works do nearly not discuss these. Some authors discuss dependencies
that we are not able to measure in our simulations such as, e.g., the
pinning force.\cite{BDG10} A decrease in the ring spacing with
decreasing distance from the probe center is sometimes described in
studies of evaporative dewetting with the meniscus technique (see
Figs.~1 and 3 of \citet{KYY11}, Figs.~2 and 3 of
\citet{LiGr05,HXL07}), and Fig.~2 of \citet{Xu06}. In all mentioned
cases this may be interpreted as observing a decrease in the period
with increasing solute concentration and/or decreasing evaporation
rate. Both of these possibilities are in accord with our results in
Figs.~\ref{f:vertcut} and \ref{f:horizcut}, respectively.  However, a
decrease in the period with decreasing concentration is found by
\citet{XXL07}, a result also reported by \citet{Xu06} when comparing
evaporating polymer solutions of different initial concentrations.
\citet{ADN95} reports a non-monotonic dependence of the line density on
the initial solute concentration.  Decreasing the concentration leads
to the line density slowly increasing before abruptly decreasing at
the edge of the patterning region. This is not unlike our findings --
see Fig.~\ref{f:vertcut}. A quantitative comparison is difficult, as
they report observing about 1--8 rings produced with a small
evaporating droplet. Related problems obstruct a detailed comparison
with many of the experiments performed with circular drops, which have
radial geometries, that always imply a drift in several system
parameters in the radial direction.  The planar geometry used by
\citet{BDG09,BDG10} allows for a more natural comparison. However,
there detailed results are given for the dependence of the pinning
force on experimental parameters whereas no such results are given for
the period of the line patterns. What is remarkable, is the agreement
for the morphology of individual lines: \citet{BDG09} shows a rather
asymmetric line profile with a negative skewness, i.e., the long tail
points away from the receding contact line, similar to the profiles
displayed in last two panels of the left column of our
Fig.~\ref{f:horizcutprof}. Such agreement also exists with results for
some unpublished experiments with nanoparticle suspensions
\cite{philnote} where line profiles having a negative skewness and
also a positive skewness were found, depending on the solvent that was
used, i.e., the long tail can point either away or towards the
receding contact line, respectively, similar to the transition we see
that is illustrated in Figs.~\ref{f:horizcut}
and~\ref{f:horizcutprof}. 

Within the limits of our numerical approach, particular care was taken
in the analysis of the onset of the formation of lines when changing
the various system control parameters. We have observed three main
types of transition at the border of the region of periodic lines: (i)
the line period diverges while the line amplitude converges to a
finite value, (ii) the line period doubles, followed by a region of
intermittent patterns, (iii) the period approaches a finite value
while the amplitude first decreases before suddenly jumping to
zero. We have also proposed a hypothetical fourth scenario -- (iv) a
case where the line period approaches a finite value while the line
amplitude approaches zero.  To relate these findings to dynamical
systems or bifurcation theory, it is helpful to consider a stationary
moving front that deposits a homogeneous layer as a steady state in a
comoving reference frame. At the onset of the line formation this
steady state becomes unstable and gives way to a time-periodic
solution (in the frame that moves with the mean speed of the
front). In this context, the above mentioned transitions can be
identified as (i) an infinite period bifurcation (either homoclinic or
Saddle-Node Infinite PERiod (SNIPER)), (iii) a subcritical Hopf, and
(iv) a supercritical Hopf bifurcation, respectively. Case (ii) is more
complicated and resembles the period doubling and intermittency route
to chaos when approaching the transition from the side of the periodic
deposits. Approaching the transition from the side of the flat
deposit, one finds that the length of the transient seems to diverge,
which is similar to the transition from an excitable medium to one
that exhibits sustained oscillations.  In particular, the homoclinic
bifurcation and the subcritical Hopf bifurcation have recently been
identified to be responsible for the transition to stripe-like
deposition patterns in the Langmuir--Blodgett transfer of a surfactant
layer onto a moving plate, where patterning is due to
substrate-mediated condensation of the surfactant.\cite{KGFC10,KGFT12}

To put these transitions in a wider context, we would like to
point out that the onset of stripe formation can also be seen as a
depinning transition: When a homogeneous layer is deposited from a
stationary moving front, the concentration profile of the solute is
pinned to this front. However, this is not the case when lines are
deposited. In this case, the lines are depinned from the average-velocity
comoving frame -- i.e.\ they stay behind the moving front. Note however that
many details of the present transition are unlike the depinning via a
homoclinic or SNIPER bifurcation described by \citet{BKHT11}
who study how
a drop depins from a substrate heterogeneity under the influence of an
external driving force, leading to the entire drop sliding along the
substrate. Here, after depinning, the concentration profile at the
moving front does not move in its entirety away from the
front. Instead, only a part of it (a line) starts to move - a process
that is then repeated periodically.  Note that for other parameter
values for the driven drop system, the depinning transition may also be
due to a Hopf or a SNIPER bifurcation.\cite{ThKn06} The present
system appears to be more complicated, as the three main types of
transition can be altered by interactions with period-doubling
bifurcations and/or intermittent behaviour.  We believe it is
necessary for reduced models to be developed in order
to investigate these issues in detail.  Although our understanding
of the onset of the patterning is still not complete, we hope that
this aspect of our investigation may prove useful in the
classification of future experimental results for the onset
of the formation of deposition patterns.

In the present work we have limited our attention to a one-dimensional
geometry, i.e., we have studied deposition from a receding straight
front with an imposed transversal stability of the front.  A fully
two-dimensional treatment is worth pursuing (but computationally not
feasible using the numerical approach taken here -- see
appendix~\ref{sec:numerics}) and highly relevant, since it is well
known that receding contact lines may be transversally unstable, in
particular, if the receding liquid is a suspension or solution with a
volatile solvent.\cite{KGMS99,GRDK02,XSDA07,XXL07,Paul08}
Experimental results on line deposition indicate that such transversal
instabilities are likely to occur on the left and/or upper border of
the region of line patterns in the phase diagram we have presented in
Fig.~\ref{f:phasediag}. In the experiments by \citet{HXL07} and
\citet{XXL07} transversal instabilities occur at very small velocities
of the receding contact line, i.e., when the evaporation is slow
and/or the solute concentration is high.  In our model, we have
observed that under these conditions the deposition process shows
period doubling and results become very sensitive to numerical
details, which suggests the system may be transversally unstable, in
agreement with these experiments.
In the system studied by \cite{YaSh05}, different types of
transversal instabilities occur at low and high solute
concentrations. However, this system represents an example of an
active geometry where the sliding velocity of an upper plate controls
the patterns that are formed. \citet{KGMS99} states that a
Plateau--Rayleigh instability of the rim at the moving front is
responsible for the fingering instability, an observation which is in
line with arguments made for receding fronts of non-volatile
\cite{BrRe92} and volatile \cite{LGP02} pure liquids. However, the
name ``Plateau--Rayleigh instability'' might be misleading in the case
of a moving front of a solution or suspension since (i) the motion of
the liquid rim itself may change the character of the
instability,\cite{ThKn03} (ii) partial wettability may trigger an
instability of a receding front,\cite{ThKn03} and, most importantly,
(iii) since the solute concentration may change along the front, new
classes of instabilities are possible, which are related to mobility
contrasts and local phase separation \cite{RAT11} or solutal Marangoni
effects. These possibilities warrant future studies employing either
geometries that allow for a more efficient numerical approach and/or
the development of reduced models in a similar spirit to the model for
Langmuir--Blodgett transfer of a surfactant layer developed by
\citet{KGFT12}.  A thin film model for this process predicts a
transition from stripes that are perpendicular to the receding front,
to parallel, as the transfer velocity is increased.\cite{KGFC10} We
expect a similar behaviour in our situation if our resulting front
speed is taken as a control parameter.

Our work may also be related to different approaches that describe
other aspects of the evaporative dewetting of solutions and
suspensions.\cite{Thie09} In particular, the role of diffusion of the
solute in suppressing the line deposition that we have described
above, is very similar to its role in the suppression of fingering
instabilities that occur at receding evaporative dewetting fronts in
ultrathin films of nanoparticle suspensions, that were experimentally
observed in by \citet{Paul08}, and subsequently modelled by a kinetic
Monte Carlo model \cite{Vanc08} and dynamical density functional
theory.\cite{ART10,RAT11} These non-hydrodynamic models are not able
to describe mesoscopic hydrodynamics, i.e., the transport of momentum,
but do allow one to incorporate the various interactions of solute and
solvent, i.e., of the complete thermodynamics of the system. These
interactions can, if strong enough, result in a local phase separation
of solute and solvent close to the moving contact line and in this
manner trigger the front instability.\cite{RAT11} The present
hydrodynamic model is not able to describe this, but it does have the
advantage that it incorporates basic wettability effects, i.e., the
substrate-film interactions. A future avenue for improvements of our
mesoscopic hydrodynamic model is to incorporate solute-solute and
solute-solvent interaction. A related thin film model for a layer of a
decomposing non-volatile binary mixture has recently been derived by
\citet{NaTh10} as a long-wave approximation to
model-H.\cite{AMW98,TMF07,BFT12} A general approach of deriving such
thin film evolution equations in the context of nonequilibrium
thermodynamics (taking the form of a gradient dynamics based on a free
energy functional), was recently proposed.\cite{Thie11b,TAP12} This
then naturally allows one to incorporate effects like solute-dependent
wettability and capillarity (including solutal Marangoni effects), and
the dependence of evaporation on the osmotic pressure. In this way,
the present model can readily be extended and may be employed to
assess the influence of additional effects on the basic mechanism that
we have describe here. This should be done for passive and active
geometries alike.

We acknowledge support by the EU via the ITN MULTIFLOW
(PITN-GA-2008-214919).

\appendix

%
%

\section{Numerical approach}\label{sec:numerics}

\subsection{Discretization and numerical scheme}
\mylab{subs:numscheme}

In order to calculate the deposit profiles $\hpfield(x,t)$
corresponding to evaporative dewetting of a thin liquid film of height
$h(x,t)$ of a solution or a suspension we perform time
simulations of the model equations
(\ref{e:tfeqh}) and (\ref{e:tfeqhp}), in one spatial dimension.
The initial film profile in the form
\begin{equation}
h(x,0) = h_1+ (h_2-h_1)H(x-x_0),
\end{equation}
which corresponds to the sharp front, located at $x=x_0$,
that subsequently moves in positive-$x$ direction, as the volatile
component evaporates. The heights $h_1$ and $h_2$ denote the
precursor and upper
stable film thicknesses, respectively, see Fig.~\ref{f:disjp}; and $H(x)$ is
the Heaviside step function.  The initial concentration profile is
constant, $\phi(x,0)=h(x,0)/\hpfield(x,0)=\phi_0$. Furthermore, to
approximate the semi-infinite drying thin film on a finite
computational domain, particular boundary conditions (BCs) are 
employed, as discussed later in the Section~\ref{subs:bc}.
The computational domain $x\in[0,L]$ is discretized employing an
inhomogeneous grid adapted to resolve the details in our unknown fields
$h(x,t)$ and $\hpfield(x,t)$.
For time stepping, we used finite differences in space and two
variable-order variable-step time-stepping methods: BDF (backward
difference formulae) as a default solver and, the much slower, Adams
method in those time intervals where there was no convergence achieved
by the default solver.

During the first few time-steps the front, originally at $x=x_0\ll L$,
is quickly smoothed by surface tension, develops a capillary
ridge and starts to move in the positive-$x$ direction, thus the
initial ``dry patch" starts to grow.
The concentration field $\phi(x,t)$ soon develops a
peak at the dewetting front that decays to the right. We need to
resolve both the long decaying right hand tail of the capillary ridge
$h(x,t)$ during the convection-dominated stage and also the
decay of the field $\phi(x,t)$.  Thus the length of the domain, $L$,
should be larger than the decay lengths of $h(x)$ and $\phi(x,t)$,
i.e.\ much larger than the lateral length scale $\alpha$.  On the other
hand, we do not need to keep the dried region in our discretized
domain where both $h(x,t)$ and $\phi(x,t)$ no longer change in
time. Thus each time the front reaches some prescribed position
$x =\xshift$, which is of the order of $\alpha$, the simulation is
automatically paused, and the computational domain is shifted (the
$x$-coordinate also is shifted) to the
right by the distance $\Pshift=\xshift-\xrst$ so that when the simulation is
restarted the front becomes positioned at a small distance $\xrst$ from the
shifted boundary $x=0$. We call this procedure {\em shifting}.
The region through which the dewetting front travels up to the
point $\xshift$ and the subsequent region where the capillary ridge needs
to be resolved have the finest and equidistant grid-spacing. Then with
increasing $x$ the grid-spacing is gradually increased in order to obtain
a more efficient numerical model.
We always use the shift period $\Pshift$, that is an integer multiple of
the finest grid-spacing so there is no interpolation involved in the
region resolving the processes at the contact line and the capillary ridge.
Several tens or hundreds of shifts usually happen during a typical
simulation that result in the deposition of 10--100 lines.
Without shifting one would need to employ
a much larger $L$ and the region with finest grid-spacing would also
need to be much larger than that we used. This would lead to a
numerically intractable problem.

After some performance and accuracy testing we determined all the
tuneable parameters of the numerics and then kept them fixed for all
simulations published in this paper and also in the previous
one.\cite{FAT11} The values of the adjustable parameters are: finest
grid-spacing $\Delta x=8.5849$, the period of shift $\Pshift=144\Delta
x$, and the domain size $L=19230.2465$. If we used the equidistant
grid over the whole domain, the total number of grid-points would be
2241. However, because with our adapted grid the number of grid-points
is 636.  It could easily be 10 times more than 2241 if we did not use
the shifting procedure.

\subsection{Boundary conditions}\mylab{subs:bc}

A semi-infinitely extended thin film evaporatively dewetting and depositing
lines has a
front that moves with an intrinsic velocity that can vary in time.
The numerical solution method involves finite differences and so we
are restricted to a finite domain. This does not cause a problem on the dried
(left) side of the solution domain, where the unknown fields $h(x,t)$ and
$\hpfield(x,t)$ essentially do not evolve in time. Thus the left BCs
at $x=0$ are simply symmetry
conditions for $h(x,t)$, i.e.\
$\partial_xh(x,t)|_{x=0}=\partial_{xxx}h(x,t)|_{x=0} = 0$; and
for $\phi(x,t)$ the fictitious additional grid point out of the
domain is fixed at the value corresponding to the last value shifted out.
These BCs are employed because when the region at the left boundary
is dried and $(\phi\geq1)$, then $h(x,t)=h_1$ and the concentration is
jammed, $\phi(x,t)=\phi(x)$.

However, one needs to be more careful regarding the right BC at $x=L$.
When we assume that the evaporation process is very slow we
can treat the dynamics of the fields $h(x,t)$ and $\phi(x,t)$ as
quasistationary, i.e.\ the front moves with a velocity that varies very
slowly in time.  Then we can reduce our Eqs.~(\ref{e:tfeqh}) and
(\ref{e:tfeqhp})
to the approximate ODEs. After linearizing about the constant film
height $h(x)=h_2$, we can solve the problem in the region far from the
dewetting front to the right (wet) side and find stationary exponentially
decaying profiles for $h(x)$ and $\phi(x)$. These profiles are used to prescribe the
boundary conditions on the right side of our solution domain, $x=L$, which
read
\begin{align}
\partial_xh(x,t) &= \lambda(h(x,t)-h_2),\label{e:decay1}\\
\partial_{xxx}h(x,t) &= \lambda^3(h(x,t)-h_2),\label{e:decay2}\\
\partial_x\phi(x,t) &= \lambda(\phi(x,t)-\phi_0),\label{e:decay3}
\end{align}
where $\phi_0$ is the initial concentration of the nanoparticles.
Initially, at the right boundary $h(L,0) = h_2$ and $\phi(L,0)=\phi_0$
and these boundary conditions reduce to
setting all odd derivatives to zero.  This is the same as in no-flux boundary
conditions (see below Eqs.~(\ref{e:symbc})).  During the simulation always
$h(L,t)>h_2$ and $\phi(L,t)>\phi_0$.  The decay rate, $\lambda$, is
\begin{equation}
\lambda = \sqrt{ \frac{\varOmega_0\eta(\phi_0)}{h_2^3}\left(
 1 + \frac{\phi_0[h(L,t) - h_2]}{h_2[\phi(L,t)-\phi_0]}\right) }
\end{equation}
so that, apart from the constant parameters, the decay rate nonlinearly
depends on the values $h(L,t)$ and $\phi(L,t)$.  This means that the
decay rate and the velocity of the moving front of a dilute suspension are
different from the ones of a more dense suspension.

Note, that the effect of an increasing average deposit thickness when
decreasing the evaporation rate (as seen, in particular, in the
calculations at fixed $\phi_0=0.31$ and varied $\varOmega_0$ in
Fig.~\ref{f:horizcut}) is due to a small systematic artefact connected
to the simple approximations employed to derive the BCs
(\ref{e:decay1})--(\ref{e:decay3}) at the right
boundary at $x=L$.  These BCs correspond to a situation with continuous
systematic supply of additional solution during the convection-dominant
regime and can be physically realised as some finite size effect.
To further minimise this effect we used a large computational domain size $L$.

Alternatively, we also performed several simulations with no-flux BCs:
\begin{equation}
\partial_xh(x,t)|_{x=L} = \partial_{xxx}h(x,t)|_{x=L}
 = \partial_x\phi(x,t)|_{x=L} = 0.
\label{e:symbc}
\end{equation}
They prescribe zero flux through the right boundary and thus correspond to
a different limit than the BCs that involve $\lambda$. Comparing results obtained
with both types of right BCs shows no qualitative difference;
see the comparison given in the next section.

\subsection{Remark on the robustness of our results}
\mylab{subs:robustness}

In order to better utilise the computational resources, a number of
numerical tricks are employed: shifting, locally refined grid, and
boundary conditions for decaying fields $h(x,t)$ and $\phi(x,t)$.
Great care is taken to avoid any artefacts in the deposition profiles
due to these tricks.  As a check, the computed pattern are in some
cases verified by recomputing them with each of these refinements
separately: (a) with different shift length, $P_\mathrm{shift}$, that
is further away from a possible resonance with the period of the
pattern (180/144 of the original $P_\mathrm{shift}$); (b) with refined
grid, where in the region with equidistant finest grid-spacing we
simply add one extra node between each two nodes of the original grid
and accordingly we modify the rest of the grid without testing whether
the grid-spacing outside the equidistant zone is optimal; (c) using
the no-flux boundary conditions at the right boundary (see
Appendix~\ref{subs:bc}); (d) larger $L$ (40/35 of the original $L$).
For these tests, we select three of the characteristic configurations
used throughout this paper e.g.\ in Figs.~\ref{f:xtph} and
\ref{f:xtphp}, namely: the standard case
$(\varOmega_0,\phi_0)=(4.64\times10^{-7},0.41)$, the case with
significant negative skewness
$(\varOmega_0,\phi_0)=(4.64\times10^{-6}, 0.41)$ and the
large-amplitude long-period case $(\varOmega_0,\phi_0)
=(4.64\times10^{-7},0.31)$.  It is found that the resulting patterns
are qualitatively independent of these refinements, however we observe
some differences in the values of the computed measures, as follows:
\begin{itemize}
\item
The smallest errors are obtained for the test using a different shift length: an error
under 1.2\%
for the amplitude and period and under 2.6\% for the relative width and skewness,
except for the large-amplitude long-period case where the error in the period
is 14\%, in the relative width is 17.6\% and in the skewness is 10.6\%.
\item
The errors from tests on the refined grid are: under 6\% for the amplitude and under 15.8\%
for the period, except for the large-amplitude long-period case, where the error in
the period is 55\%. The errors for the relative width and the skewness are somewhat
larger.
\item
The remaining tests (using symmetric BC and larger $L$) give in some
cases larger errors in the measures but changes in the BC or $L$ may be thought
of as varying the finite size effect.
\end{itemize}
Thus, the numerical parameters: $P_\mathrm{shift}$, the grid-spacing,
BCs with $\lambda$
and $L$ that were used for all presented simulations were fine
tuned, so as to give the most reliable results possible, subject to the
computational resources available.

\section{Measures for periodic patterns}\mylab{sec:measures}

To study the changes in the morphology of the deposited pattern of periodic lines,
$\hpfield(x)$, we computed the following measures in a selected
spatial region containing the fully developed periodic pattern:
The basic measures are the amplitude, $a$, and the period, $P$,
\begin{align}
a &= \frac{1}{n}\sum_{i=1}^n a_i =\frac{1}{n}\sum_{i=1}^n (\hpfield(y_i) -
\hpitilde),\label{e:amplit}\\
P &= \frac{1}{n}\sum_{i=1}^n P_i =\frac{1}{n}\sum_{i=1}^n (x_{i+1} -x_i),
\end{align}
where $x_i$ [$y_i$] is the position of $i$th subsequent minimum [maximum]
of $\hpfield(x)$, respectively;
$\hpitilde = \frac{1}{2}[\hpfield(x_i) + \hpfield(x_{i+1})]$ is the minimum
height of the $i$th line (peak). By $i$th line we mean the portion
of $\hpfield(x)$ that is defined on the interval $[x_i,x_{i+1}]$.
The remaining measures are integral ones
\begin{align}
A_i &= \int_{x_i}^{x_{i+1}}\hpihat(x)\diff x,\\
\bar x_i &= \frac{1}{A_i}\int_{x_i}^{x_{i+1}}\hpihat(x)x\diff x,\\
\sigma_i &=
  \sqrt{\frac{1}{A_i}\int_{x_i}^{x_{i+1}}\hpihat(x)(x-\bar x_i)^2\diff x},\\
\kappa_i &= 2\sigma_i/P_i,\\
\gamma_{3i} &= \frac{1}{A_i\sigma_i^3}\int_{x_i}^{x_{i+1}}\hpihat(x)(x-\bar x_i)^3\diff x,\\
\gamma_{4i} &= \frac{1}{A_i\sigma_i^4}\int_{x_i}^{x_{i+1}}\hpihat(x)(x-\bar x_i)^4\diff x,\\
\hpioverb &= \frac{1}{P_i}\int_{x_i}^{x_{i+1}}\hpfield(x)\diff x.
\end{align}
The given quantities for the $i$th peak are: $A_i$ is the area of the
excess nanoparticle film, $\bar x_i$ is the peak's centroid, $\sigma_i$ is its
standard deviation, $\kappa_i$ is its normalised width, $\gamma_{3i}$ is its
skewness, $\gamma_{4i}$ is its kurtosis, and $\hpioverb$ is its
average height. The final measures (that are used throughout this paper)
$\tilde\hpfield$, $A$, $\bar x$, $\sigma$, 
$\kappa$, $\gamma_3$, $\gamma_4$, and $\overline{\hpfield}$ are obtained by
averaging over $n$
subsequent lines as is shown in the case of amplitude in
Eq.~(\ref{e:amplit}).

Note, that higher order moments such as $\sigma$, $\gamma_3$,
$\gamma_4$ give too much weight to the tail of a peak, especially in
the case of large periods. For these, small numerical variations in
the tail of a peak profile are greatly amplified in the value.  Since
our measures shall focus on the peak characteristics and not on the
shallow valleys between them, we cut off the shallow part of the
tails, i.e.\ we set
\begin{equation}
\hpihat(x) = \max\left[\hpfield(x) - \hpitilde - ca_i, 0\right]
\end{equation}
where the cutoff parameter, $c=0.05$, is used in computing the measures
displayed in the main text.

\footnotesize{
\bibliographystyle{rsc} 
\bibliography{FAT12} 

\begin{mcitethebibliography}{109}
\bibitem[FAT()]{FAT12_materials}
Examples of combinations of materials that are used are: charge-stabilized
  polystyrene (PS) microspheres in water on glass \cite{Deeg97} and on
  mica;\cite{Deeg00} Rings were also found using metal, polyethylene, roughened
  Teflon, ceramic, and silicon substrates with acetone, methanol, toluene, and
  ethanol as solvents.\cite{Deeg00b} As solutes they used sugar and dye
  molecules, 10$\mu$m PS spheres; 144nm PS particles in water on
  glass;\cite{ADN95} 15nm silica particles in water on glass (partially
  wetting, $\theta_\mathrm{eq}=40$);\cite{PaAl96,PaAl97} 4nm CdS particles in
  pyridine and 6nm CdSe/CdS core-shell particles in water on glass
  \cite{MDSY99}; 90nm silica particles in pH-adjusted water on
  glass;\cite{RDLL06} chloroform solutions of PS and poly(3-hexylthiophene)
  (PHT) on glass;\cite{YaSh05} benzene and chloroform solutions of PS and
  chloroform solution of a polyion complex on glass or mica.\cite{KGMS99}\relax
\mciteBstWouldAddEndPuncttrue
\mciteSetBstMidEndSepPunct{\mcitedefaultmidpunct}
{\mcitedefaultendpunct}{\mcitedefaultseppunct}\relax
\EndOfBibitem
\end{mcitethebibliography}
}

\end{document}